\documentclass[twocolumn]{aastex62}
\shorttitle{Kinematic temperature of IRDC clump}
\shortauthors{Izumi et al.}

\begin{document}

\title{The ALMA Survey of 70 $\mu$m Dark High-mass Clumps in Early Stages (ASHES). X: Hot Gas Reveals Deeply Embedded Star Formation}

\author[0000-0003-1604-9127]{Natsuko Izumi}
\affiliation{Institute of Astronomy and Astrophysics, Academia Sinica, No. 1, Section 4, Roosevelt Road, Taipei 10617, Taiwan}

\author[0000-0002-7125-7685]{Patricio Sanhueza}
\affiliation{National Astronomical Observatory of Japan, National Institutes of Natural Sciences, 2-21-1 Osawa, Mitaka, Tokyo 181-8588, Japan}
\affiliation{Astronomical Science Program, The Graduate University for Advanced Studies, SOKENDAI, 2-21-1 Osawa, Mitaka, Tokyo 181-8588, Japan}

\author[0000-0003-2777-5861]{Patrick M. Koch}
\affiliation{Institute of Astronomy and Astrophysics, Academia Sinica, No. 1, Section 4, Roosevelt Road, Taipei 10617, Taiwan}

\author[0000-0003-2619-9305]{Xing Lu}
\affiliation{Shanghai Astronomical Observatory, Chinese Academy of Sciences, 80 Nandan Road, Shanghai 200030, People's Republic of China}

\author[0000-0003-1275-5251]{Shanghuo Li}
\affiliation{Max Planck Institute for Astronomy, Konigstuhl 17, D-69117 Heidelberg, Germany}

\author[0000-0002-6428-9806]{Giovanni Sabatini}
\affiliation{INAF, Osservatorio Astrofisico di Arcetri, Largo E. Fermi 5, I-50125, Firenze, Italy}

\author[0000-0002-8250-6827]{Fernando A. Olguin}
\affiliation{Institute of Astronomy, National Tsing Hua University, Hsinchu 30013, Taiwan}

\author[0000-0003-2384-6589]{Qizhou Zhang}
\affiliation{Center for Astrophysics \textbar Harvard \& Smithsonian, 60 Garden Street, Cambridge, MA 02138, USA}

\author[0000-0001-5431-2294]{Fumitaka Nakamura}
\affiliation{Department of Astronomy, Graduate School of Science, The University of Tokyo, 7-3-1 Hongo, Bunkyo-ku, Tokyo 113-0033, Japan}
\affiliation{National Astronomical Observatory of Japan, National Institutes of Natural Sciences, 2-21-1 Osawa, Mitaka, Tokyo 181-8588, Japan}
\affiliation{Astronomical Science Program, The Graduate University for Advanced Studies, SOKENDAI, 2-21-1 Osawa, Mitaka, Tokyo 181-8588, Japan}

\author[0000-0002-8149-8546]{Ken'ichi Tatematsu}
\affiliation{National Astronomical Observatory of Japan, National Institutes of Natural Sciences, 2-21-1 Osawa, Mitaka, Tokyo 181-8588, Japan}
\affiliation{Astronomical Science Program, The Graduate University for Advanced Studies, SOKENDAI, 2-21-1 Osawa, Mitaka, Tokyo 181-8588, Japan}

\author[0000-0002-6752-6061]{Kaho Morii}
\affiliation{Department of Astronomy, Graduate School of Science, The University of Tokyo, 7-3-1 Hongo, Bunkyo-ku, Tokyo 113-0033, Japan}
\affiliation{National Astronomical Observatory of Japan, National Institutes of Natural Sciences, 2-21-1 Osawa, Mitaka, Tokyo 181-8588, Japan}

\author[0000-0003-4521-7492]{Takeshi Sakai}
\affiliation{Graduate School of Informatics and Engineering, The University of Electro-Communications, Chofu, Tokyo 182-8585, Japan}

\author[0000-0002-2149-2660]{Daniel Tafoya}
\affiliation{Department of Space, Earth and Environment, Chalmers University of Technology, Onsala Space Observatory, SE-439 92 Onsala, Sweden}

\begin{abstract}
Massive infrared dark clouds (IRDCs) are considered to host the earliest stages of high-mass star formation.
In particular, 70 $\micron$ dark IRDCs are the colder and more quiescent clouds. 
At a scale of about 5000 au using formaldehyde (H$_2$CO) emission, we investigate the kinetic temperature of dense cores in 12 IRDCs obtained from the pilot ALMA Survey of 70 $\micron$ dark High-mass clumps in Early Stages (ASHES). Compared to 1.3 mm dust continuum and other molecular lines, such as C$^{18}$O and deuterated species, we find that H$_2$CO is mainly sensitive to low-velocity outflow components rather than to quiescent gas expected in the early phases of star formation.
The kinetic temperatures of these components range from 26 to 300 K. 
The Mach number reaches about 15 with an average value of about 4, suggesting that the velocity distribution of gas traced by H$_2$CO is significantly influenced by a supersonic non-thermal component. 
In addition, we detect warm line emission from HC$_3$N and OCS in 14 protostellar cores,
which requires high excitation temperatures (E$_u/k \sim100$ K). These results show that some of the embedded cores in the ASHES fields are in an advanced evolutionary stage, previously unexpected for 70 $\micron$ dark IRDCs.  
\end{abstract}

\section{Introduction}\label{sec:intro}
High-mass stars play a major role in the energy budget of galaxies via their radiation, wind, and supernova events, but the picture of their formation remains unclear.
Massive infrared dark clouds (IRDCs), dark silhouettes against the Galactic 8 $\micron$ mid-infrared background in Galactic plane surveys \citep[e.g., ][]{Simon2006, Peretto2009},
are suitable targets for studying the earliest stages of high-mass star formation {\citep{Rathborne2006,Chambers2009,Sanhueza2012,Bovino2019,Sabatini2020,Radaelli2021}. 
Among IRDCs, those that are 70 $\micron$ dark are especially important because they are the coldest, most quiescent clouds \citep[e.g.,][]{Guzman2015}.
They are considered to trace the earliest stages of high-mass star formation \citep[e.g.,][]{Zhang2009,Tan2013,Sanhueza2013,Sanhueza2017,Contreras2018,Sabatini2019}.
However, previous studies have found that not all cores in 70 $\micron$ dark IRDCs are in the prestellar phase. 
Some evidence of active star formation is revealed by the detection of molecular outflows 
\citep[e.g.,][]{Sanhueza2010,Wang2011,Traficante2015,Feng2016,Svoboda2019,Tan2016,Li2019,Kong2019,Morii2021,Tafoya2021,Radaelli2022,Sabatini2023,Jiao2023}. 

We have carried out the pilot Atacama Large Millimeter/submillimeter Array (ALMA) Survey of 70 $\micron$ dark High-mass clumps in Early Stages (ASHES) toward 12 IRDC clumps.
The sample was selected by combining the ATLASGAL survey \citep{schuller2009,Contreras2013}
and a series of studies from the MALT90 survey \citep{Foster2011,Foster2013,Jackson2013,Hoq2013,Guzman2015,Rathborne2016,Contreras2017,Whitaker2017}.
The source selection is described in \citet{Sanhueza2019} and \citet{Morii2023}\footnote{
We note that \citet{Morii2023} include the complete sample toward 39 targets for this survey.
}.
The names of the 12 targets are listed in Table \ref{tab:ratio}.
From the pilot survey, 294 dense cores were detected in continuum emission using the astropy astrodendro package \citep{Rosolowsky2008,AstropyColl2013}.
Among them, 97 cores are thought to be in the protostellar stage (hereafter, protostellar cores) based on the detection of molecular outflows and/or emission from any of the three lines requiring warm excitation temperatures:
methanol (CH$_3$OH) $J_{K_AK_C}$ = 4$_{2,2}$--3$_{1,2}$ ($E_u/k$ = 45.46 K), formaldehyde (H$_2$CO) $J_{K_AK_C}$ = 3$_{2,2}$--2$_{2,1}$
 ($E_u/k$ = 68.09 K),
 and H$_2$CO $J_{K_AK_C}$ = 3$_{2,1}$--2$_{2,0}$ ($E_u/k$ = 68.11 K).
The other 197 dense cores are prestellar candidates (hereafter, prestellar cores).
In addition, 97 protostellar cores are classified into three evolutionary stages based on searching molecular outflows and/or warm-core line emission. They are named ``outflow core", ``warm core'', and ``warm and outflow core''.
The detailed explanation of the core classification is summarized in \citet{Li2022}.
Using these data, we have investigatedthe fragmentation process \citep{Sanhueza2019}, outflows \citep{Li2020,Tafoya2021}, chemical properties \citep{Sakai2022,Sabatini2022,Li2022}, active star formation signatures \citep{Morii2021}, and dynamical properties \citep{Li2023}.
The summary of the whole ASHES sample has been recently presented in \citep{Morii2023}.

In this paper, we inspect H$_2$CO, which is a slightly asymmetric rotor molecule and ubiquitous
in the interstellar medium
\citep[e.g.,][]{Downes1980,Bieging1982,Henkel1991,Zylka1992,Liszt2006,Mangum2008,Mangum2013b,Mangum2019,Ao2013,Tang2013,Tang2014,Ginsburg2015,Ginsburg2016,Guo2016,Yan2019}.
Since the relative population of the $K_a$ ladders of H$_2$CO are almost exclusively determined by collisional processes \citep{Mangum1993},
the line ratios involving different $K_a$ ladders are good tracers of the kinetic temperatures \citep{Mangum1993,Muhle2007}.
Particularly useful are the three transitions of para-H$_2$CO ($J_{K_AK_C}$ = 3$_{0,3}$--2$_{0,2}$, 3$_{2,2}$--2$_{2,1}$, and 3$_{2,1}$--2$_{2,0}$;
hereafter, H$_2$CO(3$_{0,3}$--2$_{0,2}$), H$_2$CO(3$_{2,2}$--2$_{2,1}$), and H$_2$CO(3$_{2,1}$--2$_{2,0}$)), 
which can be measured simultaneously at about 218 GHz with a bandwidth of 1 GHz and whose relative strengths
(H$_2$CO(3$_{2,2}$--2$_{2,1}$)/H$_2$CO(3$_{0,3}$--2$_{0,2}$) 
and H$_2$CO(3$_{2,1}$--2$_{2,0}$)/H$_2$CO(3$_{0,3}$--2$_{0,2}$)) provide a sensitive thermometer.
The $E_u/k$ of H$_2$CO(3$_{2,2}$--2$_{2,1}$) is 21.0, while that of the other two lines is about 68.1 as aforementioned.
Therefore, the line ratios are sensitive to gas kinetic temperatures of less than 100 K, and the uncertainty in gas temperature is relatively small for a measured line ratio at temperatures of less than 100 K \citep{Mangum1993} in the case of optically thin H$_2$CO emission \citep{Ao2013}.

The metastable lines of ammonia (NH$_3$) are frequently used as a standard molecular cloud thermometer within our Galaxy and also in external galaxies
\citep[e.g.,][]{Ho1983,Walmsley1993,Danby1988,Friesen2009,Mangum2013a,Keown2019}.
However, the NH$_3$ abundance can vary strongly in different molecular environments
(e.g., 10$^{-5}$ in dense molecular ``hot cores'' around newly formed massive stars, \citealt{Mauersberger1987}; 10$^{-8}$ in dark clouds, \citealt{Benson1983}; \citealt{Chira2013};
$\sim$ 10$^{-10}$ in the Orion Bar PDR, \citealt{Batrla2003}) and furthermore, is can be extremely affected by a high UV flux.
In contrast to this, H$_2$CO has a relatively constant fractional abundance, with variations rarely exceeding one order of magnitude during different stages of star formation
{\citep[e.g.,][]{Mangum1990,Mangum1993,Caselli1993,Johnstone2003,Gerner2014,Tang2017a,Tang2017b,Tang2018b,Zhu2020,Sabatini2021}.
For instance, the H$_2$CO abundance is the same in the hot core and in the compact ridge of the Orion nebula \citep{Caselli1993,Mangum1993}.
The NH$_3$ (2,2)/(1,1) ratio is sensitive to gas temperatures with temperatures of less than 50 K \citep{Mangum2013a,Gong2015}.
This is similar to the kinetic temperature range that the H$_2$CO ratio is most sensitive to \citep{Mangum1993}.
The NH$_3$ lines have lower effective excitation densities than the H$_2$CO transitions by a few orders of magnitude,
$n_{\rm eff}$(NH$_3$(1,1)) $\sim$ 10$^3$ cm$^{-1}$ while $n_{\rm eff}$(H$_2$CO(3$_{0,3}$--2$_{0,2}$)) $\sim$ 10$^5$ cm$^{-1}$ \citep{Shirley2015}.
Previous observations reported that the temperatures derived from NH$_3$ reflect an average temperature of more extended and cooler gas
\citep{Henkel1987,Ginsburg2016} while H$_2$CO traces denser and hotter regions \citep{Ginsburg2016, Tang2017a, Tang2018b, Tang2018a}.
From the above, the three transitions of H$_2$CO allow us to probe the kinetic temperature of IRDCs and unveil potential precursors of future massive hot cores.

The paper is organized as follows.
The observations are presented in Section 2.
Section 3 describes the H$_2$CO results and analysis in this IRDC sample.
In Section 4, we discuss the kinetic temperatures derived from H$_2$CO.
Our main conclusions are summarized in Section 5.

\section{Observations}\label{sec:obs}
Observations of twelve 70 $\mu$m IRDC clumps
were performed with ALMA in Band 6 ($\sim$224 GHz; 1.34 mm)
using the main 12 m array, the 7 m array, and the total power (TP) array (Project ID: 2015.1.01539.S, PI: P. Sanhueza).
The mosaic observations were carried out with the 12 m array and 7 m array to cover a significant portion of the clumps, 
as defined by single-dish continuum images.
The same correlator setup was used for all sources.
More details on the observations have been provided in earlier articles of the ASHES series \citep{Sanhueza2019, Li2020}.

The data calibration was performed using the CASA software package version 4.5.3, 4.6, and 4.7,
while imaging was carried out using CASA 5.4 \citep{Mcmullin2007}.
The data cubes for lines were produced using the automatic masking procedure YCLEAN \citep{Contreras2018}.
Since some sources were observed with different array configurations \citep{Sanhueza2019}, 
a uv-taper was used for such sources in order to obtain a similar synthesized beam of  $\sim1\farcs2$
for all sources.
We adopted a Briggs robust weighting of 0.5 and 2 for the visibilities of continuum and lines in the imaging process, respectively.
This yielded an averaged 1$\sigma$ root mean square (rms) noise level of $\sim$ 0.1 mJy beam$^{-1}$ for continuum images.
For H$_2$CO lines, the sensitivity is $\sim$ 3.3 mJy beam$^{-1}$ per 0.67 km s$^{-1}$.

At the time of observations,
the ALMA TP antennas did not provide continuum observations.
Therefore, our continuum and molecular line analyses are mostly focused on combined 12 m and 7 m images (hereafter 12m7m),
whereas the combined 12 m, 7 m, and TP data (hereafter 12m7mTP) are also used to assess the missing flux in images without TP data.
The mean ratio of 12m7m flux over 12m7mTP flux in the whole mapping area where the emission is more than 3 $\sigma$
for H$_2$CO(3$_{0,3}$--2$_{0,2}$) emission is 0.66.
While the ratio for H$_2$CO(3$_{2,2}$--2$_{2,1}$) and H$_2$CO(3$_{2,1}$--2$_{2,0}$) emission is
1.00 and 0.96, respectively.
This indicates that H$_2$CO(3$_{2,2}$--2$_{2,1}$) and H$_2$CO(3$_{2,1}$--2$_{2,0}$) trace more spatially compact emission 
as compared to H$_2$CO(3$_{0,3}$--2$_{0,2}$).

\section{Results} \label{sec:res} 
Figure \ref{Spectrum} shows an example of an observed spectrum including the H$_2$CO transition lines. 
The peak intensities, line widths, and integrated intensities of the para-H$_2$CO emission were derived using  Gaussian fitting (red curve in Figure \ref{Spectrum}).

\subsection{Spatial distribution} \label{sec:res_1}
Figure \ref{MOM0} shows the integrated intensity distributions of three transitions of H$_2$CO in two representative sources of the sample
(G010.991--00.082 and G337.541--00.082), while the other 10 sample is summarised in the appendix (Section \ref{sec:a-1} and Figure \ref{A-MOM0}).
H$_2$CO(3$_{0,3}$--2$_{0,2}$) is more spatially extended than
H$_2$CO(3$_{2,2}$--2$_{2,1}$) and H$_2$CO(3$_{2,1}$--2$_{2,0}$).
This result is consistent with the amount of resolved out flux reported in Section \ref{sec:obs}.
We note that H$_2$CO(3$_{2,2}$--2$_{2,1}$) and H$_2$CO(3$_{2,1}$--2$_{2,0}$)
are detected at less than 3$\sigma$ in the whole mapping area of
G332.969-00.029, G340.179-00.242, and G340.222-00.167.
The spatial distributions of all H$_2$CO transition lines differ from the material traced by the 1.3 mm dust continuum emission.
This finding is inconsistent with previous results reporting that H$_2$CO is associated well with the dense gas traced by 450 and 850 $\micron$ continuum in high-mass star-forming regions (e.g., IRDCs, \ion{H}{2} regions) using single-dish telescopes, including APEX and JCMT \citep[e.g.,][]{Tang2017a,Tang2018b}.

\subsection{Core detection rate} \label{sec:res_2} 
Among all the 294 cores identified in the pilot ASHES study, 
the detection rate of H$_2$CO(3$_{0,3}$--2$_{0,2}$) is 65\% (192/294)
\footnote{
We have also examined the detection rate assuming a simple elliptical core structure
(core size is referenced from Table 3 in \citet{Sanhueza2019},
corresponding to half of the geometric mean between the deconvolved major and minor axes of the ellipse determined via dendrograms). 
The detection rate based on this simplified structure is 60 \% (175/294), which is not a significant difference
from the results presented in the main text, which use dendrogram leaves.
Moreover, we have also investigated the H$_2$CO properties of cores based on this simplified structure 
(e.g., 1.3 mm dust continuum, velocity distribution,
and temperatures), and the main trends align with those described in the main text.
}.
For H$_2$CO(3$_{2,2}$--2$_{2,1}$) and H$_2$CO(3$_{2,1}$--2$_{2,0}$),
which require a warmer excitation temperature, the detection rates are 15\% (45/294) and 15\% (45/294), respectively.
The simultaneous detection of H$_2$CO(3$_{2,2}$--2$_{2,1}$) and H$_2$CO(3$_{2,1}$--2$_{2,0}$) is of 14\% (40/294).
For H$_2$CO(3$_{0,3}$--2$_{0,2}$), as \citet{Li2022} already reported, the detection rate of protostellar cores is 92\% (89/97),
which is about twice as large as the detection rate of prestellar candidates (52\%;  103/197).

For H$_2$CO(3$_{2,2}$--2$_{2,1}$) and H$_2$CO(3$_{2,1}$--2$_{2,0}$),
the detection rates of protostellar cores are both 46\% (45/97).
These values are about half of the one for H$_2$CO(3$_{0,3}$--2$_{0,2}$).
We note that the detection threshold is 3$\sigma$.
Figure \ref{Hist_DR} shows the relation between 1.3 mm flux density ($S_{\rm 1.3mm}$) and the detection rate of H$_2$CO(3$_{0,3}$--2$_{0,2}$)
for protostellar and prestellar core candidates.
The plot indicates that the detection rate of both protostellar and prestellar core candidates grows with increasing $S_{\rm 1.3mm}$.
However, a larger sample is needed for a fully robust statistical assessment.
The details on the detection rates are summarized in Table \ref{tab:ratio}.

\subsection{Correlation with 1.3mm dust continuum} \label{sec:res_3} 
Figure \ref{int-relation_core} shows the relationship between the 1.3mm flux density ($S_{\rm 1.3mm}$) and the 
H$_2$CO(3$_{0,3}$--2$_{0,2}$) integrated intensity ($I_{\rm H_2CO(3_{0,3}-2_{0,2})}$) for all cores.
As \citet{Li2022} already reported, we confirm that $I_{\rm H_2CO(3_{0,3}-2_{0,2})}$ in protostellar cores is relatively higher than in prestellar candidates.
We also confirm a weak positive correlation between $I_{\rm H_2CO(3_{0,3}-2_{0,2})}$ and $S_{\rm 1.3mm}$ 
only in protostellar cores.
The protostellar cores have a Spearman rank-order correlation (Spearman's correlation)
\footnote{Spearman's correlation is a nonparametric measure of the strength and direction of association that exists between two variables measured on at least an ordinal scale.
The Spearman's coefficient, $r$, ranges from -1 to 1, with 0 indicating no correlation.
The value of 1 implies an exact increasing monotonic relation between two quantities, while -1 implies an exact decreasing monotonic relation.
The probability value, $p$-value, obtained from the calculator is a measure of how likely or probable it is that any observed correlation is due to chance.
The $p$-value ranges between 0 (0 \%) to 1 (100 \%).
The $p$-value close to 1 suggests no correlation other than due to chance and that the null hypothesis assumption is correct.
While the $p$-value close to 0 suggests that the observed correlation is unlikely to be due to chance, and there is a very high probability that the null hypothesis is wrong. 
}
coefficient of $r$ = 0.25 with a $p$-value of 0.02,
while the prestellar candidates have $r$ = 0.07 and $p$-value = 0.49.
In particular, correlations are seen in the cores with warm lines, which include both ``warm core" and ``warm \& outflow core",
while there is no clear correlation in ``outflow core".
The Spearman's correlation coefficient for each classification of cores is summarized in Table \ref{CC-Ih2co-S1mm}.
We note that previous IRDC studies using single-dish telescopes --
which could not resolve the IRDC clumps into individual cores -- reported a positive correlation 
between H$_2$CO(3$_{0,3}$--2$_{0,2}$) and dust continuum \citep[e.g.,][]{Tang2017a, Tang2018b}.
Therefore, the targets in the aforementioned results are expected to be dominated by evolved regions that have already progressed into the proto-stellar phase.

Figure \ref{int-relation_core_f2f3} presents the relation between $S_{\rm 1.3mm}$ and H$_2$CO(3$_{2,2}$--2$_{2,1}$)
integrated intensity ($I_{\rm H_2CO(3_{2,2}-2_{2,1})}$; left panel)
and between $S_{\rm 1.3mm}$ and H$_2$CO(3$_{2,1}$--2$_{2,0}$)} integrated intensity
($I_{\rm H_2CO(3_{2,1}-2_{2,0})}$; right panel) for all protostellar cores.
Weak correlations
are also seen between $S_{\rm 1.3mm}$ and $I_{\rm H_2CO(3_{2,2}-2_{2,1})}$ and
between $S_{\rm 1.3mm}$ and $I_{\rm H_2CO(3_{2,1}-2_{2,0})}$.
The $r$ and $p$-value between $S_{\rm 1.3mm}$ and $I_{\rm H_2CO(3_{2,2}-2_{2,1})}$ is 0.18 and 0.24, respectively.
The $r$ and $p$-value between $S_{\rm 1.3mm}$ and $I_{\rm H_2CO(3_{2,1}-2_{2,0})}$ is 0.16 and 0.30, respectively.
However, a larger sample is needed for more robust statistical conclusions.

\subsection{Velocity distribution} \label{sec:velo} \label{sec:res_4} 
The H$_2$CO(3$_{0,3}$--2$_{0,2}$) intensity-weighted mean velocity (moment 1; left panel),
and velocity dispersion (moment 2; right panel) maps
for two representative IRDC clumps: G010.991-00.082 and G337.541-00.082 are shown in Figure \ref{MOM12}.
The velocity range used to generate these moment maps for the 12 IRDC clumps is approximately 5--25 km s$^{-1}$ .
The central velocity values are provided in Table 1 in \citet{Sanhueza2019}.
We note that these velocity ranges were determined through visual inspection of the spectra for each clump, ensuring that they encompassed all of the line emission.
The figure reveals complex structures, including different velocity components and large velocity dispersions.
For instance, the difference between the largest and smallest peak velocities estimated from the moment 1 map reaches to more than 5 km s$^{-1}$.
The largest velocity dispersion estimated from the moment 2 map is more than 2 km s$^{-1}$.

Figure \ref{Hist_dv} shows the number distribution of H$_2$CO(3$_{0,3}$--2$_{0,2}$) velocity dispersion
($\sigma_{\rm H_2CO(3_{0,3}-2_{0,2})}$)
for pixels in each IRDC clump.
The pixel scale was re-binned to 0$\farcs$6, approximately half of the synthesized beam, from the original value (0$\farcs$2).
This figure indicates that the $\sigma_{\rm H_2CO(3_{0,3}-2_{0,2})}$ is up to $\sim$ 10 km s$^{-1}$.
This is much larger than the velocity dispersions of other molecular lines, such as C$^{18}$O, DCO$^+$, and N$_2$D$^+$ \citep{Li2022},
which have a typical velocity dispersion of $\sim$1 km s$^{-1}$.
Figure \ref{dv-relation} presents the relation between the velocity dispersion of C$^{18}$O(2-1) ($\sigma_{\rm C^{18}O}$) and H$_2$CO
($\sigma_{\rm H_2CO(3_{0,3}-2_{0,2})}$) for cores and pixels.
We note that the spatial distribution of H$_2$CO(3$_{0,3}$--2$_{0,2}$) does not  agree well with those of the C$^{18}$O line \citep{Li2022}.
Therefore, we only investigate the core and pixel where both lines are detected above the 3$\sigma$ level. 
The pixels includes  regions where the continuum emission is not detected.
Most of the protostellar cores ($\sim$73\%) have larger $\sigma_{\rm H_2CO(3_{0,3}-2_{0,2})}$ than $\sigma_{\rm C^{18}O}$.
About half of the prestellar candidates ($\sim$49\%)  have larger $\sigma_{\rm H_2CO(3_{0,3}-2_{0,2})}$ than $\sigma_{\rm C^{18}O}$.
There is no clear correlation between $\sigma_{\rm C^{18}O}$ and $\sigma_{\rm H_2CO(3_{0,3}-2_{0,2})}$ in both the prestellar candidate and protostellar cores.
The protostellar cores  show a Spearman rank-order correlation coefficient of $r$ = 0.10 with $p$-value = 0.37.
The prestellar candidate shows $r$ = -0.04 with $p$-value = 0.68.
The relation between $\sigma_{\rm C^{18}O}$ and H$_2$CO(3$_{0,3}$--2$_{0,2}$) in each pixel is similar to that in each core.
There is no clear correlation between $\sigma_{\rm C^{18}O}$ and H$_2$CO(3$_{0,3}$--2$_{0,2}$), and 
62 \% of the pixel have larger H$_2$CO(3$_{0,3}$--2$_{0,2}$) than $\sigma_{\rm C^{18}O}$.
These results suggest that H$_2$CO and C$^{18}$O trace different components in the IRDC clumps.
The H$_2$CO might not trace the quiescent gas.

The blue- and red-shifted components of H$_2$CO(3$_{0,3}$--2$_{0,2}$) are shown in Figure \ref{WING} and appendix
(Section \ref{sec:a-1} and Figure \ref{A-WING} ).
The velocity range of the blue- and red-shifted components, which are selected manually by visually inspecting the spectra,
are written in the upper-right on Figure \ref{WING}.
In G340.179-00.0242 and G340.222-00.167, blue- and red-shifted components are not detected (emission is less than 3$\sigma$).
All components seem to be associated with protostellar cores.
The direction of the blue- and red-shifted components roughly agrees with these components detected in CO emission that is considered to trace
molecular outflows, as investigated by  \cite{Li2020} in the ASHES survey.
The velocity and spatial range of these components detected in H$_2$CO(3$_{0,3}$--2$_{0,2}$) emission (1.5--20 km s$^{-1}$, 0-0.3 pc)
is smaller than those detected in CO emission \citep[9--95 km s$^{-1}$, 0.04--0.45 pc;][]{Li2020}.
Therefore, these H$_2$CO velocity components are considered to be tracing low-velocity outflows. 
This consideration is consistent with the result of a recent study on chemical diagnostics of protostellar sources in the low-mass star-forming regions with ALMA \citep{Tychoniec2021}.
Among the 59 outflows identified by  \citet{Li2020}, 30 outflows are detected in 
H$_2$CO(3$_{0,3}$--2$_{0,2}$) emission (solid arrows in Figure \ref{WING}).
In G343.489-00.416, two candidate of blue-shifted outflows are newly detected by H$_2$CO(3$_{0,3}$--2$_{0,2}$)
emission (black arrows in Figure \ref{WING}).
Figure \ref{PV} shows examples of the position-velocity map of the outflows.

\subsection{Kinetic temperature from H$_2$CO lines} \label{sec:temp} \label{sec:res_5} 
We derived the kinetic temperature of the molecular gas in the IRDC clumps
traced by H$_2$CO lines using two methods:
(1) under the assumption of local thermodynamic equilibrium (LTE)
and (2) with the RADEX non-LTE modeling program \citep{Vandertak2007}.
Temperatures were derived only for those cores and pixels where all three transitions of H$_2$CO were detected at more than 5$\sigma$.
Therefore, we could not derive the temperatures for the three IRDCs G332.969--00.029, G340.179--00.242, and G340.222--00.167 because the two transitions 
H$_2$CO(3$_{2,2}$--2$_{2,1}$) and H$_2$CO(3$_{2,1}$--2$_{2,0}$) are undetected at the 5$\sigma$ level over the whole mapping area.
Under this 5$\sigma$ threshold, we could derive the temperature toward 26 cores among all the 294 cores.
We note that among all the other 268 cores, 97 cores show more than 5 $\sigma$ detection of H$_2$CO(3$_{0,3}$--2$_{0,2}$), 
and 10 cores show more than  5 $\sigma$ detection of H$_2$CO(3$_{0,3}$--2$_{0,2}$), and either H$_2$CO(3$_{2,2}$--2$_{2,1}$) or H$_2$CO(3$_{2,1}$--2$_{2,0}$).
In terms of pixel, among all pixels where H$_2$CO(3$_{0,3}$--2$_{0,2}$)  is detected above 5 $\sigma$, we could derive temperature for 13 \% of pixels.

\subsubsection{LTE assumption} \label{sec:res_51} 
We fit the three H$_2$CO lines assuming LTE conditions and estimate best-fitting parameters.
Under this assumption, the fitting routine constructs line profiles with a range of input parameters, including peak velocities, temperatures, line-widths, and column densities, following \citet{Mangum2015}.
The peak velocities of the lines can be selected manually and are fixed according to their known rest frequencies.
The code
then minimizes the differences between the constructed and observed lines, using the non-linear least-squares python fitting routine {\it lmfit}  \footnote{https://pypi.org/project/lmfit/}.
The best fit is returned, along with the estimated temperature, line-width, column density, and associated errors, which we then take as the optimized fitting results
\footnote{Python code for fitting 218 GHz H$_2$CO lines can be found at \url{https://github.com/xinglunju/FFTL}}.
The fitted temperatures, column densities, and line-widths are in the range of 
10 -- 300 K, 1.0 $\times$ 10$^{10}$ -- 1.0 $\times$ 10$^{20}$ cm$^{-2}$, and 0.14 -- 1.37 km s$^{-1}$, respectively.
When we derive the temperature for each pixel, the peak velocities are automatically selected as the position of  H$_2$CO(3$_{0,3}$--2$_{0,2}$) peak intensity and are fixed according to their known rest frequencies.

\subsubsection{Radex modeling} \label{sec:res_52} 
We employ the Radex code \citep{Vandertak2007}, by means of the ndRADEX tool\footnote{\url{https://pypi.org/project/ndradex/}}, 
to generate a set of different models covering a wide range of conditions.
As pointed out by \citet{Ao2013} and appendix (Section \ref{sec:a-2} and Figure \ref{Radex-Model}), the H$_2$CO(3$_{2,1}$--2$_{2,0}$)/H$_2$CO(3$_{0,3}$--2$_{0,2}$) ratio
is more sensitive to densities than the H$_2$CO(3$_{2,2}$--2$_{2,1}$)/H$_2$CO(3$_{0,3}$--2$_{0,2}$) ratio, 
and is therefore used to derive kinetic temperatures in the following.
The Radex code needs five input parameters:
1) the background temperature,
2) the kinetic temperature,
3) the H$_2$ volume density, 
4) the H$_2$ (or H$_2$CO) column density,
and 5) the line width.
For the background temperature, we adopt 2.73 K.
The line width is derived from the Gaussian fitting.
For the H$_2$ density and H$_2$ (or H$_2$CO) column density, we adopt the values derived from the 1.3 mm dust continuum \citep{Sanhueza2019}.
The fractional abundance of  N(H$_2$CO)/N(H$_2$) was adopted as 1.0 $\times$ 10$^{-10}$ from \citet{Tang2017a}, which investigated the fractional abundance in IRDCs.
When we derive the temperature of each pixel, we use the averaged values of H$_2$ density, column density, and linewidth over the whole IRDC for simplicity.
With the above parameters and the H$_2$CO(3$_{2,1}$--2$_{2,0}$)/H$_2$CO(3$_{0,3}$--2$_{0,2}$) integrated intensity ratio derived from the Gaussian fitting,
we then compute the temperatures using a model grid for the H$_2$CO lines encompassing 580 temperatures ranging from 10 to 300 K.
The temperature error is propagated from the optimum of the Gaussian fitting.

\subsubsection{Comparison between LTE temperature and Radex temperature} \label{sec:res_53} 
Figure \ref{LTE-Radex_core} displays 
temperatures 
derived under the LTE assumption ($T_{\rm LTE}$)
versus temperatures from Radex modeling ($T_{\rm Radex}$)
for all protostellar cores.
Due to the 5$\sigma$ detection limit, we derived temperatures for 26 among the total of 97 protostellar cores.
Figure \ref{TEMP-1} shows the $T_{\rm LTE}$ and $T_{\rm Radex}$ distribution
in two examples of IRDC clumps.
For both $T_{\rm LTE}$ and $T_{\rm Radex}$, the temperature error becomes larger than 20 K for temperatures of more than 100 K.
The $T_{\rm LTE}$}and $T_{\rm Radex}$ are roughly similar,
but in almost all cases, the $T_{\rm Radex}$ is slightly higher than the $T_{\rm LTE}$.
The difference likely results from
(1) the difference between fitting the H$_2$CO(3$_{2,2}$--2$_{2,1}$)/H$_2$CO(3$_{0,3}$--2$_{0,2}$) integrated intensity ratio for the Radex modeling and the LTE assumption,
(2) the assumption of the H$_2$ volume (n(H$_2$)) and column densities(N(H$_2$)) for the Radex modeling,
and
(3) the LTE assumption.
We note that the effect from the assumption of the N(H$_2$) is negligible compared to the assumption of the n(H$_2$) for the Radex modeling (see Section \ref{sec:a-2} and Figure \ref{Radex-Model}).
These results indicate that the temperature determination can vary significantly depending on the assumed models and n(H$_2$).
Figure \ref{LTE-Radex_ratio} shows the relation between $T_{\rm LTE}$ and $T_{\rm Radex}$ and the H$_2$CO(3$_{2,1}$--2$_{2,0}$)/H$_2$CO(3$_{0,3}$--2$_{0,2}$) integrated intensity ratio.
This plot indicates that the temperature determination can vary significantly depending on the assumed models and n(H$_2$).
To investigate the uncertainty, we calculate the minimum and maximum temperature (gray filled area in Figure \ref{LTE-Radex_ratio}) based on the assumption of possible n(H$_2$) and N(H$_2$) ranges:
n(H$_2$) = 10$^5$ -- 2.0 $\times$ 10$^7$ cm$^{-3}$ and N(H$_2$) = 10$^{22}$ -- 10$^{24}$ cm$^{-2}$.
These ranges come from the minimum and maximum values of the cores derived from the 1.3 mm continuum emission \citep{Sanhueza2019}, which include both protostellar cores and prestellar candidates.
The uncertainty increases with growing H$_2$CO(3$_{2,1}$--2$_{2,0}$)/H$_2$CO(3$_{0,3}$--2$_{0,2}$) integrated intensity ratios.
For instance, the uncertainty at the ratio of 0.4 is about 100 K, which is roughly 5 times larger than the one at the ratio of 0.2.
In all models and under all assumptions, more than half of the regions traced by H$_2$CO show high temperatures of more than 50 K.

In order to investigate the impact of the missing flux on the line emission, we compare the temperatures derived from 12m7m and 12m7mTP data.
The ratio of $T_{\rm LTE}$ derived from 12m7m per 12m7mTP data is 0.74--2.06 with an average of 1.03$\pm$0.06. 
The ratio of $T_{\rm Radex}$ is 0.19--6.52 with an average of 1.04$\pm$0.17.
Compared to the uncertainty from the assumption, the impact of the missing flux is likely negligible.

\section{Discussion} \label{sec:dis}  

\subsection{Comparison with previous observations of H$_2$CO} \label{sec:dis_1} 
The $T_{\rm LTE}$ and $T_{\rm Radex}$ values across all cores range from  
36 to 300 K (average: 84$\pm$52 K)\footnote{The error values of average values in this paper indicates the standard deviation (1 $\sigma$)}
and from 37 to 292 K (average: 98$\pm$57 K), respectively.
The $T_{\rm LTE}$ and $T_{\rm Radex}$ temperatures measured in each pixel range from 
29 to 300 K (average: 75$\pm$50 K) and from 26 to 300 K (average: 91$\pm$60 K), respectively.
We stress that these values are derived where all three H$_2$CO lines are detected above 5$\sigma$.
Therefore, this also includes regions outside of cores.
Previous studies reported that the temperatures derived from H$_2$CO lines in IRDCs and outflow objects range between
40 and 60 K and between 31 and 100 K, respectively \citep{Tang2017a}.
These previous observations were performed by single-dish telescopes, such as the JCMT, with spatial resolutions around 20--30\arcsec \citep[corresponding to $\sim$ 0.5 pc;][]{Tang2017a}.
This means that their temperatures are considered to be an average temperature of the entire area of the object.
Owing to the much higher resolution in the work here, we could resolve the IRDCs into several components, including jets, outflows, and cores \citep{Tychoniec2021}.
Therefore, the temperature range of our results is much wider than what is seen in the previous studies with single-dish telescopes.
We note that a recent ALMA study targeting the IRDC G10.21-0.31, with a spatial resolution of about 5000 au, 
reported temperatures derived from H$_2$CO lines in two dense cores as 67.7 K and 83.0 K (Jian et al., 2023),
consistent with the range of values found in the ASHES sample.

\subsection{Comparison of temperatures derived from H$_2$CO and NH$_3$} \label{sec:dis_2} 
For this comparison,
the NH$_3$ (1,1) and (2,2) lines in IRDCs are obtained from the Complete ATCA (The Australia Telescope Compact Array) Census of High-Mass
Clumps survey (CACHMC survey; Allingham et al., 2023 in prep) at an angular resolution of $\sim$5$\arcsec$. 
In order to compare the kinetic temperatures derived from para-H$_2$CO lines and NH$_3$ lines \citep[][Allingham et al., 2023 in prep]{Li2022},
we smoothed the H$_2$CO data cube to a half-power beam width (HPBW) of $\sim$5$\arcsec$ with a 2D Gaussian function.

\begin{figure*}
\gridline{\fig{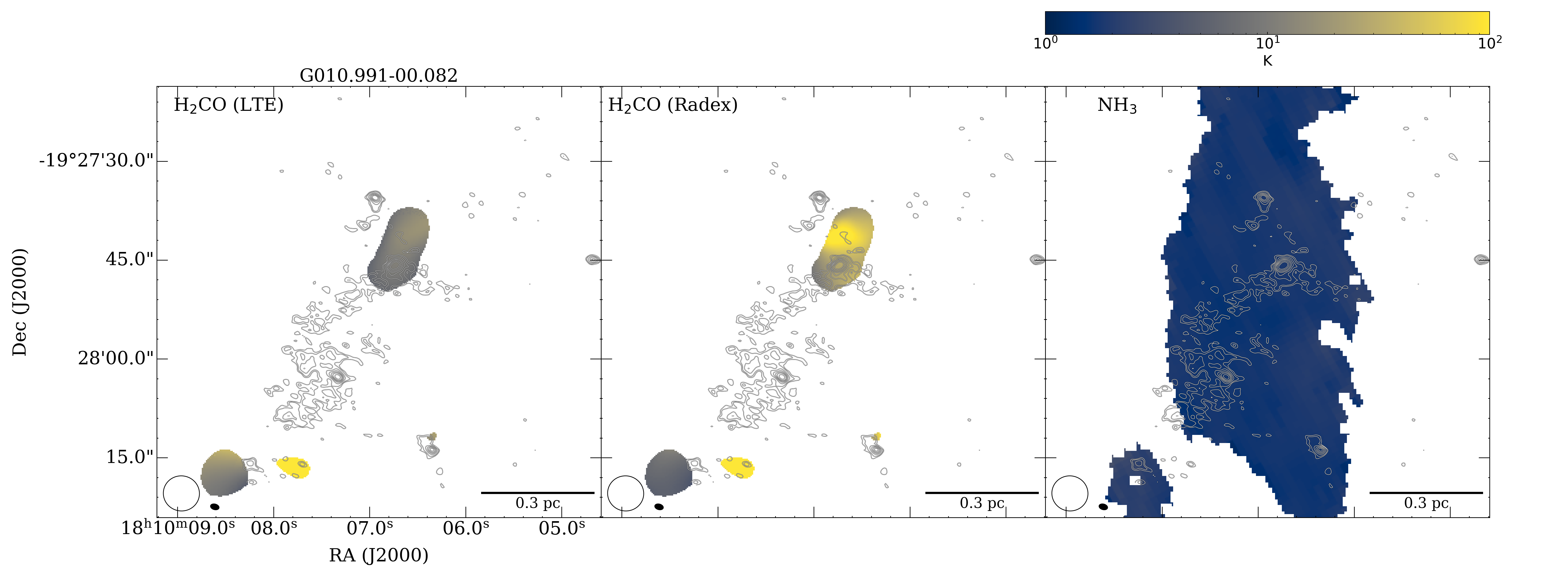}{0.9\textwidth}{}}
\vspace{-0.75cm}
\gridline{\fig{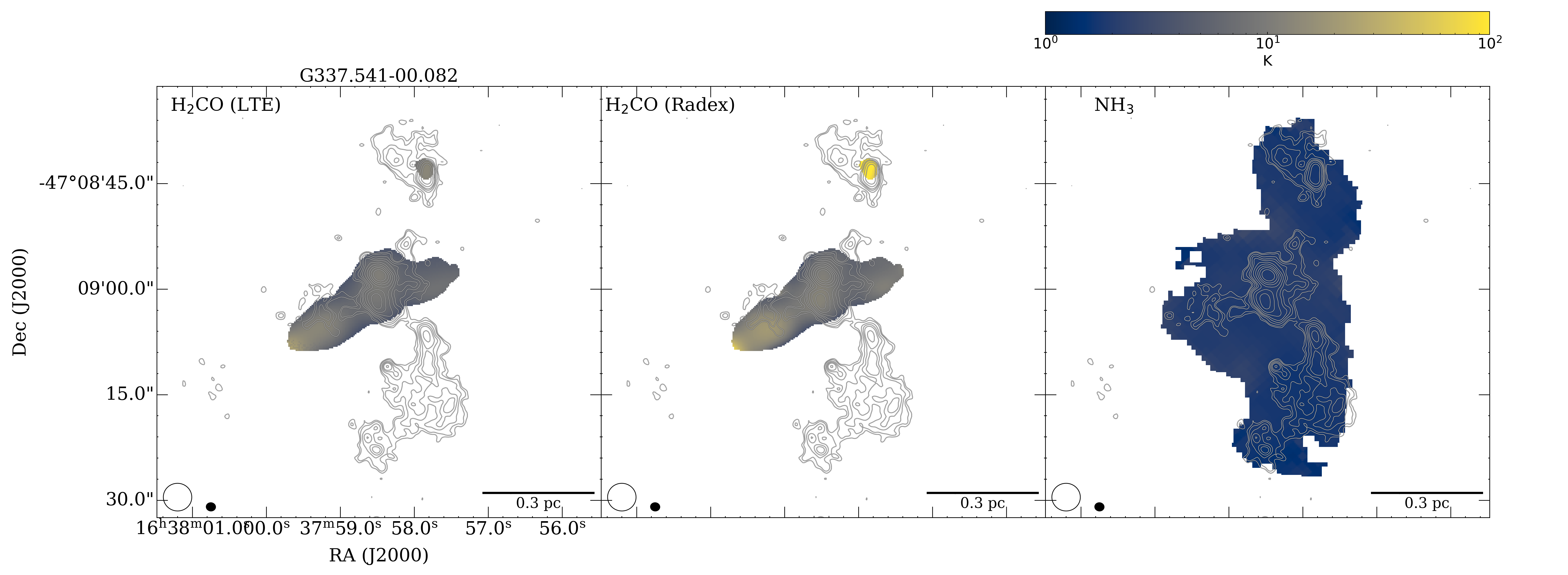}{0.9\textwidth}{}}
\vspace{-0.75cm}
\caption{
Kinetic temperature distributions (Left: LTE, Middle: Radex, Right: NH$_3$) for two IRDC clumps.
The black contours are the 1.3mm dust continuum.
Contour levels are 3, 4, 5, 7, 10, 14, and 20 $\times$ $\sigma$, where
$\sigma$ = 0.115 mJy beam$^{-1}$ for G010.991-00.082
(1$\farcs$1 resolution);
and 3, 4, 6, 8, 10, 14, 20, 30, 45, and 75 $\times$ $\sigma$, where $
\sigma$ = 0.068 mJy beam$^{-1}$ for G337.541-00.082
(1$\farcs$2 resolution).
Synthesized beams are displayed at the bottom left of each panel
(open ellipses: H$_2$CO and NH$_3$; filled ellipses: continuum).
}
\label{TEMP5sec-1} 
\end{figure*}

\begin{figure*}
\epsscale{0.8}
\plotone{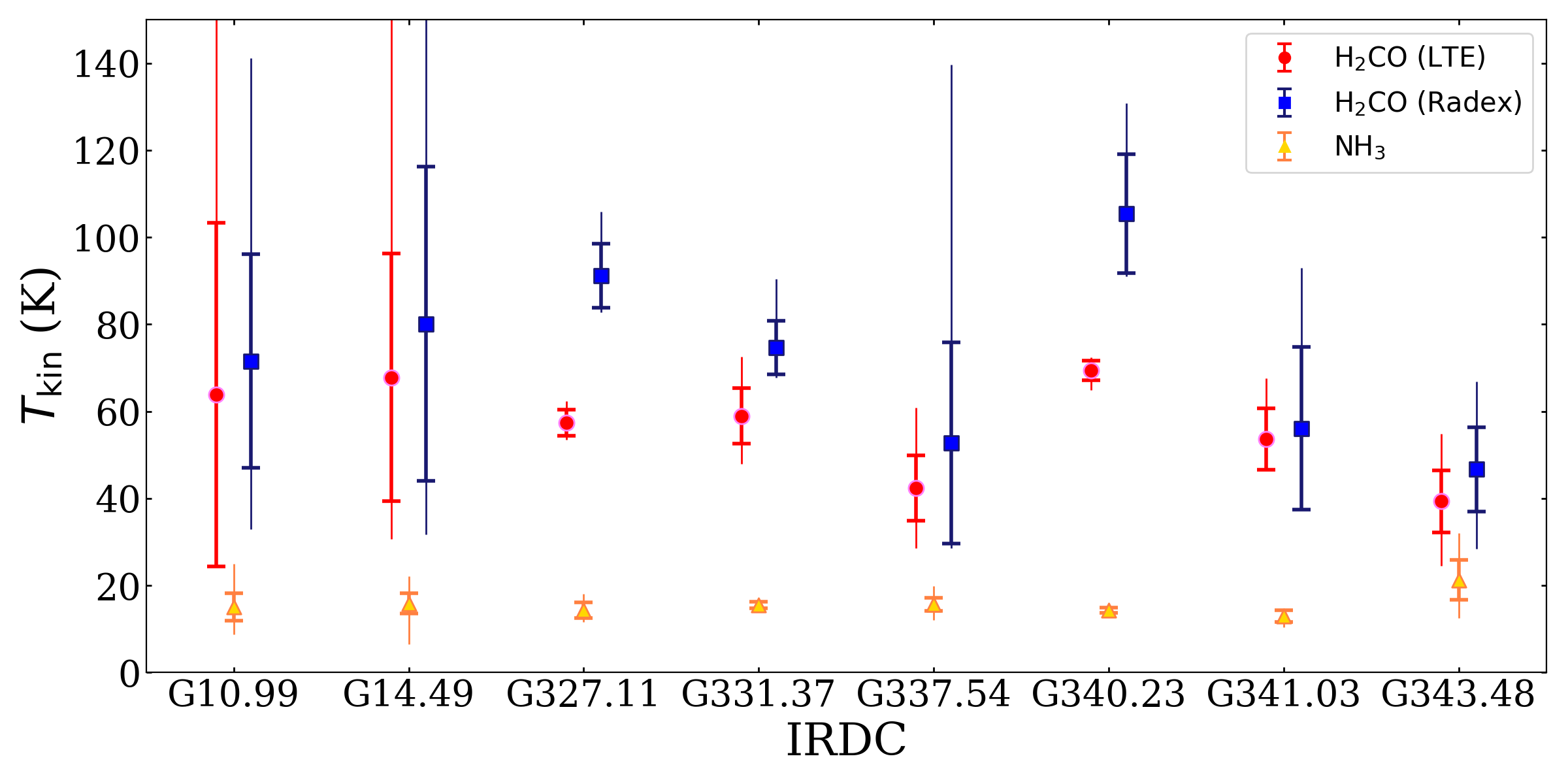}
\caption{
Comparison of kinetic temperatures ($T_{\rm kin}$) derived from H$_2$CO (red circle: LTE, blue square: Radex) and NH$_3$ (yellow triangle) for each IRDC clump.
The error bars indicate the standard deviation of $T_{\rm kin}$ (1 $\sigma$).
The extended bars indicate the minimum-maximum range of $T_{\rm kin}$.
To derive these values, the pixel scale was re-binned to 2$\farcs$4, approximately half of the angular resolution of NH$_3$ data ($\sim$ 5$\arcsec$), from the original value (0$\farcs$2).
}
\label{Temp_h2co-nh3} 
\end{figure*}

Figure \ref{TEMP5sec-1} shows examples of
spatial temperature distributions derived from H$_2$CO and NH$_3$ with a resolution of $\sim$5$\arcsec$.
The spatial distributions differ significantly.
NH$_3$ shows a much more extended distribution than H$_2$CO. 
The $T_{\rm LTE}$, $T_{\rm Radex}$, and NH$_3$ temperatures across all IRDCs are
24--300 (average: 78$\pm$65), 29--300 (average: 89$\pm$66), and 7--32 (average: 14$\pm$5), respectively.
Figure \ref{Temp_h2co-nh3} compares temperatures derived from H$_2$CO and NH$_3$ for each IRDC.
We only compare the H$_2$CO and NH$_3$ temperatures in those region where both molecules are detected.
The H$_2$CO temperature of G028.273--00.167 is not derived in the smoothed data because the 
H$_2$CO components in G028.273--00.167 detected with a 1$\farcs$2 resolution (see Figure \ref{A-MOM0} in appendix) are diluted at a  
5$\arcsec$ beam.
These results from the comparison of H$_2$CO and NH$_3$ are similar to those for outflows and IRDCs from \citet{Tang2017a}.
However, the H$_2$CO temperature range in our results is much larger than
in \citet{Tang2017a}, and most likely due to our originally higher spatial resolution, despite the factor of smoothing.
The NH$_3$ temperature range is similar to the previous observations with the 
Effelsberg telescope \citep[$\sim$40$\arcsec$;][]{Tang2017a}, despite the resolution being higher 
by a factor of about 10.
These results indicate that in these IRDCs H$_2$CO traces components that are different from what is seen in NH$_3$.
Moreover, even if at the same resolution of 5\arcsec\, the variation of the temperatures derived from H$_2$CO is much larger than the variation of the NH$_3$ temperatures.
This finding might indicate that H$_2$CO is more sensitive to star-formation activities, including cores, outflows, and jets,
than NH$_3$ in IRDCs. 

\subsection{Thermal and non-thermal motions} \label{sec:dis_3} 
We calculated thermal linewidth ($\sigma_T$), non-thermal velocity dispersion ($\sigma_{NT}$),
thermal sound speed ($a_{\rm s}$), ratio of thermal to non-thermal pressure ($R_{\rm p}$), and Mach number ($\mathcal{M}$).
The $\sigma_T$, $\sigma_{NT}$, $a_{\rm s}$, $R_{\rm p}$, and $\mathcal{M}$ are given by
\begin{eqnarray}
    \sigma_T &=& \sqrt{kT_{\rm kin}/m_x} \\
    \sigma_{NT} &=& \sqrt{(\Delta V^2/(8\ln 2)) - \sigma_T^2} \\
    a_{\rm s} &=& \sqrt{kT_{\rm kin}/(\mu m_{\rm H})} \\
    R_{\rm p} &=& a_{\rm s}^2/\sigma_{\rm NT}^2 \\
    \mathcal{M} &=& \sigma_{NT} / a_{\rm s}
\end{eqnarray}
where $k$ is the Boltzmann constant, $T_{\rm kin}$ the kinetic temperature of the gas ($T_{\rm LTE}$ or $T_{\rm Radex}$), $m_x$ the mass of the relevant molecule,
$\Delta V$ the measured FWHM linewidth, $\mu$ $=$ 2.37 the mean molecular weight for molecular clouds,
and $m_{\rm H}$ the mass of the hydrogen atom.
The FWHM linewidth is measured by two methods: fitting with the LTE assumption (see Section \ref{sec:res_51}) and the Gaussian fitting.
The errors of these quantities are derived by varying the assumed  $T_{\rm kin}$ from minimum to maximum values.
We note that we only considered the uncertainties (i.e., from minimum to maximum values) of $T_{\rm kin}$ 
because the uncertainties for the measured FWHM are comparatively small and they can be neglected. 

Table \ref{tab:pro_H2CO-1} and \ref{tab:pro_H2CO-2} summarize the averaged values of
$T_{\rm kin}$,  $\sigma_T$, $\sigma_{NT}$, $a_{\rm s}$, $R_{\rm p}$,and $\mathcal{M}$ over all pixels in the IRDCs derived from H$_2$CO(3$_{0,3}$--2$_{0,2}$). 
The $\sigma_{NT}$ is much larger than $\sigma_T$ ($R_{\rm p}$ $<<$ 1.0), and the values of $\sigma_{NT}$ are almost identical to the linewidth ($\sigma_{\rm H_2CO(3_{0,3}-2_{0,2})}$).
The $\mathcal{M}$ derived from the $T_{\rm LTE}$ and 
$T_{\rm Radex}$ is 1.0--15.0 (average: 4.1$\pm$1.7) and 
0.8--14.6 (average: 3.8$\pm$1.8), respectively.
These results suggest that $\sigma_{\rm H_2CO(3_{0,3}-2_{0,2})}$ is strongly affected by non-thermal motions.
This is consistent with our finding that H$_2$CO mainly traces the low-velocity outflow components rather than the quiescent gas (see Section \ref{sec:velo}).

\subsubsection{Non-thermal velocity dispersion} \label{sec:dis_31} 
\begin{figure*}
\gridline{\fig{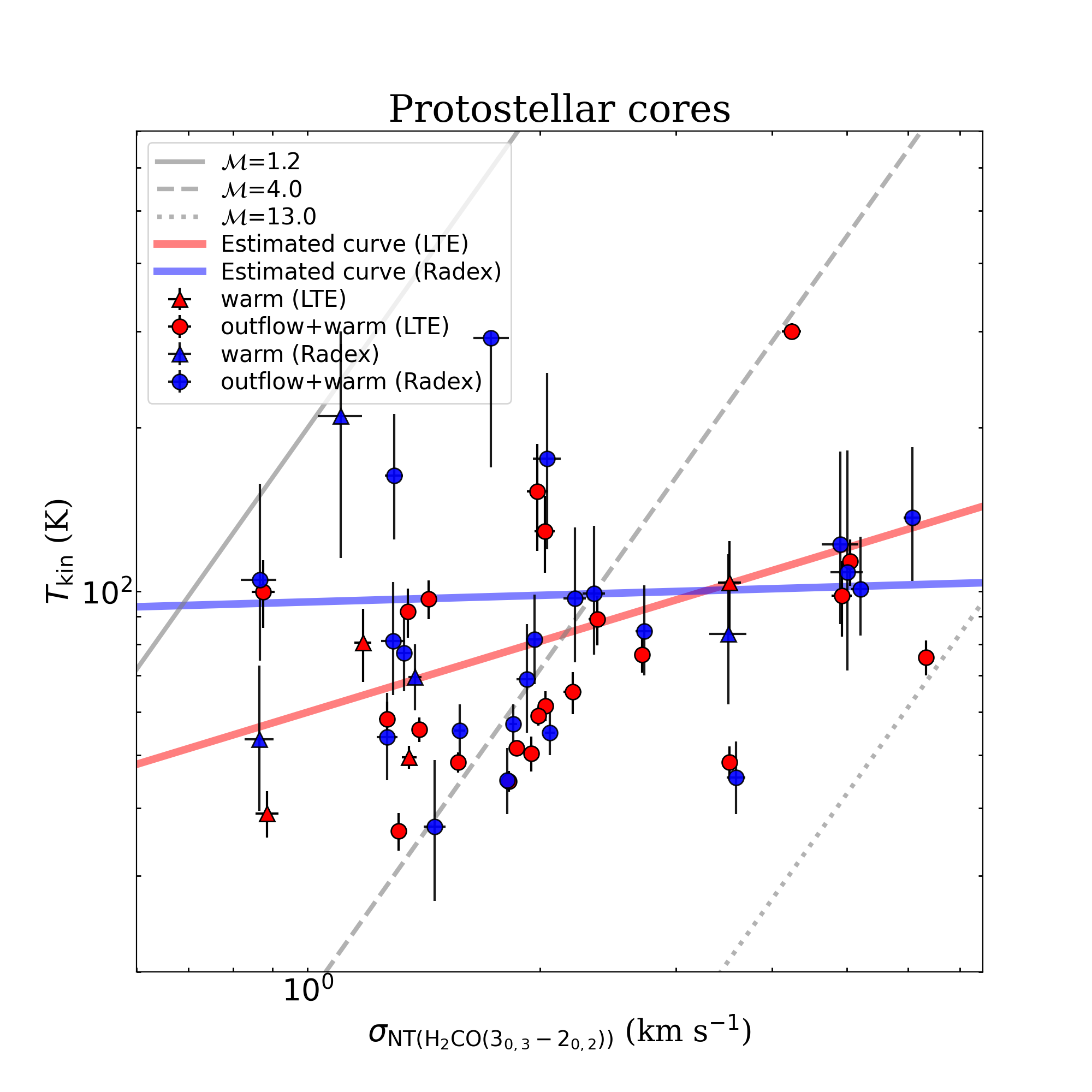}{0.45\textwidth}{}
\fig{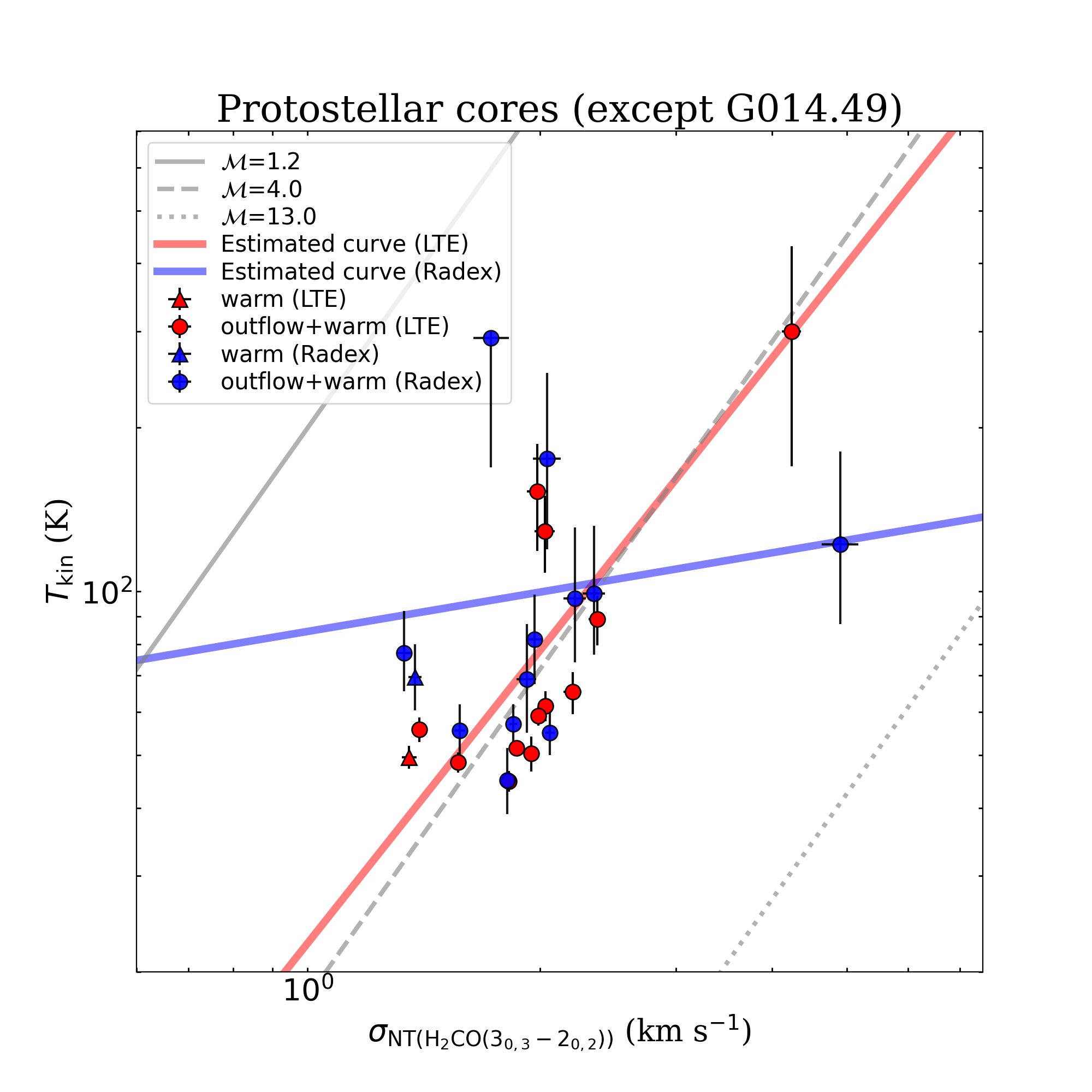}{0.45\textwidth}{}}
\caption{
Left: Relation between H$_2$CO($3_{0,3}-2_{0,2}$) non-thermal velocity dispersion ($\sigma_{\rm NT(H_2CO(3_{0,3}-2_{0,2}))}$)
and kinetic temperature ($T_{\rm kin}$) for  protostellar cores.
The red circles and triangles are for warm cores and warm \& outflow cores, respectively, based on the LTE assumption.
The blue circles and triangles are from 
Radex modeling.
The red (LTE) and blue (Radex) lines are the least-squares fitting results:
$T_{\rm LTE}$ = 60.0 ($\pm$ 13.7) $\times$ $\sigma_{\rm NT(H_2CO(3_{0,3}-2_{0,2}))}^{0.43(\pm 0.21)}$
and 
$T_{\rm Radex}$ = 95.6 ($\pm$ 19.0) $\times$ $\sigma_{\rm NT(H_2CO(3_{0,3}-2_{0,2}))}^{0.04(\pm 0.22)}$, respectively.
Right: Identical to the left panel but excluding the cores in G014.49.
The red (LTE) and blue (Radex) lines for the least-squares fitting are:
$T_{\rm LTE}$ = 60.3 ($\pm$ 15.8) $\times$ $\sigma_{\rm NT(H_2CO(3_{0,3}-2_{0,2}))}^{0.43(\pm 0.23)}$
and 
$T_{\rm Radex}$ = 85.0 ($\pm$ 19.0) $\times$ $\sigma_{\rm NT(H_2CO(3_{0,3}-2_{0,2}))}^{0.16(\pm 0.23)}$, respectively.
The gray solid, dashed, and dotted lines in both panels indicate Mach numbers $\mathcal{M}$ of 1.2, 4.0, and 13.0.
The error bars in both panels are derived by varying the assumed $T_{\rm kin}$ from minimum to maximum values.
We note that we only considered the uncertainties (i.e., from minimum to maximum values) of $T_{\rm kin}$ 
because the uncertainties for the measured FWHM are comparatively small and they can be neglected. 
}
\label{linewidth-temp_core} 
\end{figure*}

The left panel of Figure \ref{linewidth-temp_core} shows the relation between temperature and $\sigma_{NT}$ derived from the 
H$_2$CO(3$_{0,3}$--2$_{0,2}$) line ($\sigma_{\rm NT(H_2CO(3_{0,3}-2_{0,2}))}$).
There is a positive correlation between them.
This result is consistent with previous observations of other IRDCs and star-forming regions using single-dish telescopes
(temperature $\propto \sigma_{\rm NT(H_2CO(3_{0,3}-2_{0,2}))}^{0.66-1.26}$),
which suggests that the gas traced by H$_2$CO is turbulent and might be subject to turbulent heating \citep[e.g.,][]{Immer2016, Tang2017a,Tang2018b,Tang2018a}.
The right panel of Figure \ref{linewidth-temp_core} shows the same plot as the left panel but excludes the data from G014.492-00.139 (hereafter, G014.49).
In this case, we find an even stronger positive correlation.
This might be due to G014.49 having the most complicated velocity structure in this sample \citep{Radaelli2022}.
The spectra for many of the cores in G014.19 are different from a single Gaussian (e.g., double-peak structure).
Therefore, the fits for the cores in G014.49 are likely more uncertain than for other cores. 
Spearman's correlation coefficients and least-square fitting results are summarized in Table \ref{Relation_Temp-SNT}.

\begin{figure*}
\gridline{\fig{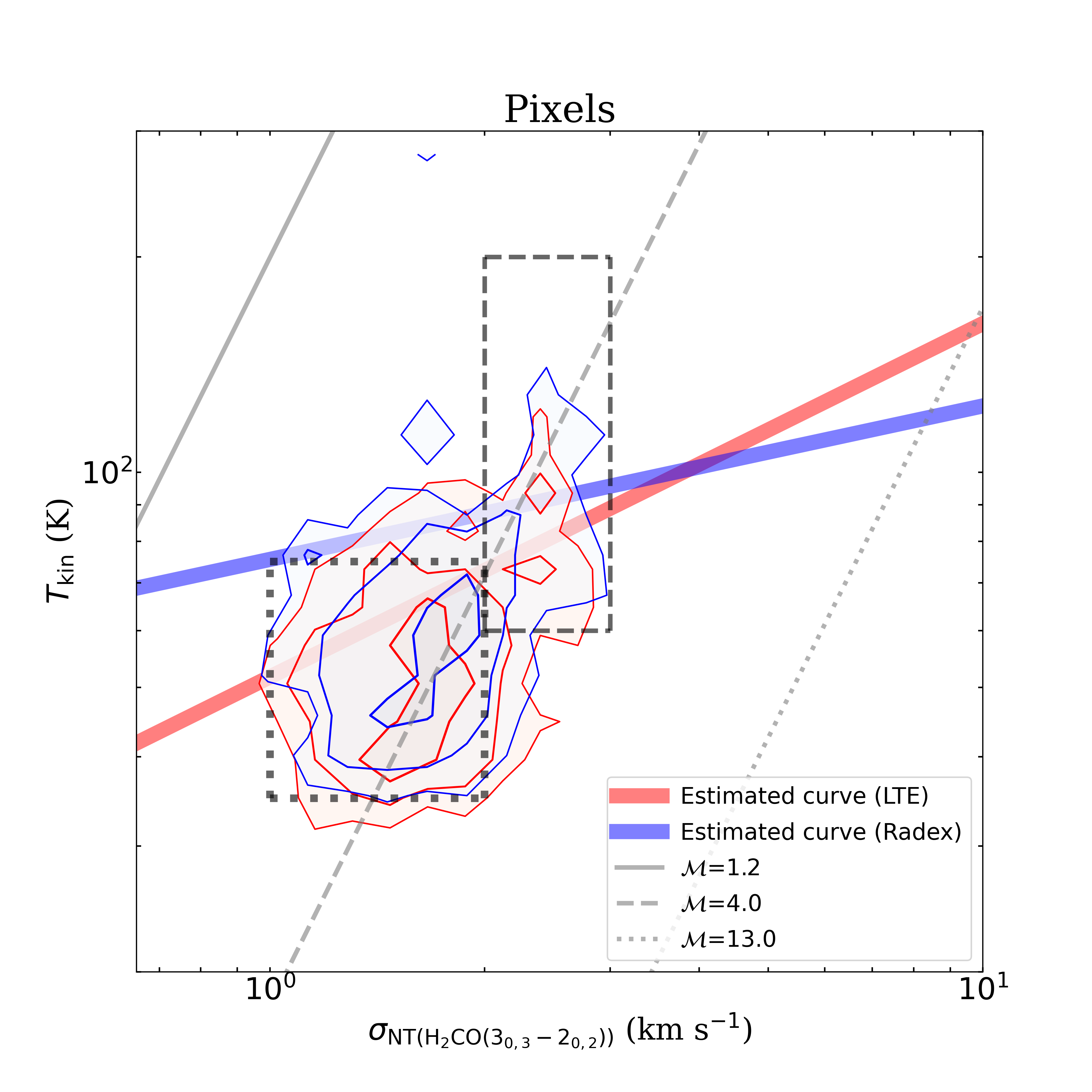}{0.45\textwidth}{}
\fig{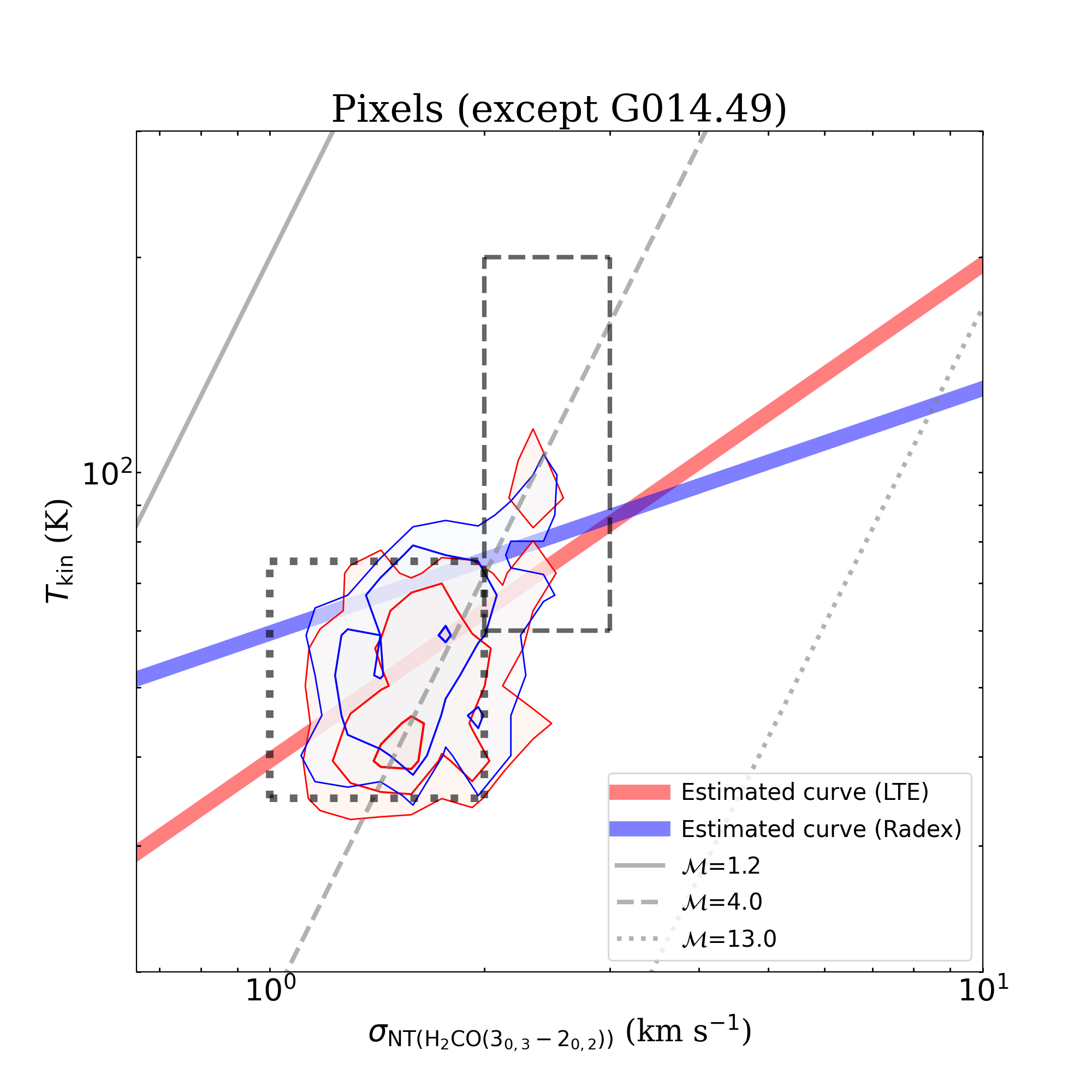}{0.45\textwidth}{}}
\caption{
Right: Pixel-by-pixel correlation between H$_2$CO($3_{0,3}-2_{0,2}$) non-thermal velocity dispersion ($\sigma_{\rm NT(H_2CO(3_{0,3}-2_{0,2}))}$)
and kinetic temperature ($T_{\rm kin}$).
The pixel scale was re-binned to 0$\farcs$6, approximately half of the synthesized beam, from the original value (0$\farcs$2).
Red and blue contours show the distributions of all pixels based on the LTE assumption and with Radex modeling, respectively
(LTE: 6, 12, and 24 points per cell, which is constructed by a 20 $\times$ 20 grid within the range of 
0.76 $\leq$ $\sigma_{\rm NT(H_2CO(3_{0,3}-2_{0,2}))}$ $\leq$ 7.6 and
28.9 $\leq$ $T_{\rm LTE}$ $\leq$ 299.3 with log scale;
Radex: 6, 12, and 24 points per cell, which is constructed by a 20 $\times$ 20 grid within the range of 
0.72 $\leq$ $\sigma_{\rm NT(H_2CO(3_{0,3}-2_{0,2}))}$ $\leq$ 8.3 and
25.5 $\leq$ $T_{\rm Radex}$ $\leq$ 296.3 with log scale).
The red (LTE) and blue (Radex) lines show the least-squares fitting results:
$T_{\rm LTE}$ = 51.7 ($\pm$ 1.6) $\times$ $\sigma_{\rm NT(H_2CO(3_{0,3}-2_{0,2}))}^{0.49(\pm 0.03)}$
and 
$T_{\rm Radex}$ = 75.5 ($\pm$ 2.4) $\times$ $\sigma_{\rm NT(H_2CO(3_{0,3}-2_{0,2}))}^{0.21(\pm 0.04)}$, respectively.
Right: Identical to the left panel but excluding pixels in G014.49
(LTE: 6, 12, and 24 points per cell, which is constructed by a 20 $\times$ 20 grid within the range of 
0.76 $\leq$ $\sigma_{\rm NT(H_2CO(3_{0,3}-2_{0,2}))}$ $\leq$ 4.9 and
29.0 $\leq$ $T_{\rm LTE}$ $\leq$ 289.8 with log scale;
Radex: 6, 12, and 24 points per cell, which is constructed by a 20 $\times$ 20 grid within the range of 
0.72 $\leq$ $\sigma_{\rm NT(H_2CO(3_{0,3}-2_{0,2}))}$ $\leq$ 5.3 and
25.5 $\leq$ $T_{\rm Radex}$ $\leq$ 296.3 with log scale).
The red (LTE) and blue (Radex) lines are the least-squares fittings, yielding
$T_{\rm LTE}$ = 39.5 ($\pm$ 1.5) $\times$ $\sigma_{\rm NT(H_2CO(3_{0,3}-2_{0,2}))}^{0.69(\pm 0.05)}$
and 
$T_{\rm Radex}$ = 59.5 ($\pm$ 2.8) $\times$ $\sigma_{\rm NT(H_2CO(3_{0,3}-2_{0,2}))}^{0.34(\pm 0.07)}$, respectively.
The gray solid, dashed, and dotted lines in both panels indicate the Mach numbers $\mathcal{M}$ of 1.2, 4.0, and 13.0.
}
\label{linewidth-temp_pix} 
\end{figure*}

Figure \ref{linewidth-temp_pix} is the same as  Figure \ref{linewidth-temp_core}, but for pixels.
We also confirm a positive correlation between $T_{\rm kin}$ and $\sigma_{\rm NT(H_2CO(3_{0,3}-2_{0,2}))}$, 
although there is large scattering.
The fitting results are also summarized in Table \ref{Relation_Temp-SNT}.
We also find a weak trend in this distribution.
The majority of the pixels is located in the area where
1.0 $\leq$ $\sigma_{\rm NT(H_2CO(3_{0,3}-2_{0,2}))}$ $\leq$ 2.0 km s$^{-1}$
and 
35.0 $\leq$ $T_{\rm kin}$ $\leq$ 75.0 K (hereafter, main-area; black dotted box in Figure \ref{linewidth-temp_pix}).
If the data are derived based on the LTE assumption or based on Radex modeling,
41 \% or 39 \% of the pixels are located in the main-area, respectively.
Discarding G014.49 (right panel of Figure \ref{linewidth-temp_pix}), it is 56 \% or 55 \%. 
In addition to the main-area, some data points are also concentrated in the area
where 
2.0 $\leq$ $\sigma_{\rm NT(H_2CO(3_{0,3}-2_{0,2}))}$ $\leq$ 3.0 km s$^{-1}$
and 
60.0 $\leq$ $T_{\rm kin}$ $\leq$ 200.0 K (hereafter, sub-area; black dashed box in Figure \ref{linewidth-temp_pix}).
In the sub-area, the $T_{\rm kin}$ rapidly increases with increasing $\sigma_{\rm NT(H_2CO(3_{0,3}-2_{0,2}))}$.
Both when derived under the LTE assumption or 
with Radex modeling, 
about 13\% of the pixels are located in the sub-area.
Without G014.49 (right panel of Figure \ref{linewidth-temp_pix}),
this becomes about 12 \% (both LTE and Radex).
In space, both data points located in the main-area and sub-area mainly trace outflow structures.
However, the data points in the sub-area tend to trace the edges of the outflows.
This result suggests that the kinematics along the edges of outflows are different from the main part of the outflows.

\subsubsection{Mach number} \label{sec:dis_32}
\begin{figure*}
\gridline{\fig{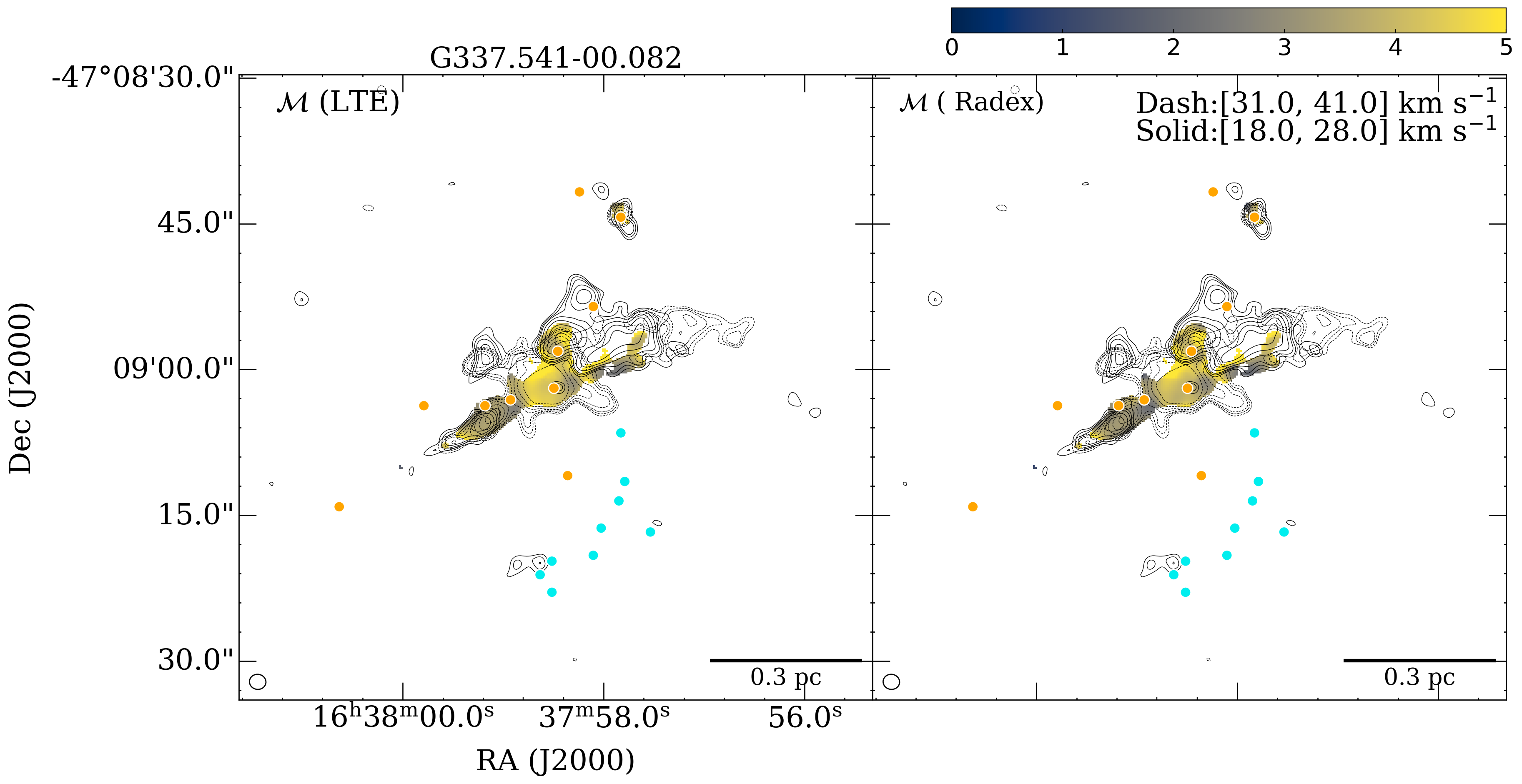}{0.9\textwidth}{}}
\vspace{-0.75cm}
\gridline{\fig{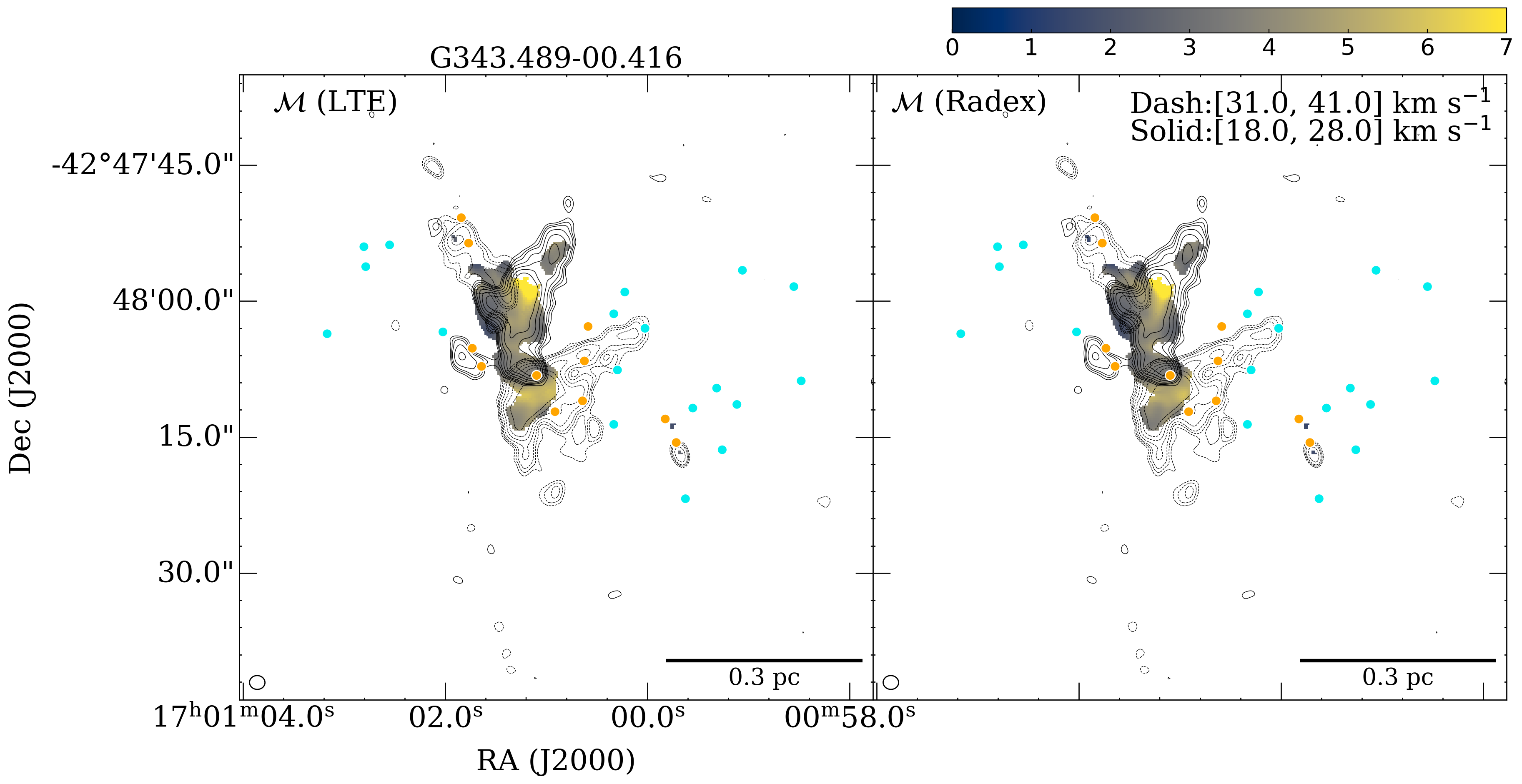}{0.9\textwidth}{}}
\vspace{-0.75cm}
\caption{
Mach number distribution (Left: LTE, Right: Radex) for 
G337.541-00.082 (top) and G343.489-00.416 (bottom).
The dashed and solid contours show the blue- and red-shifted components of H$_2$CO(3$_{0,3}$-3$_{0,2}$).
The cyan and orange circles indicate the positions of prestellar candidates and protostellar cores, respectively,
identified in ASHES \citep{Li2022}.
Synthesized beams are displayed at the bottom left corner in each panel.
}
\label{M-1} 
\end{figure*}

Figure \ref{M-1} shows the spatial distribution of $\mathcal{M}$
for two examples of IRDC clumps.
The range of $\mathcal{M}$ derived from LTE and Radex temperatures is 
1.5--12.3 (average: 4.5$\pm$2.4) and 
1.3--9.0 (average: 4.3$\pm$2.4), respectively (e.g., gray solid, dashed, and dotted lines in Figures \ref{linewidth-temp_core} and \ref{linewidth-temp_pix}).
These average values reveal a supersonic environment, while the largest values point at a hypersonic setting.
This finding suggests that the velocity distributions of the gas traced by H$_2$CO are 
significantly influenced by supersonic non-thermal components 
(e.g., turbulent motions, infall, outflows, shocks, and/or magnetic fields).
Furthermore, we find that $\mathcal{M}$ is larger at the edges of the outflows in G337.54 and G343.48 (Figure \ref{M-1}).
The values reach to more than 5, corresponding to hypersonic conditions.
Such high Mach numbers might be caused by shocks between outflows and the surrounding gas.
We note that high values for $\mathcal{M}$ are also detected in G014.49, which has several outflow components.
However, the values for G014.49 might be more uncertain 
due to the complicated velocity structures, as discussed in Section \ref{sec:dis_31}.

\subsection{Warm core candidates} \label{sec:dis_4}
Among all 26 cores with temperatures derived from H$_2$CO, 14 cores also show warmer line emission from 
HC$_3$N $J$ = 24-23 (hereafter, HC$_3$N(24-23)) with $E_u/k$ = 130 K and OCS $J$ =18-17 (hereafter, OSC(18-17)) with $E_u/k$ = 100 K (e.g., Figure \ref{Spectrum}).
Table \ref{warmline} summarizes the properties of these cores and their detected lines.
Both HC$_3$N(24-23) and OCS(18-17) are considered to trace hot cores \citep[e.g.,][]{Chapman2009, Silva2017, Tychoniec2021}.
We note that some cores also have c-C$_3$H$_2$ $J_{K_AK_C}$ = $5_{1,4}-4_{2,3}$ ($E_u/k$ = 35.4 K), which can trace outflow cavities \citep[e.g.,][]{Tychoniec2021}.
The $T_{\rm LTE}$ and $T_{\rm Radex}$ ranges of cores with warmer lines are 45--300 (average: 103$\pm$64) and 45--292 (average: 115$\pm$64), respectively, while
these range for cores without warmer lines are 36--104 (average: 61$\pm$22) and 37--210 (average: 79$\pm$46).
The $\sigma_{\rm H_2CO(3_{0,3}-2_{0,2})}$ to derive $T_{\rm LTE}$ and $T_{\rm Radex}$ for cores with warmer lines ranges from 
1.4 to 6.4 (average: 2.9$\pm$1.6) and 1.3 to 6.1 (average: 2.9$\pm$1.7).
The corresponding ranges for cores without warmer lines are
0.0---3.6 (average: 1.8$\pm$0.9) and 0.9--3.6 (average: 1.8$\pm$0.9).
These results suggest that cores with relatively warmer temperatures and larger $\sigma_{\rm H_2CO(3_{0,3}-2_{0,2})}$ tend to have warmer lines.

\subsection{Global connection to star formation} \label{sec:dis_5}
In summary, by utilizing the three transitions of H$_2$CO, we have successfully derived 
kinetic temperature distributions for 9 out of 12 70 $\micron$ dark IRDCs (26 out of 294 embedded cores in IRDCs), which are regarded as the coldest and most quiescent clouds \citep[e.g.,][]{Guzman2015}.
The temperature range spans from 30 K to 300 K, with more than half of the region exhibiting temperatures higher than 50 K.
Such high-temperature gas is considered to be associated with a low-velocity outflow component, which is significantly influenced by supersonic non-thermal motions.
Furthermore, 14 out of 294 cores show detections in HC$_3$N and/or OCS, which require excitation tempertures of more than  100 K
These results suggest that some of the embedded cores in 70 $\micron$ dark IRDCs have already entered the protostellar phase and may serve as hosts for hot core precursors.

\section{Summary}
We report kinetic temperatures of 12 IRDC clumps derived from the three transitions of H$_2$CO
with a resolution of about 5000 au.
The main results are as follows.
\begin{enumerate}
\item The spatial distribution of the H$_2$CO(3$_{0,3}$--2$_{0,2})$ transition is more extended than those from the  
H$_2$CO(3$_{2,2}$--2$_{2,1}$) and H$_2$CO(3$_{2,1}$--2$_{2,0}$) transitions.
The distributions of all three H$_2$CO lines is different from that of the 1.3 mm dust continuum emission.  
This result is not consistent with previous observations of IRDCs from single-dish telescopes with a resolution of about 0.5 pc \citep[e.g., ][]{Tang2017a, Tang2018b}.

\item The population of protostellar cores shows a weak positive correlation between the 1.3 mm flux density and
the H$_2$CO(3$_{0,3}$--2$_{0,2}$) integrated intensity.
In contrast, the results obtained for prestellar cores show no clear correlation.
Therefore, earlier results reporting a positive correlation based on single-dish telescopes
\citep[e.g., ][]{Tang2017a, Tang2018b} suggest that their targets are dominated by evolved regions which have already entered the protostellar phase.

\item The velocity dispersion of H$_2$CO is much larger than that of 
other molecular lines, such as C$^{18}$O, DCO$^+$, and N$_2$D$^+$ \citep{Li2022}.
We confirmed that H$_2$CO mainly traces the low-velocity outflow components rather than the quiescent gas in the early phase of high-mass star formation.
This is consistent with results for low-mass star-forming regions \citep[e.g., ][]{Tychoniec2021}.

\item The kinetic temperatures across all IRDCs range from 26 to 300 K. This is much larger than what is reported in  previous studies using single-dish telescopes \citep[e.g., ][]{Tang2017a}.
This temperature range is also larger than the temperatures derived from NH$_3$ with the same resolution.
This result suggests that H$_2$CO is more sensitive to star-formation activity than NH$_3$.

\item The Mach numbers derived from H$_2$CO reach about 15 with an average of about 4.
This indicates that the velocity distribution of gas traced by H$_2$CO is significantly influenced by supersonic non-thermal components.
High Mach numbers of more than 5 are mainly detected at the edges of outflows. 
These high numbers might be caused by shocks between outflows and surrounding material.

\item 14 protostellar cores show emission from warmer lines, such as HC$_3$N and OCS.
This suggests that some cores embedded in 70 $\mu$m dark IRDCs may be developing hot core precursors given than 
these lines are known hot-core tracers \citep[e.g.,][]{Chapman2009,Tychoniec2021}.

\end{enumerate}

Altogether, these results confirm that some of the embedded cores in 70 $\mu m$ dark IRDCs have already entered the protostellar phase.

\begin{acknowledgments}
PS was partially supported by a Grant-in-Aid for Scientific Research (KAKENHI Number JP22H01271 and JP23H01221) of JSPS. 
PMK acknowledges support from the National Science and Technology Council (NSTC) in Taiwan through grants
NSTC 112-2112-M-001-049 -, NSTC 111- 2112-M-001-070-, and NSTC 110-2112-M-001-057-.
GS acknowledges the projects PRIN-MUR 2020 MUR BEYOND-2p (``Astrochemistry beyond the second period elements'', Prot. 2020AFB3FX) and INAF-Minigrant 2023 TRIESTE (``TRacing the chemIcal hEritage of our originS: from proTostars to planEts''; PI: G. Sabatini).
KT was supported by JSPS KAKENHI (Grant Number JP20H05645). This paper makes use of the following ALMA data: ADS/JAO.ALMA\#2015.1.01539.S (PI: P.~Sanhueza). ALMA is a partnership of ESO (representing its member states), NSF (USA) and NINS (Japan), together with NRC (Canada), MOST and ASIAA (Taiwan), and KASI (Republic of Korea), in cooperation with the Republic of Chile. The Joint ALMA Observatory is operated by ESO, AUI/NRAO and NAOJ.
Data analysis was in part carried out on the Multi-wavelength Data Analysis System operated by the Astronomy Data Center (ADC), National Astronomical Observatory of Japan.
\end{acknowledgments}
\clearpage
\begin{figure*}
\epsscale{1.3}
\plotone{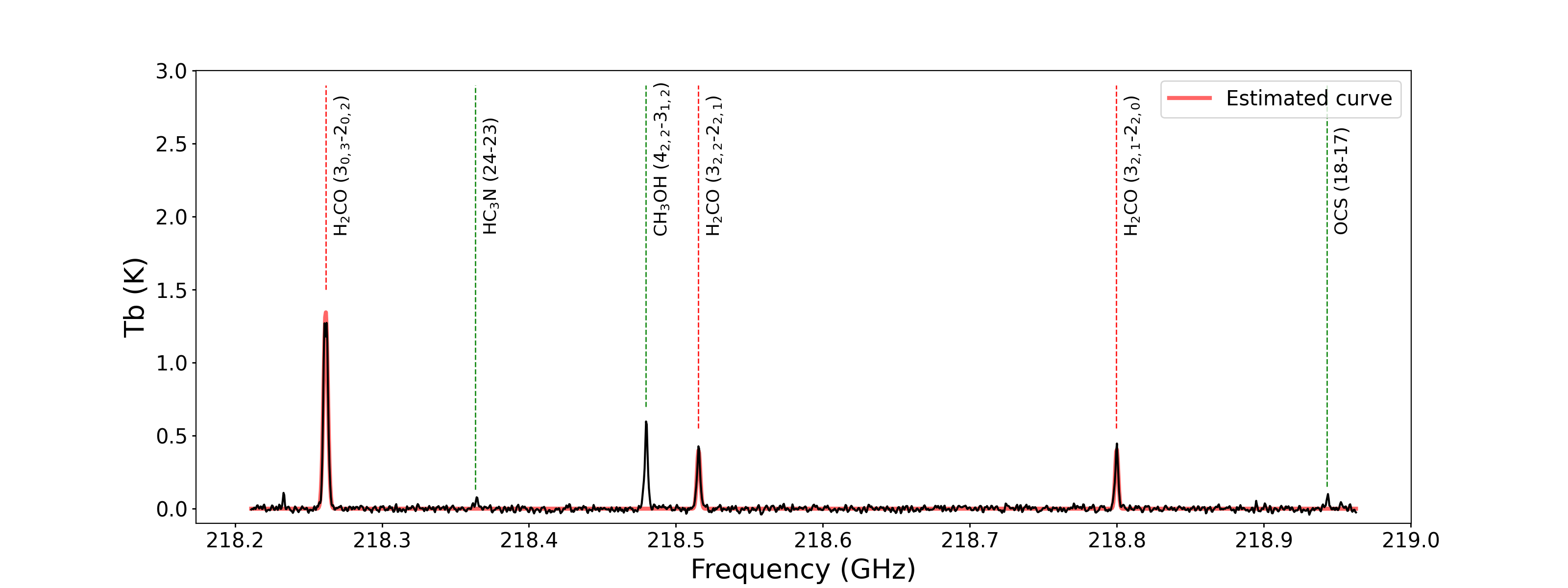}
\caption{
Example of detected spectrum in IRDC clumps (clump: G337.541-00.082, core: ALMA1).
The black curve is the observed spectrum.
The red curve indicates the results of the Gaussian fitting.
}
\label{Spectrum} 
\end{figure*}
\begin{longrotatetable}
\begin{deluxetable*}{ccc|ccc|ccc|ccccc|ccccc}
\tablecaption{Core detection rate of three H$_2$CO emission lines.} \label{tab:ratio}
\tablehead{
& \colhead{IRDC} &&& \colhead{Prestellar candidates\tablenotemark{a}} &&& \multicolumn{10}{c}{Protostellar cores\tablenotemark{a}} \\
\cline{7-19}
& \colhead{} &&& \colhead{}  &&& \colhead{outflow} &&& \multicolumn3c{warm} &&& \multicolumn3c{outflow \& warm} \\
\cline{7-9}\cline{10-14}\cline{15-19}
& \colhead{} &&& \colhead{f1\tablenotemark{b}}  &&& \colhead{f1}  &&&
\colhead{f1}  & \colhead{f2\tablenotemark{b}}  &  \colhead{f3\tablenotemark{b}} &&&
\colhead{f1}  & \colhead{f2}  &  \colhead{f3} \\
& \colhead{} &&& \colhead{(\%)}  &&& \colhead{(\%)}  &&&
\colhead{(\%)}  & \colhead{(\%)}  &  \colhead{(\%)} &&&
\colhead{(\%)}  & \colhead{(\%)}  &  \colhead{(\%)} 
} 
\startdata
&G010.991-00.082  &&& 73 (11/15)\tablenotemark{c}   &&&  80 (4/5)   &&& 100 (4/4)  & 0 (0/4)    & 25 (1/4)   &&& 100 (4/4)  & 100 (4/4)  & 100 (4/4)  \\
&G014.492-00.139  &&& 56 (5/9)     &&& 0 (0/1)    &&& 89 (8/9)   & 67 (6/9)   & 67 (6/9)   &&& 94 (17/18) & 89 (16/18) & 83 (15/18) \\
&G028.273-00.167  &&& 11 (4/9)     &&& 100 (1/1)  &&& 100 (2/2)  & 0 (0/2)    & 0 (0/2)    &&& 100 (1/1)  & 100 (1/1)  & 100 (1/1)  \\
&G327.116-00.294  &&& 24 (4/17)    &&& 100 (1/1)  &&& --- (0/0)  & --- (0/0)  & --- (0/0)  &&& 100 (3/3)  & 67 (2/3)   & 100 (3/3)  \\
&G331.372-00.116  &&& 61 (19/31)   &&& 60 (3/5)   &&& 100 (2/2)  & 0 (0/2)    & 0 (0/2)    &&& 100 (1/1)  & 0 (0/1)    & 0 (0/1)    \\
&G332.969-00.029  &&& 39 (7/18)    &&& --- (0/0)  &&& 100 (2/2)  & 50 (1/2)   & 0 (0/2)    &&& --- (0/0)  & --- (0/0)  & --- (0/0)  \\
&G337.541-00.082  &&& 78 (7/9)     &&& 100 (1/1)  &&& 80 (4/5)   & 20 (1/5)   & 40 (2/5)   &&& 100 (4/4)  & 100 (4/4)  & 100 (4/4)  \\
&G340.179-00.242  &&& 57 (8/14)    &&& --- (0/0)  &&& 100 (2/2)  & 0 (0/2)    & 0 (0/2)    &&& --- (0/0)  & --- (0/0)  & --- (0/0)  \\
&G340.222-00.167  &&& 48 (10/21)   &&& --- (0/0)  &&& --- (0/0)  & --- (0/0)  & --- (0/0)  &&& --- (0/0)  & --- (0/0)  & --- (0/0)  \\
&G340.232-00.146  &&& 64 (7/11)    &&& --- (0/0)  &&& 100 (3/3)  & 33 (1/3)   & 0 (0/3)    &&& 100 (2/2)  & 100 (2/2)  & 100 (2/2)  \\
&G341.039-00.114  &&& 52 (13/25)   &&& 86 (6/7)   &&& --- (0/0)  & --- (0/0)  & --- (0/0)  &&& 100 (3/3)  & 67  (2/3)  & 33  (1/3)  \\
&G343.489-00.416  &&& 44 (8/18)    &&& --- (0/0)  &&& 100 (8/8)  & 25 (2/8)   & 38 (3/8)   &&& 100 (3/3)  & 100 (3/3)  & 100 (3/3)  \\ \hline
&Total            &&& 52 (103/197) &&& 76 (16/21) &&& 95 (35/37) & 30 (11/37) & 32 (12/37) &&& 97 (38/39) & 87 (34/39) & 85 (33/39) \\
\enddata
\tablenotetext{a}{
Detection rates for prestellar candidates and ``outflow" protostellar cores are only derived with respect to  H$_2$CO(3$_{0,3}$--2$_{0,2}$) because other H$_2$CO lines are not detected in these cores due to the definition of cores.
}
\tablenotetext{b}{
f1, f2, and f3 stand for H$_2$CO(3$_{0,3}$--2$_{0,2}$), H$_2$CO(3$_{2,2}$--2$_{2,1}$), and H$_2$CO(3$_{2,1}$--2$_{2,0}$), respectively.
}
\tablenotetext{c}{
Values inside the parentheses indicate the ratio of detected cores to the total
number of cores.
}
\end{deluxetable*}
\end{longrotatetable}
\begin{figure*}
\gridline{\fig{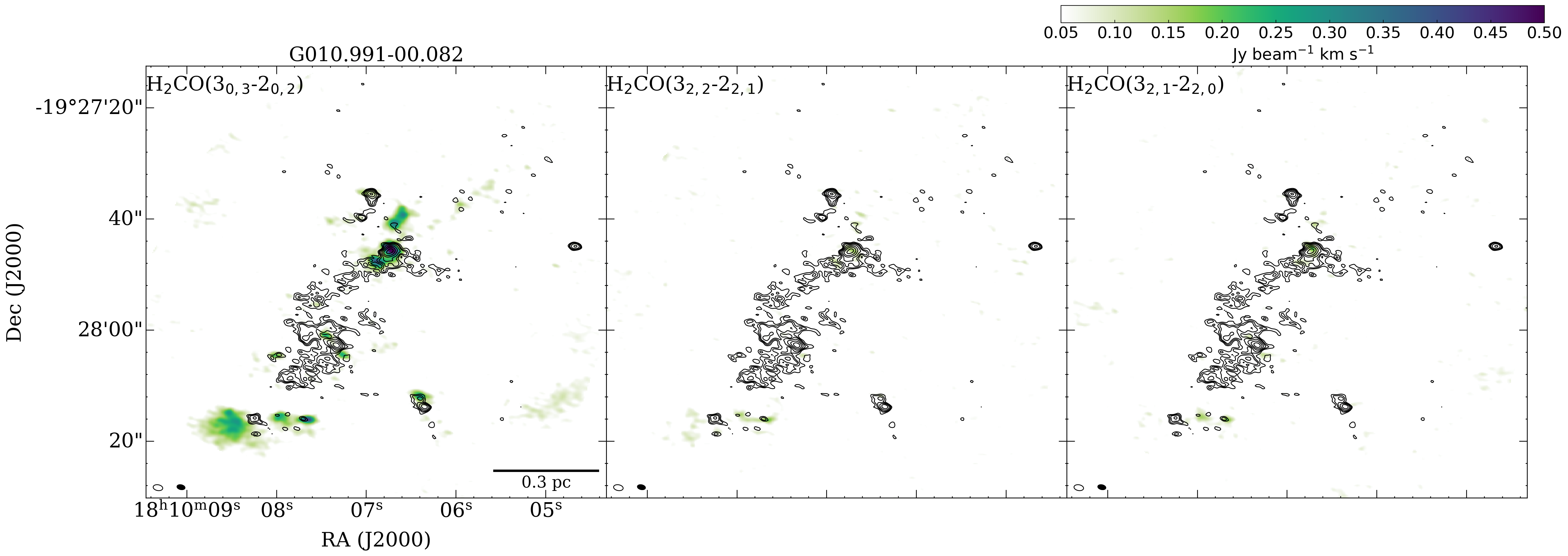}{0.95\textwidth}{}}
\vspace{-0.75cm}
\gridline{\fig{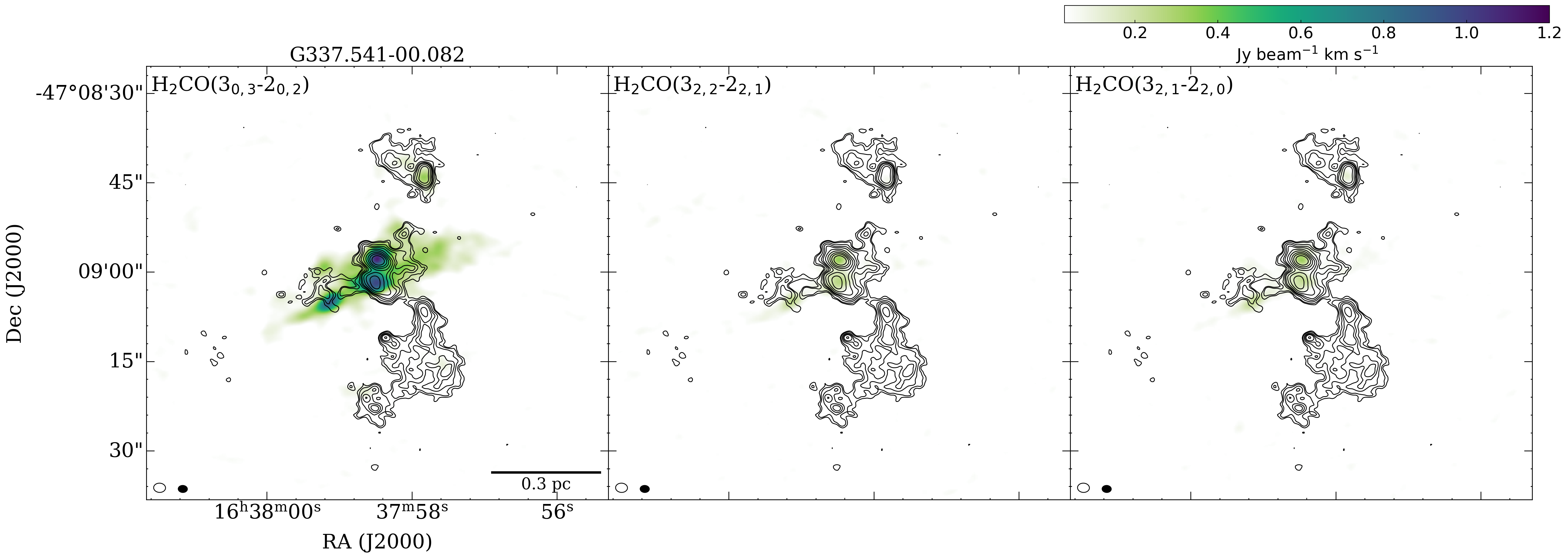}{0.95\textwidth}{}}
\vspace{-0.75cm}
\caption{
Integrated intensity distribution of the three observed transitions of H$_2$CO
(Left: H$_2$CO(3$_{03}$-3$_{02}$),
Middle: H$_2$CO(3$_{22}$-3$_{21}$),
Right: H$_2$CO(3$_{21}$-3$_{20}$))
for two representative clumps: G10.991-00.082 (top) and G337.541-00.082 (bottom).
The black contours show the 1.3 mm dust continuum.
Contour levels are 3, 4, 5, 7, 10, 14, and 20 $\times$ $\sigma$, where
$\sigma$ = 0.115 mJy beam$^{-1}$ for G010.991-00.082
(1$\farcs$1 resolution) and 
3, 4, 6, 8, 10, 14, 20, 30, 45 and 75 $\times$ $\sigma$, where $
\sigma$ = 0.068 mJy beam$^{-1}$ for G337.541-00.082
(1$\farcs$2 resolution).
Synthesized beams are displayed at the bottom left in each panel
(open ellipses: H$_2$CO; filled ellipses: continuum).
The images of the other 10 IRDC clumps are presented in the appendix (Figure \ref{A-MOM0}).
}
\label{MOM0} 
\end{figure*}
\begin{figure*}
\epsscale{1.0}
\plotone{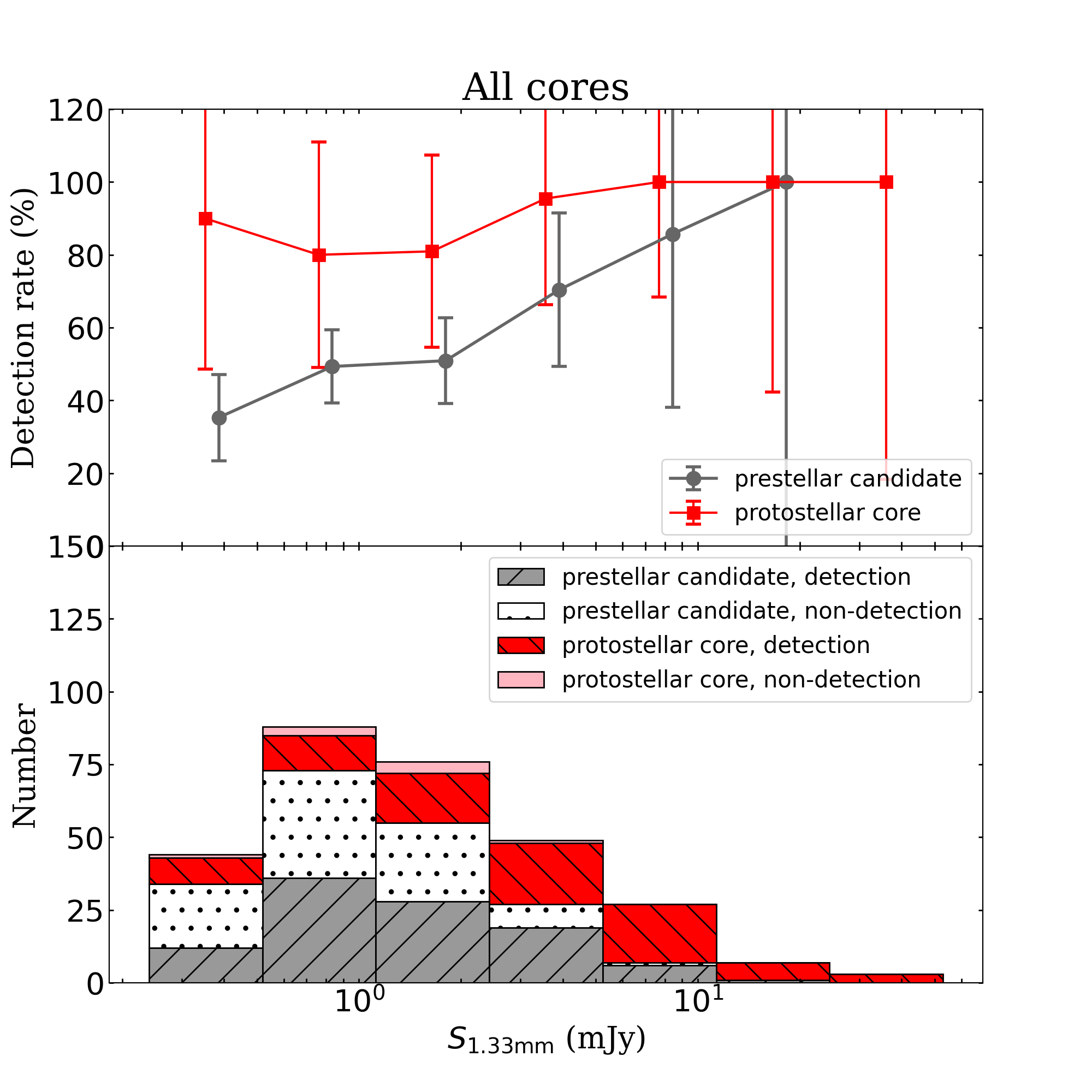}
\caption{
Top: Relation between the 1.3 mm continuum flux ($S_{\rm 1.3mm}$) and core detection rate of H$_2$CO(3$_{0,3}$--2$_{0,2}$).
The gray-filled circles and red-filled squares indicate prestellar candidates and protostellar cores, respectively.
The error bars show the Poisson counting uncertainties (1 $\sigma$), which described by 1 / $\sqrt{n}$ (n: number).
Bottom: Relation between $S_{\rm 1.3mm}$ and core numbers.
The gray, white, red, and light-magenta histograms indicate the number of prestellar candidates with H$_2$CO(3$_{0,3}$--2$_{0,2}$) detected, 
prestellar candidates with H$_2$CO(3$_{0,3}$--2$_{0,2}$) not detected,
protostellar cores with H$_2$CO(3$_{0,3}$--2$_{0,2}$) detected, 
and protostellar cores with H$_2$CO(3$_{0,3}$--2$_{0,2}$) not detected.
}
\label{Hist_DR} 
\end{figure*}
\begin{figure*}
\gridline{\fig{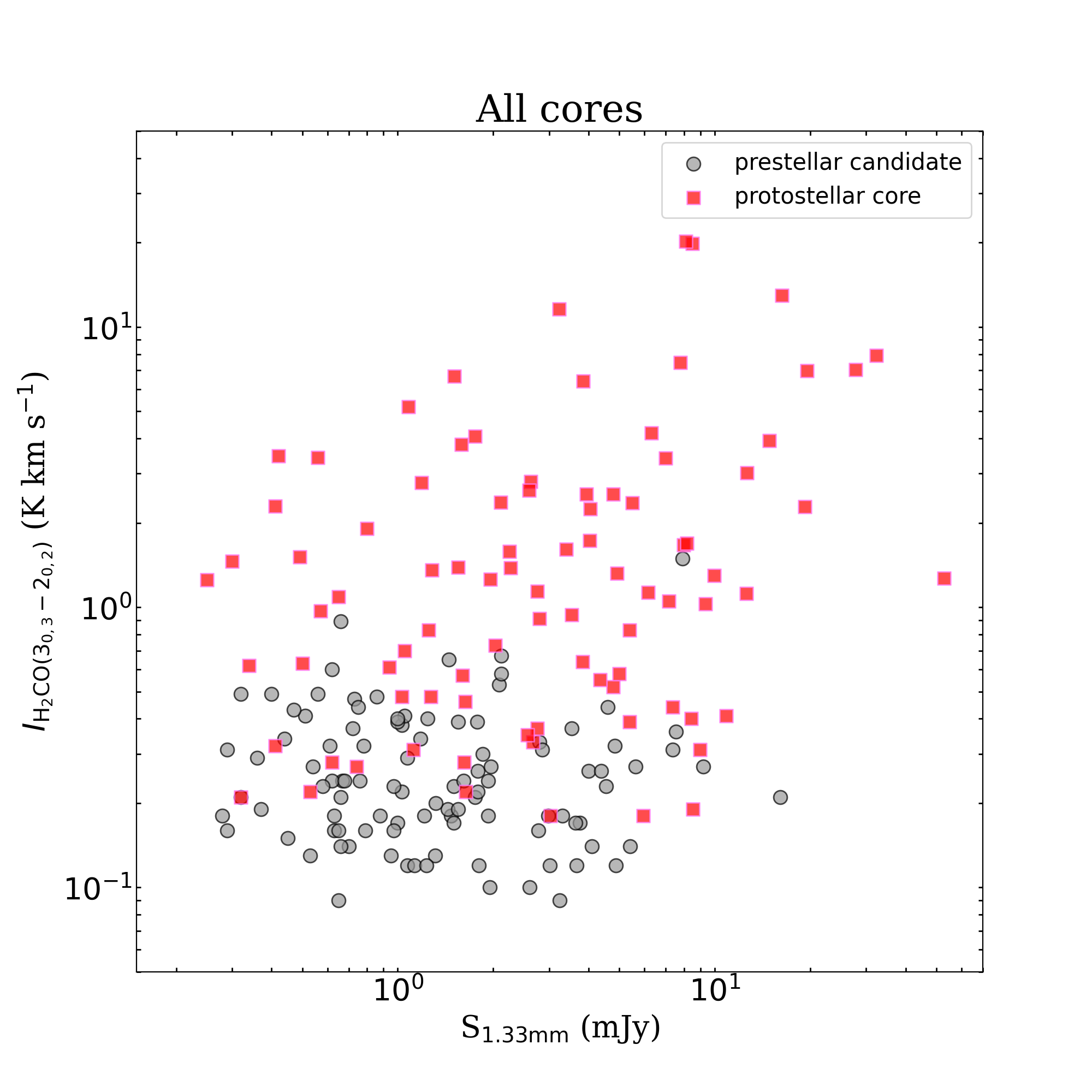}{0.45\textwidth}{}
\fig{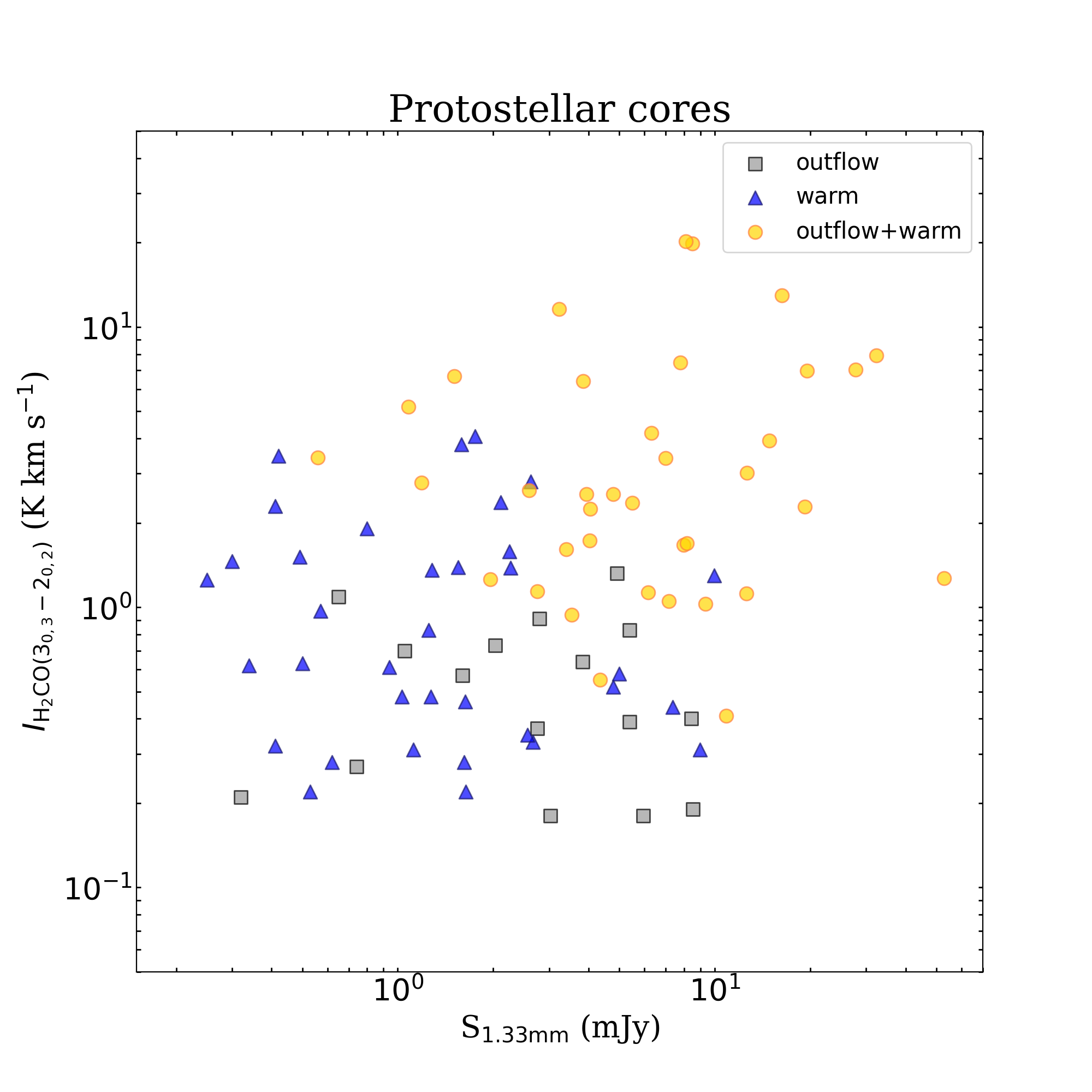}{0.45\textwidth}{}}
\caption{
Relation between 1.3mm continuum flux density ($S_{\rm 1.3mm}$) and H$_2$CO(3$_{0,3}$--2$_{0,2}$) integrated intensity ($I_{\rm H_2CO(3_{0,3}-2_{0,2})}$)
of cores identified in ASHES (Left: all cores, Right: only protostellar cores).
The gray circles and red squares in the left panel indicate the prestellar candidates and protostellar cores, respectively.
The gray squares, blue triangles, and yellow circles in the right panel indicate the three classified protostellar cores: outflow core, warm core, and warm and outflow core, respectively.
}
\label{int-relation_core} 
\end{figure*}
\begin{figure*}
\gridline{\fig{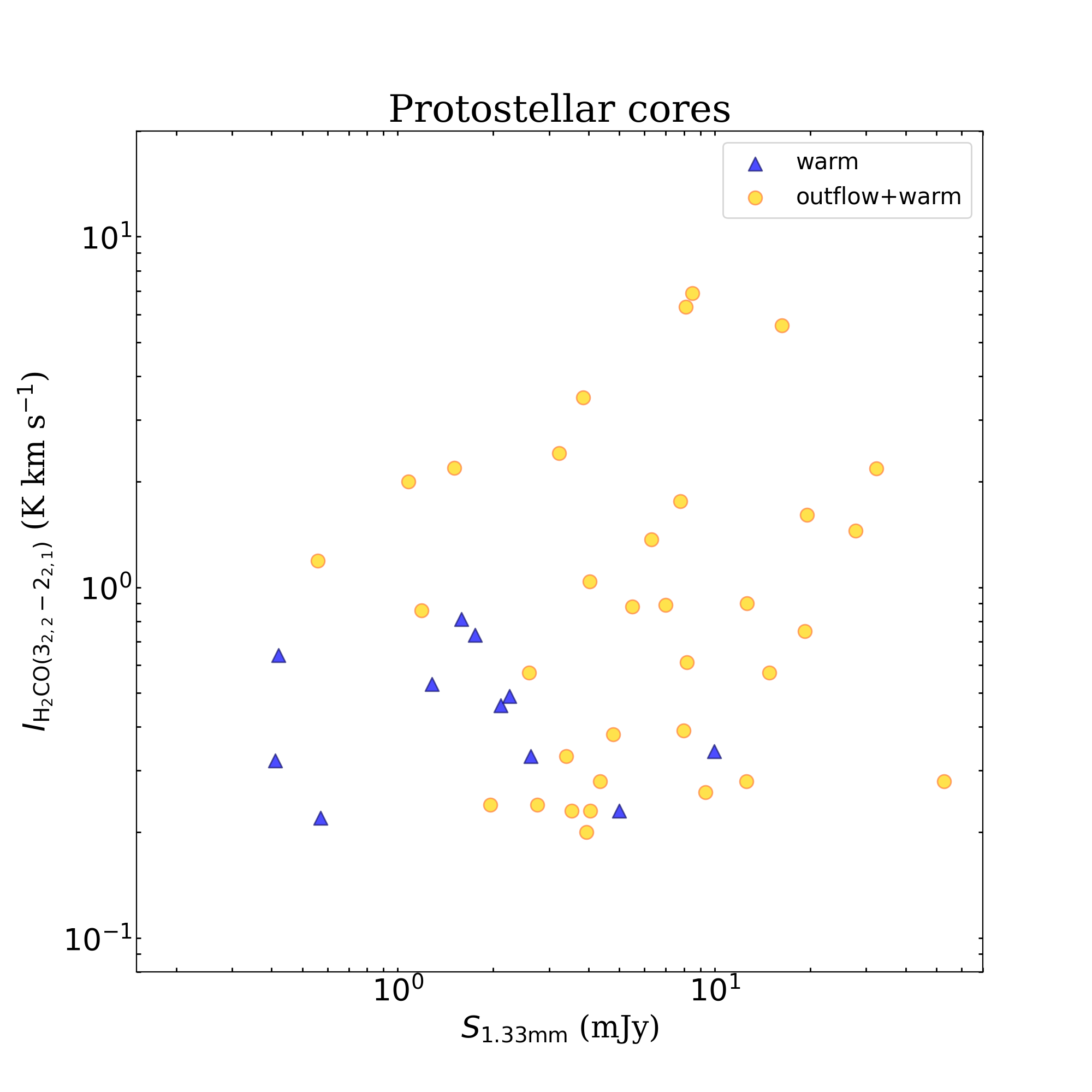}{0.45\textwidth}{}
 \fig{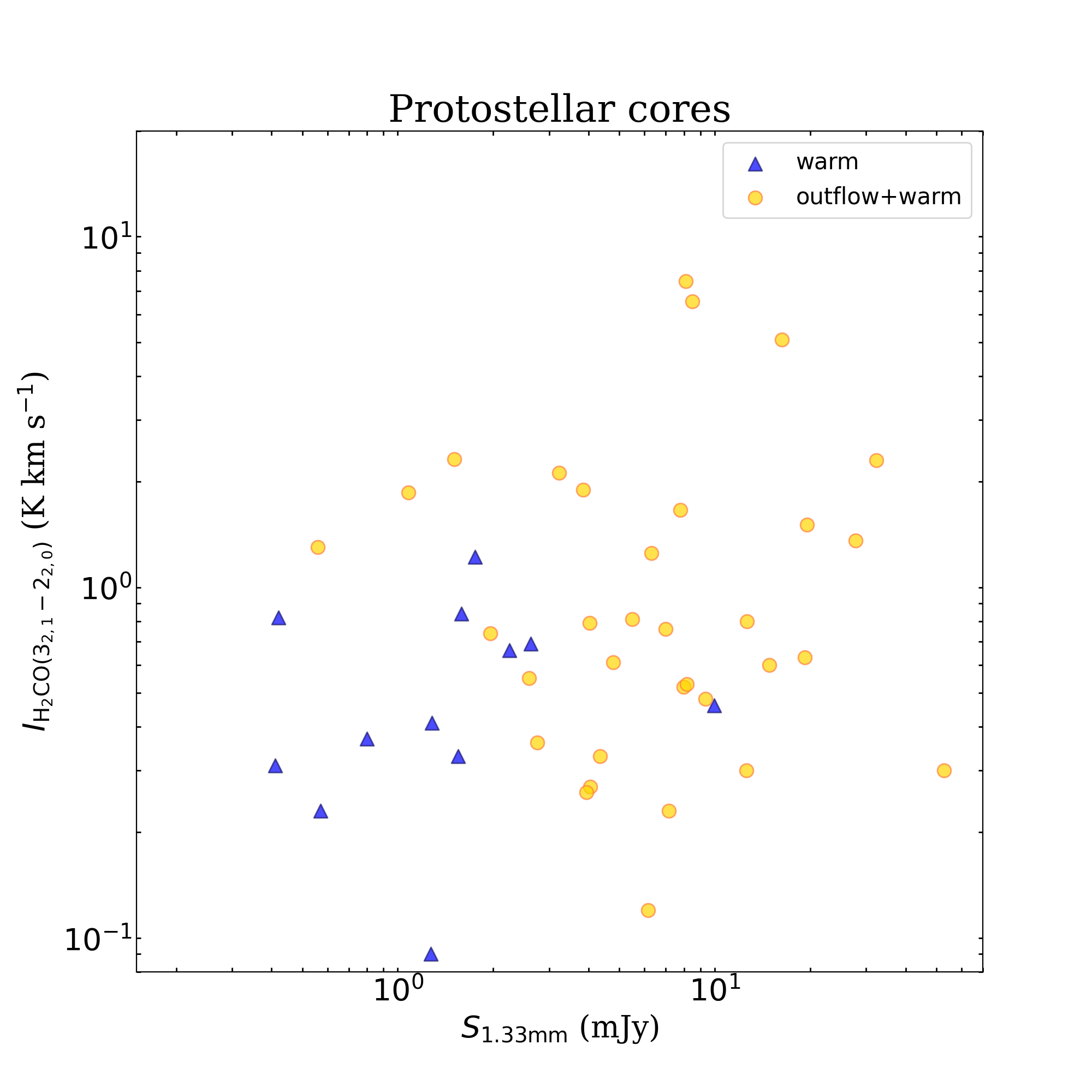}{0.45\textwidth}{}}
\caption{
Relation between 1.3mm continuum flux density ($S_{\rm 1.3mm}$) and H$_2$CO(3$_{2,2}$--2$_{2,1}$) integrated intensity ($I_{\rm H_2 CO(3_{2,2}-2_{2,1})}$: Left)
and H$_2$CO(3$_{2,1}$--2$_{2,0}$) integrated intensity ($I_{\rm H_2 CO(3_{2,1}-2_{2,0})}$: Right) of protostellar cores identified in ASHES.
The blue triangles and yellow circles indicate the two classified protostellar cores: warm core and warm and outflow core, respectively.
We note that there are no data of the other classified protostellar cores: outflow core due to the definition of core.
}
\label{int-relation_core_f2f3} 
\end{figure*}
\begin{deluxetable*}{cccc}
\tablecaption{Spearman rank-order coefficients for relations between $I_{\rm H_2CO 3_{0,3}-2_{0,2}}$ and $S_{\rm 1.3mm}$. \label{CC-Ih2co-S1mm}}
\tablewidth{0pt}
\tablehead{
\multicolumn2c{Core} & \colhead{$r$-value} &  \colhead{$p$-value}
}
\startdata
\multicolumn2c{Prestellar candidate}                      & 0.07  & 0.49   \\ \hline
Protostellar core & all                                   & 0.25  & 0.02  \\
                  & outflow core                          & -0.19 & 0.49   \\
                  & warm core                             & -0.11 & 0.52   \\
                  & warm and outflow core                 & 0.10  & 0.55   \\  
                  & outflow core + warm core              & -0.19 & 0.18   \\
                  & warm core + warm and outflow core     & 0.28  & 0.04 \\
                  & outflow core + warm and outflow core  & 0.31 & 0.01 \\
\enddata
\end{deluxetable*}
\begin{figure*}
\gridline{\fig{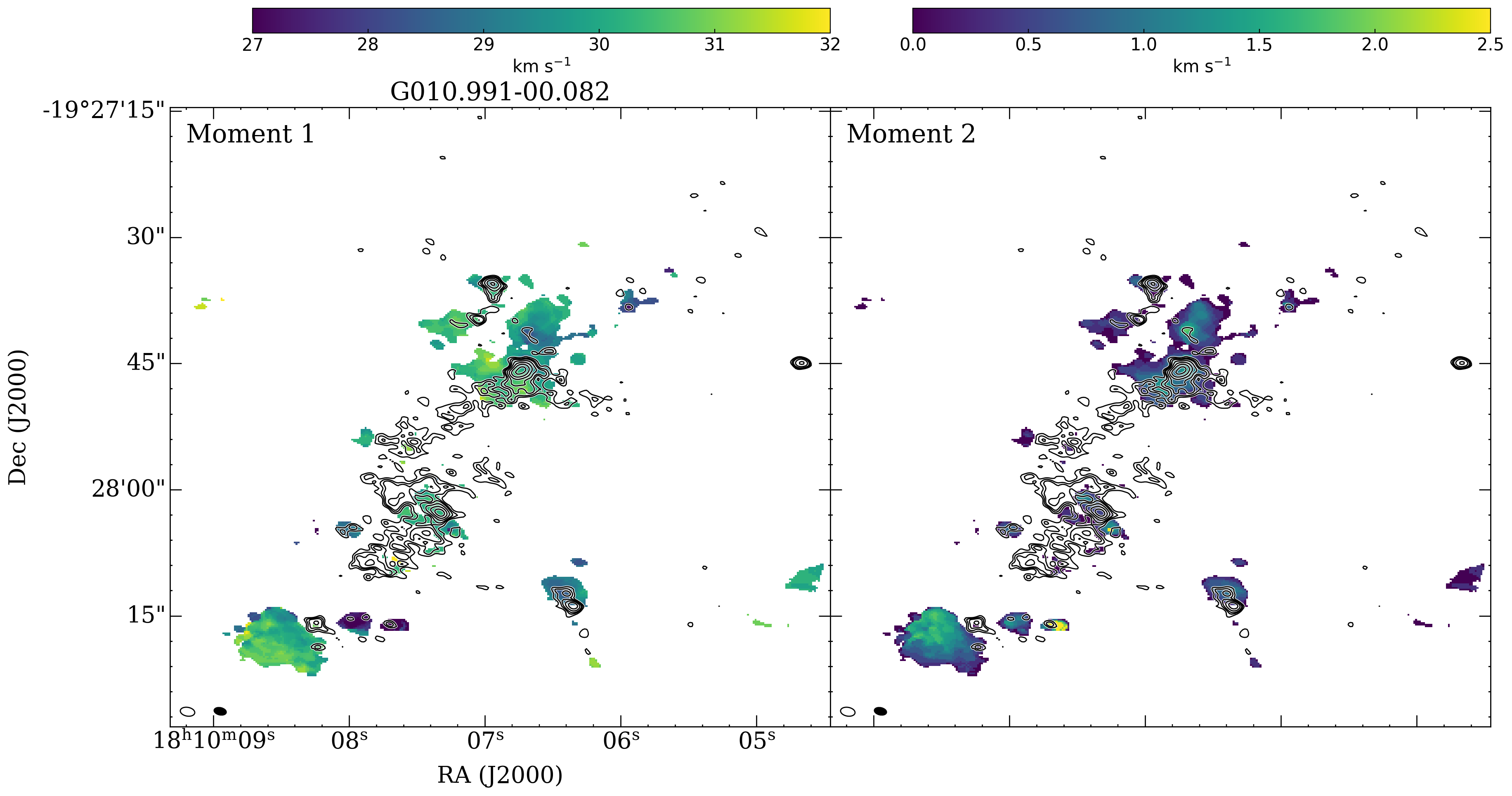}{1.0\textwidth}{}}
\vspace{-0.75cm}
\gridline{\fig{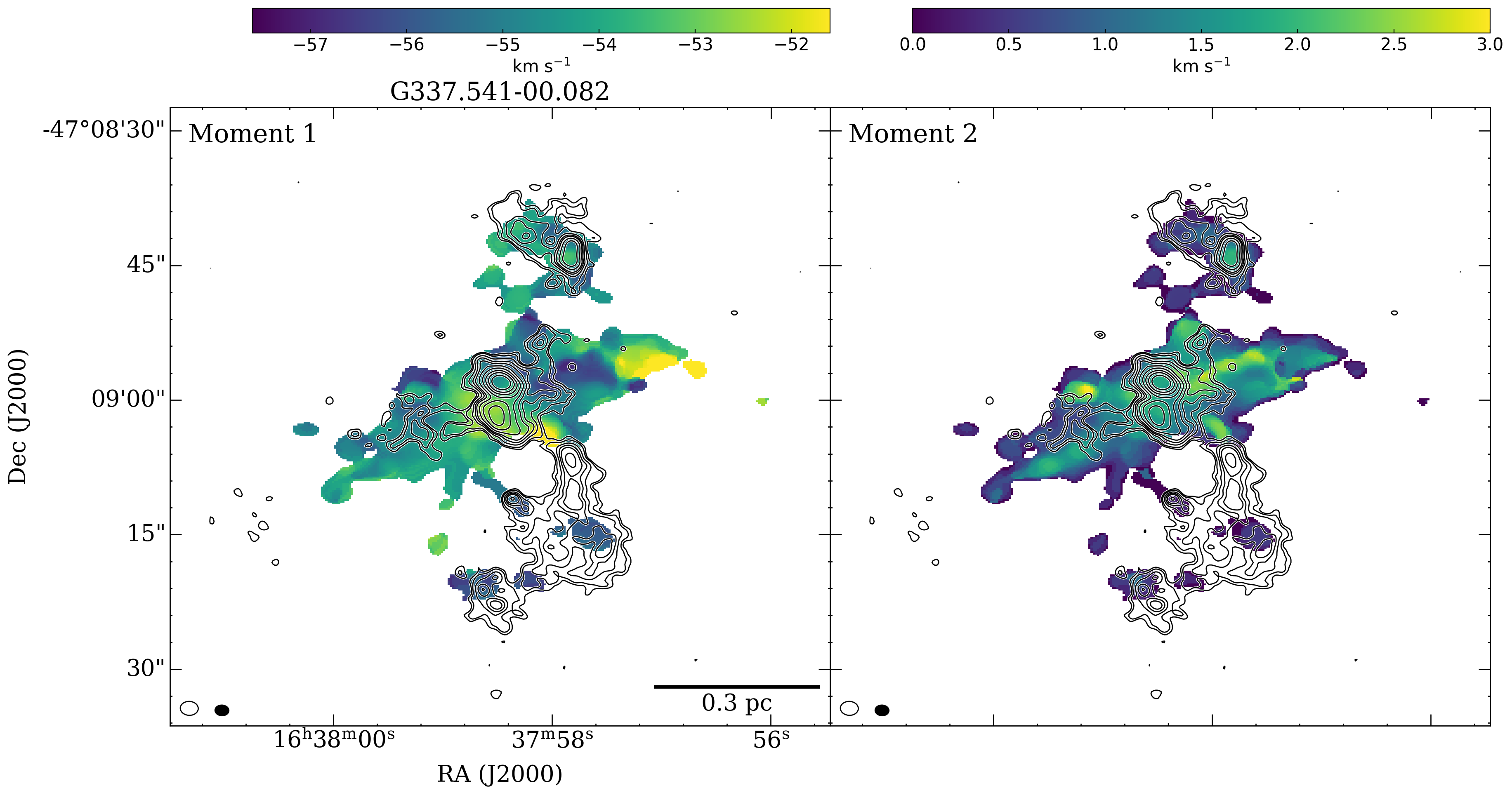}{1.0\textwidth}{}}
\vspace{-0.75cm}
\caption{
H$_2$CO(3$_{0,3}$--2$_{0,2}$)
intensity-weighted mean velocity map (moment 1, Left)
and velocity dispersion map (moment 2, Right) for two representative IRDC clumps.
H$_2$CO(3$_{0,3}$--2$_{0,2}$) emission weaker than 5 $\sigma$ is clipped.
The black contours are 1.3mm dust continuum.
Contour levels are 3, 4, 5, 7, 10, 14, and 20 $\times$ $\sigma$, where
$\sigma$ = 0.115 mJy beam$^{-1}$ for G010.991-00.082
(1$\farcs$1 resolution) and 
3, 4, 6, 8, 10, 14, 20, 30, 45, and 75 $\times$ $\sigma$, where $
\sigma$ = 0.068 mJy beam$^{-1}$ for G337.541-00.082
(1$\farcs$2 resolution).
Synthesized beams are displayed at the bottom left of each panel
(open ellipses: H$_2$CO; filled ellipses: continuum).
}
\label{MOM12} 
\end{figure*}
\begin{figure*}
\epsscale{1.0}
\plotone{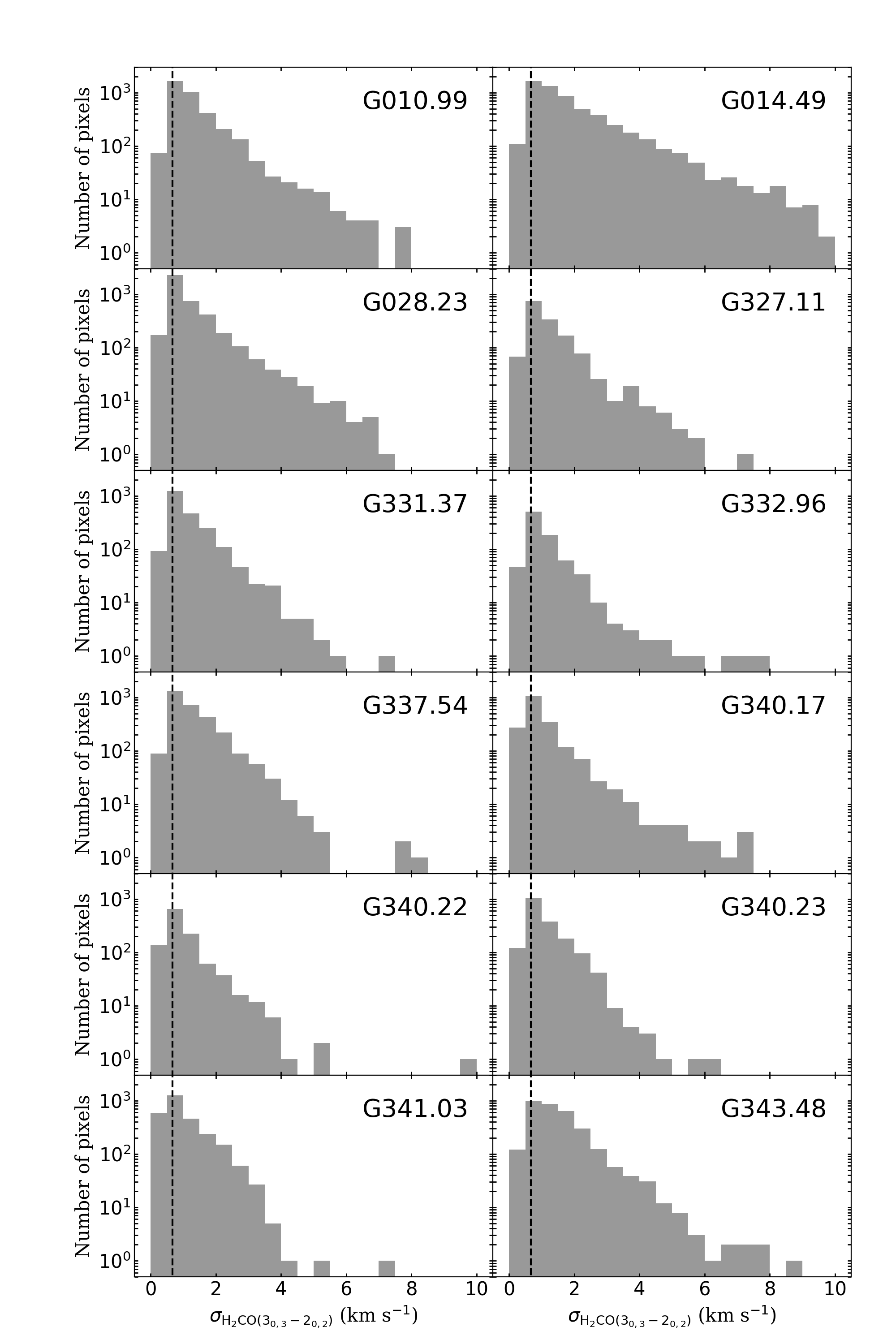}
\caption{
Number distribution of H$_2$CO(3$_{0,3}$--2$_{0,2}$) velocity dispersion ($\sigma_{\rm H_2CO(3_{0,3}-2_{0,2})}$) for pixels in each IRDC clump. 
The pixel scale was re-binned to 0$\farcs$6, approximately half of the synthesized beam, from the original value (0$\farcs$2).
H$_2$CO(3$_{0,3}$--2$_{0,2}$) emission weaker than 3$\sigma$ is clipped.
The black-dashed lines indicate the channel size (0.671 km s$^{-1}$).
}
\label{Hist_dv} 
\end{figure*}
\begin{figure*}
\gridline{\fig{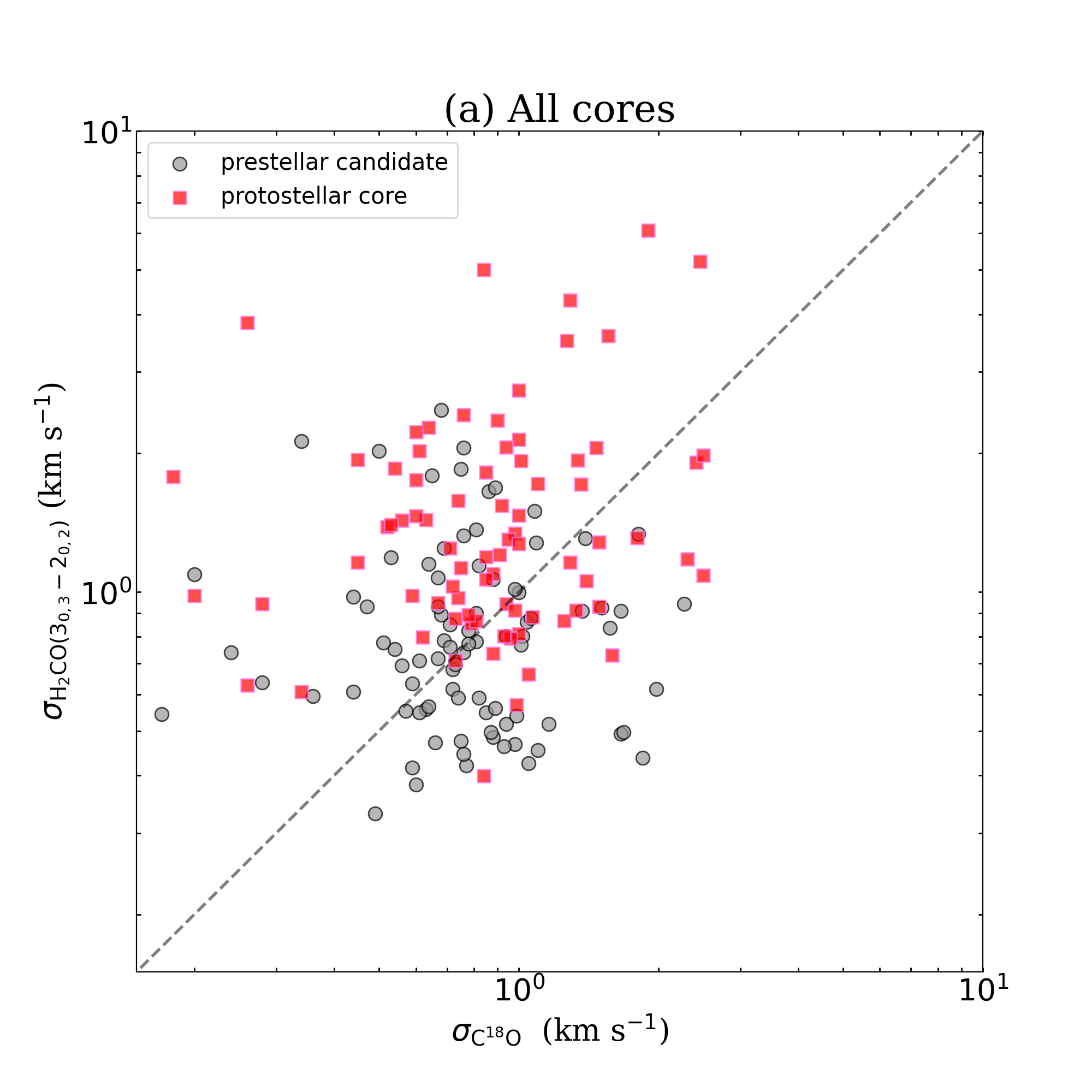}{0.33\textwidth}{}
\fig{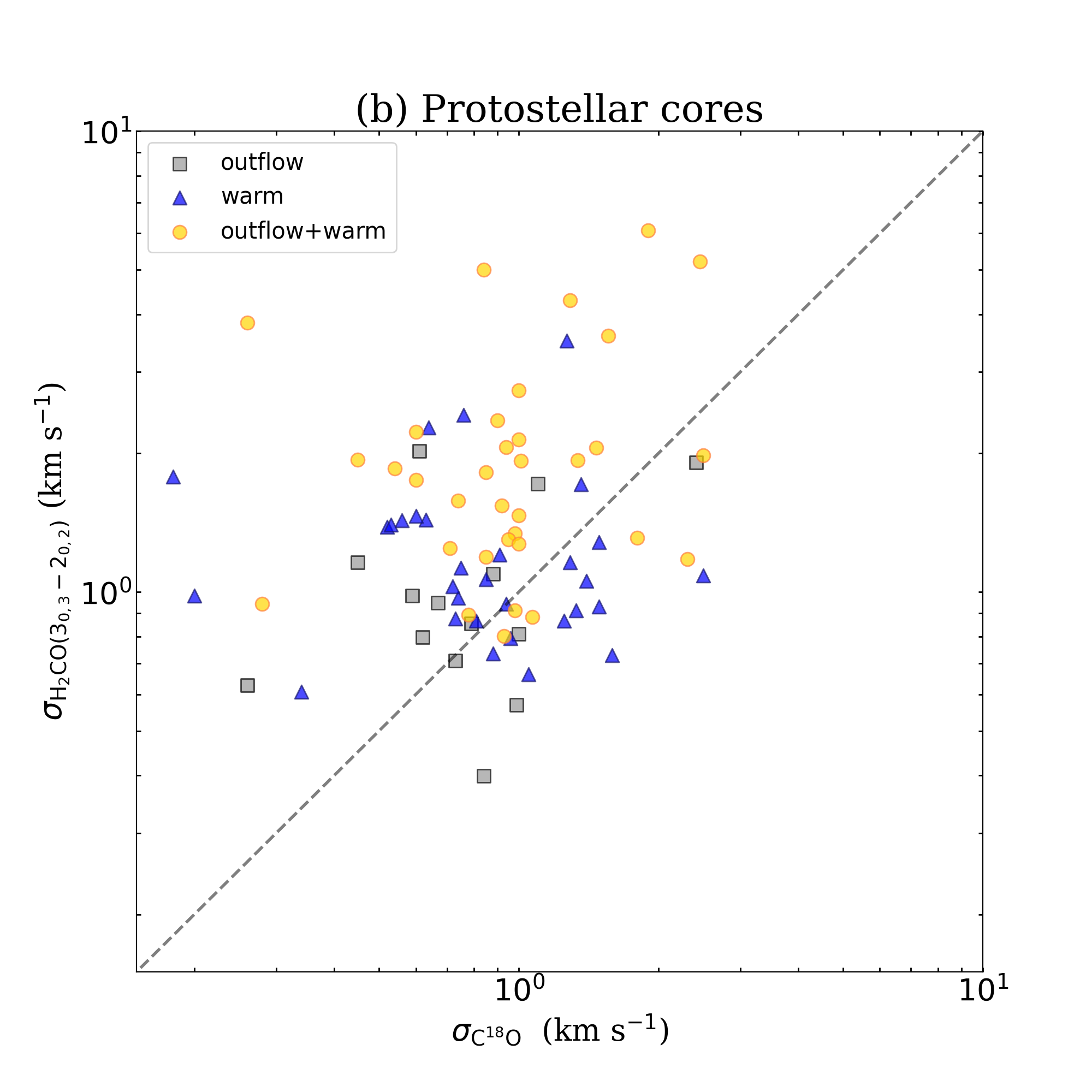}{0.33\textwidth}{}
\fig{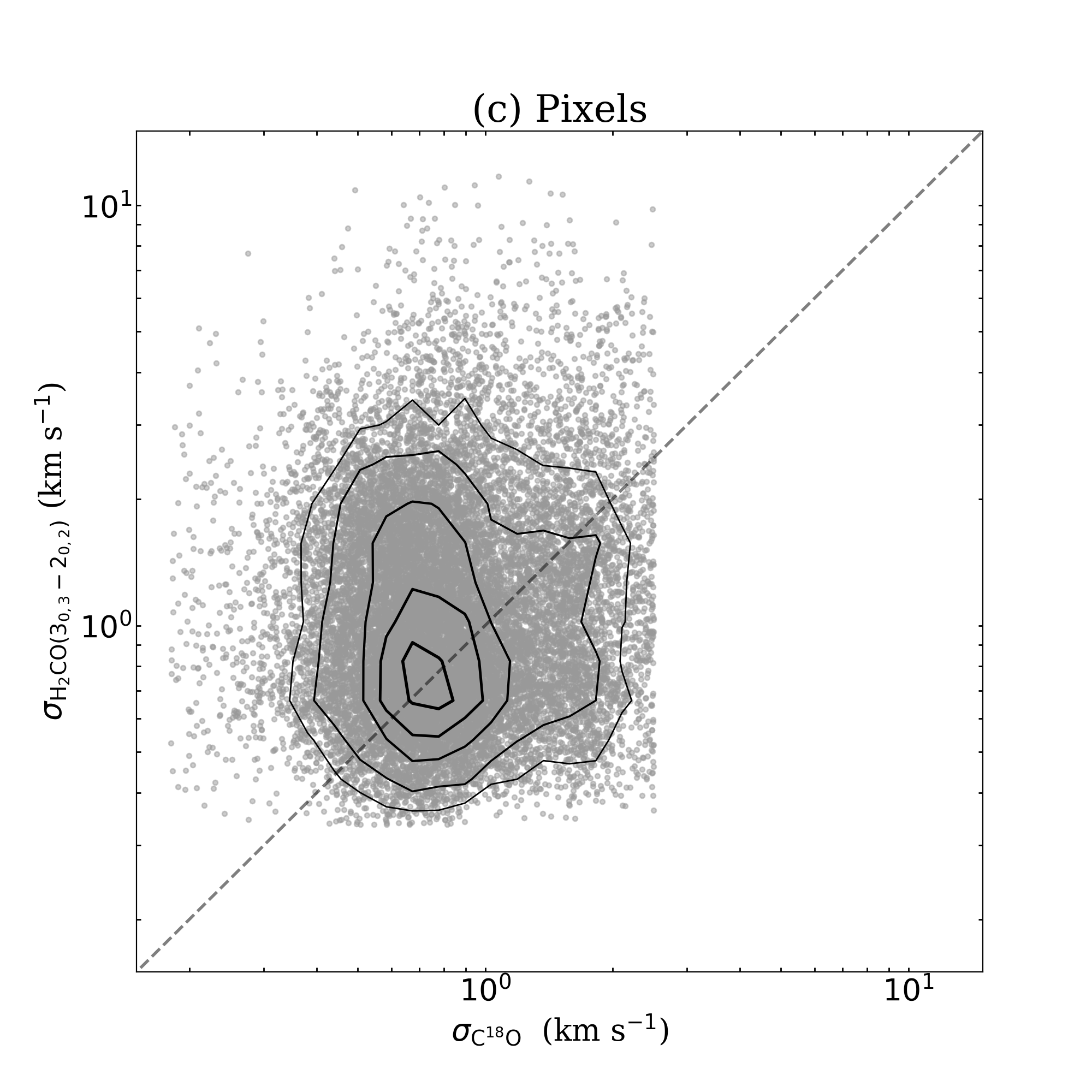}{0.33\textwidth}{}}
\caption{
Relation between the velocity dispersion of C$^{18}$O(2-1) ($\sigma_{\rm C^{18}O}$) and H$_2$CO ($\sigma_{\rm H_2CO(3_{0,3}-2_{0,2})}$) for all IRDC clumps
((a): all cores, (b) only protostellar cores, and (c) pixels).
The gray circles and red squares in the left panel indicate the prestellar candidates and protostellar cores, respectively.
The gray squares, blue triangles, and yellow circles in the right panel indicate the three classified protostellar cores: outflow core, warm core, and warm and outflow core, respectively.
The pixel scale was re-binned to 0$\farcs$6, approximately half of the synthesized beam, from the original value (0$\farcs$2).
}
\label{dv-relation} 
\end{figure*}
\begin{figure*}
\gridline{\fig{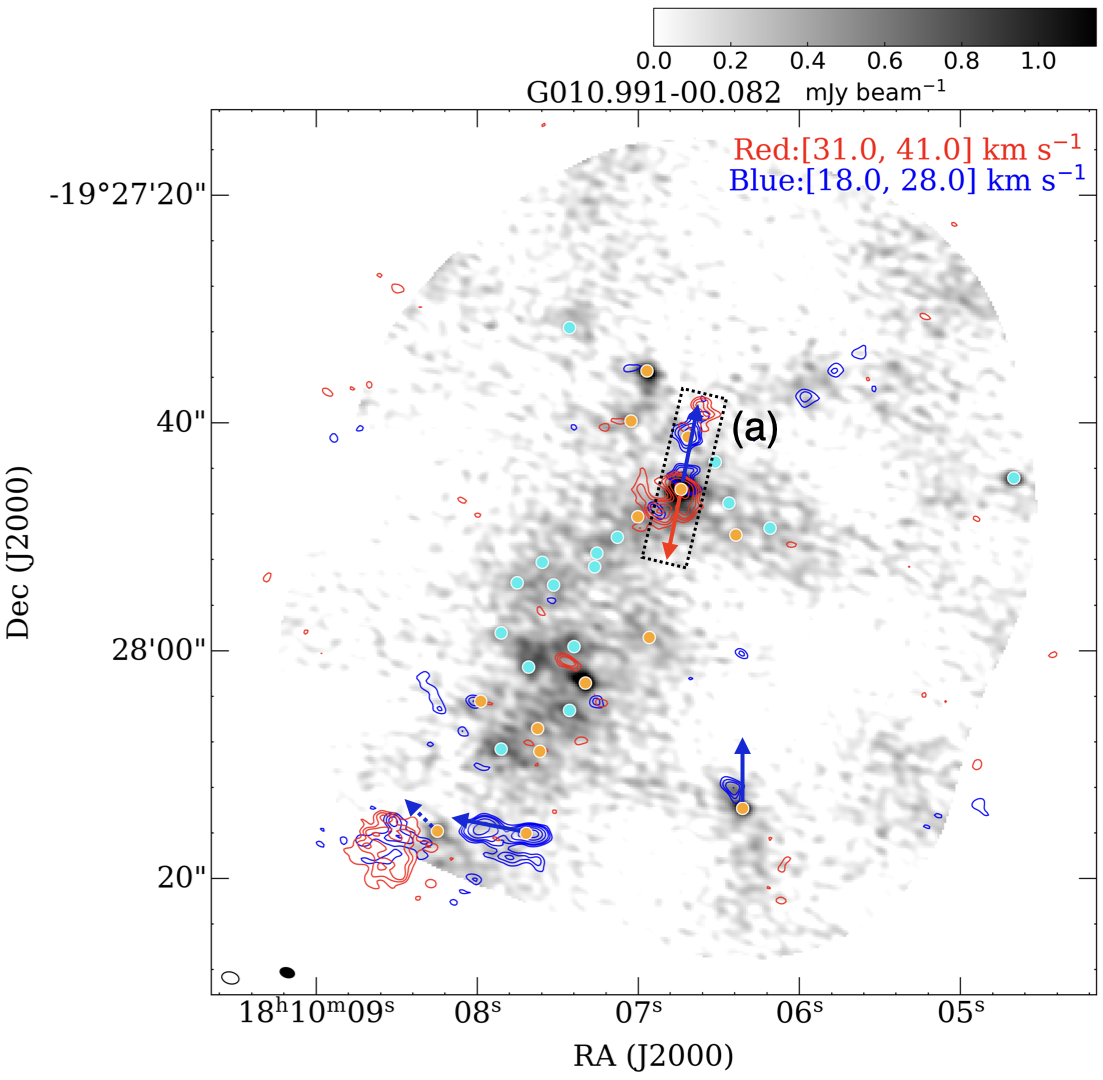}{0.45\textwidth}{}
          \fig{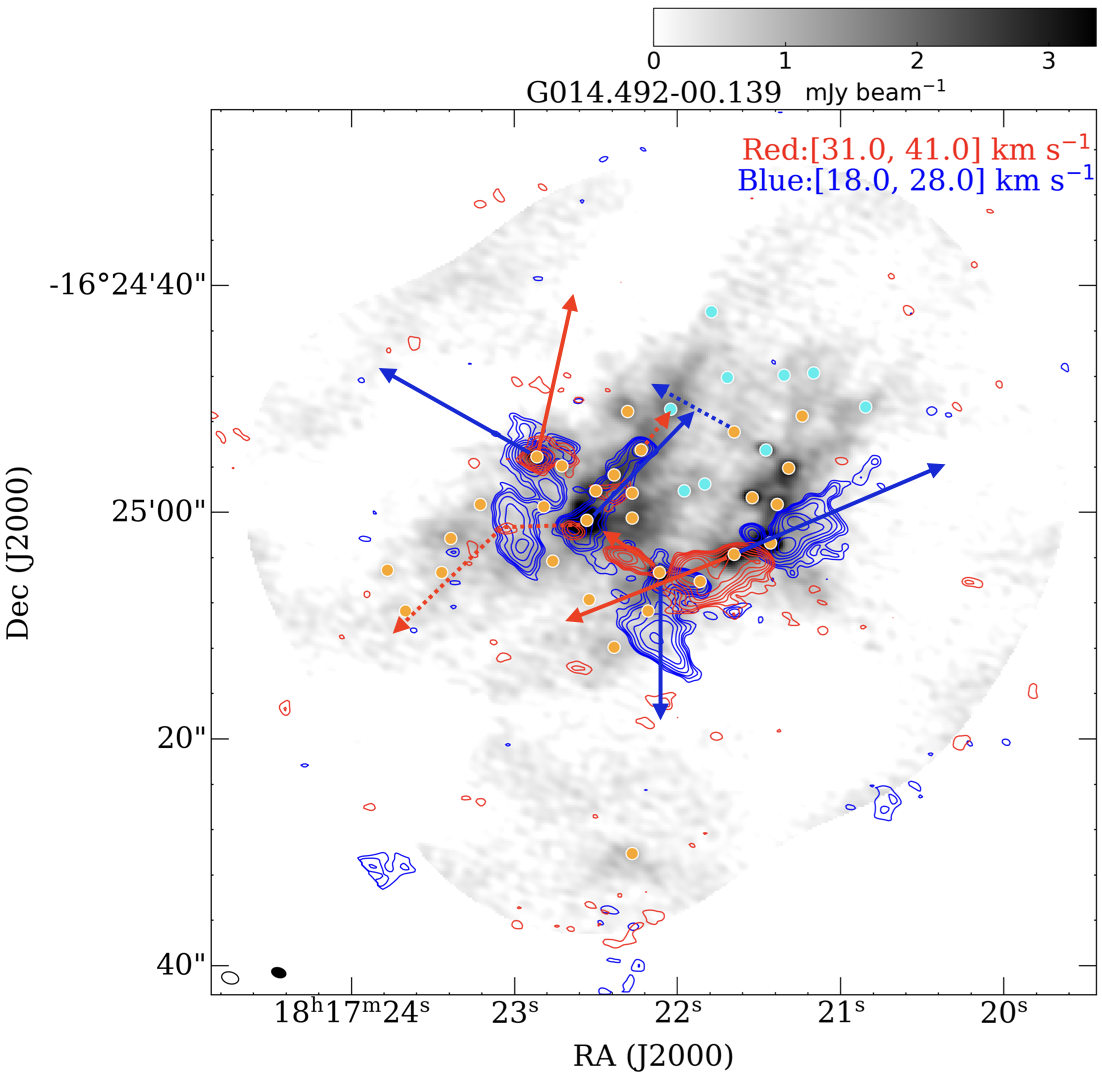}{0.45\textwidth}{}}
\vspace{-0.75cm}
\gridline{\fig{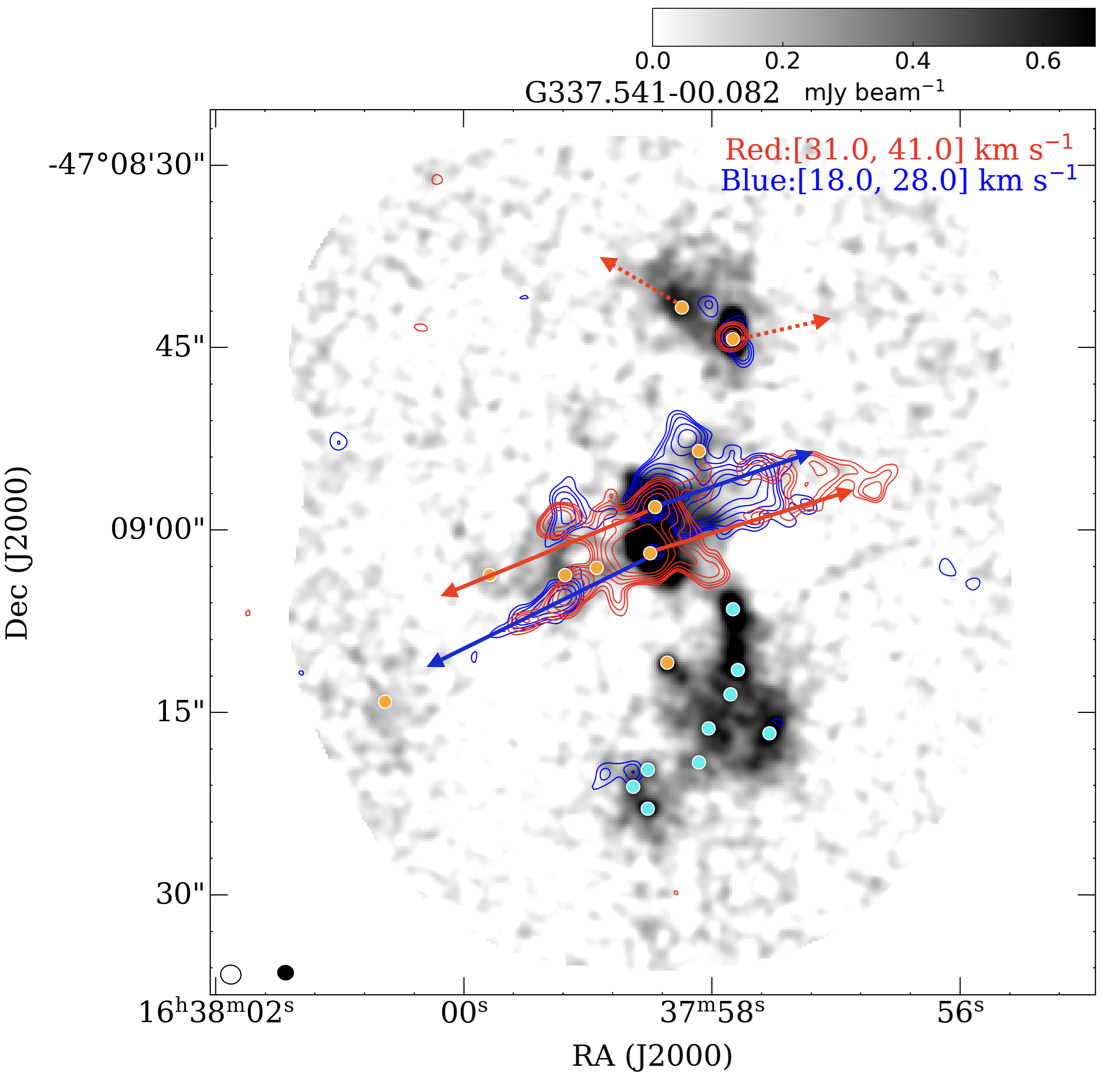}{0.45\textwidth}{}
          \fig{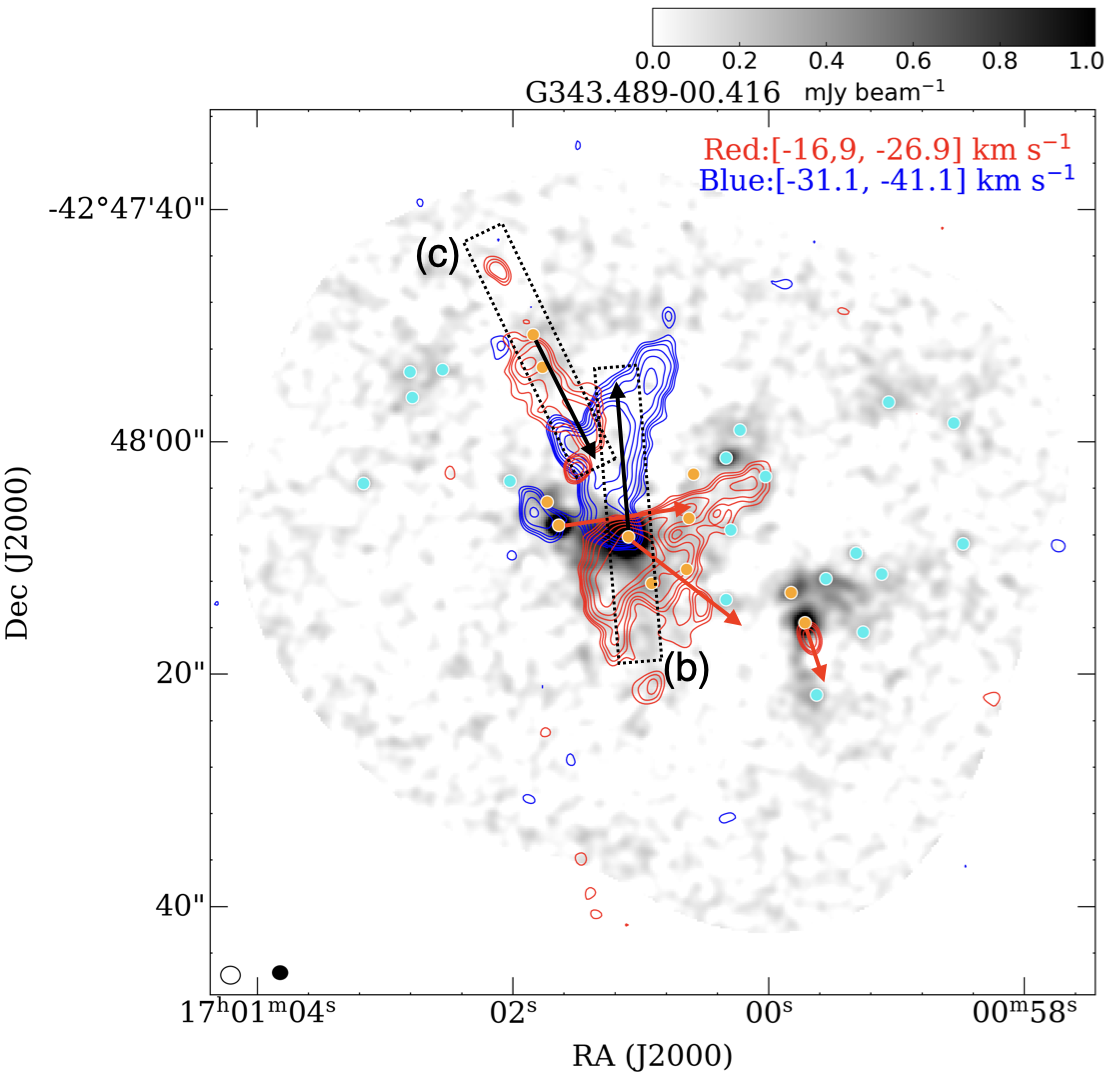}{0.45\textwidth}{}}
\caption{
Blue- (blue contours) and red-shifted (red contours) components of H$_2$CO(3$_{0,3}$-3$_{0,2}$) for four
representative IRDCs.
The gray-scale background shows the 1.3 mm dust continuum.
The cyan and orange circles indicate the positions of prestellar candidates and protostellar cores, respectively,
identified in ASHES \citep{Li2022}.
Synthesized beams are displayed at the bottom left of each panel
(open ellipses: H$_2$CO; filled ellipses: continuum).
The red and blue arrows indicate the outflow directions identified by CO and SiO emission lines in \citet{Li2020},
which investigates the outflow structure with the same ASHES data.
(solid arrows: also detected in H$_2$CO(3$_{0,3}$-3$_{0,2}$), dotted arrows: not detected in H$_2$CO(3$_{0,3}$-3$_{0,2}$)).
The black arrows indicate new outflow candidates detected in H$_2$CO(3$_{0,3}$-3$_{0,2}$), which are not seen in CO or SiO in \citet{Li2020}.
The black dotted boxes mark the regions for the position-velocity diagrams in Figure \ref{PV}.
The images of the other 5 IRDC clumps are presented in the appendix (Figure \ref{A-WING}).
}
\label{WING} 
\end{figure*}
\begin{figure*}
\gridline{\fig{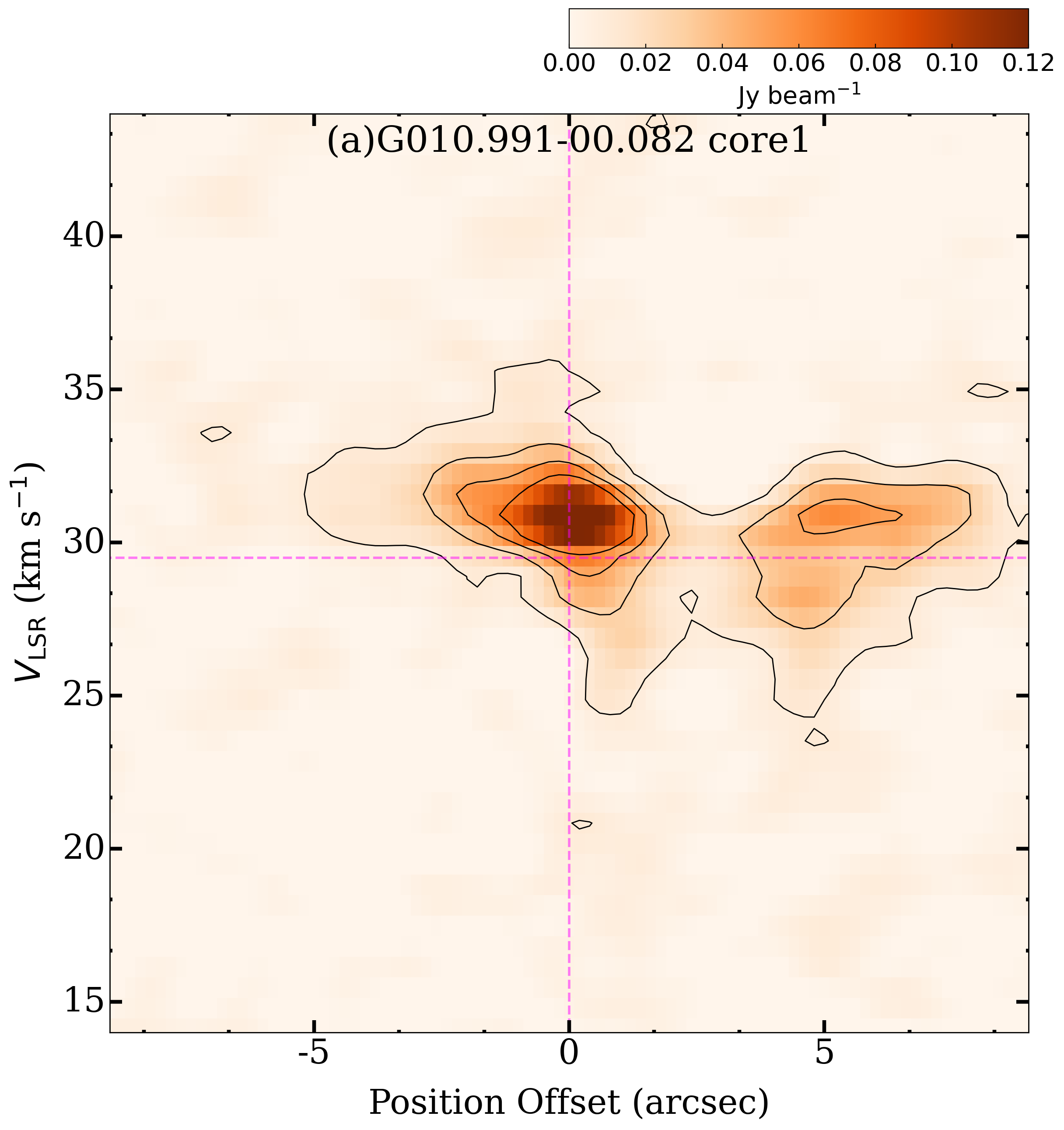}{0.3\textwidth}{}
          \fig{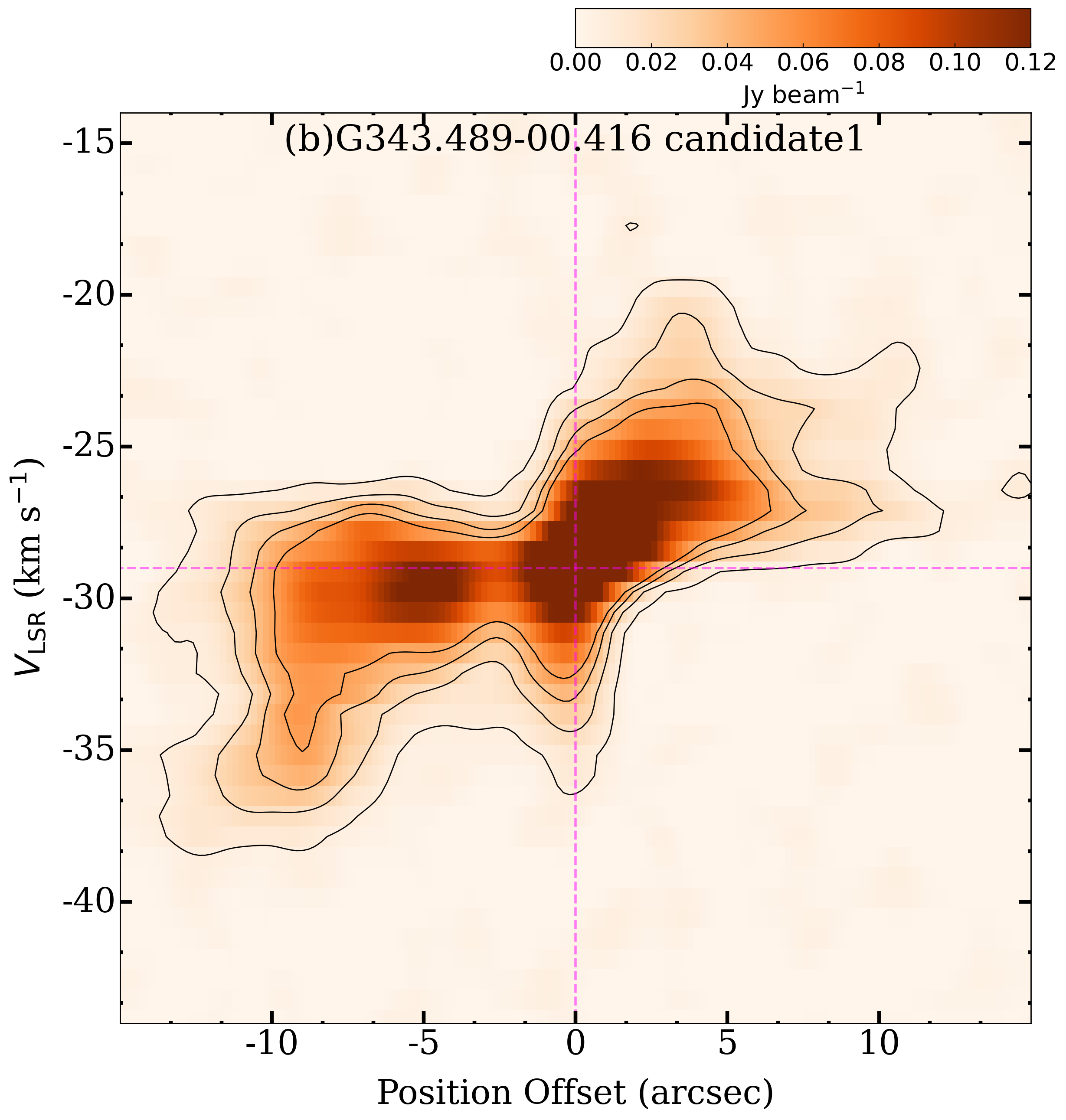}{0.3\textwidth}{}
          \fig{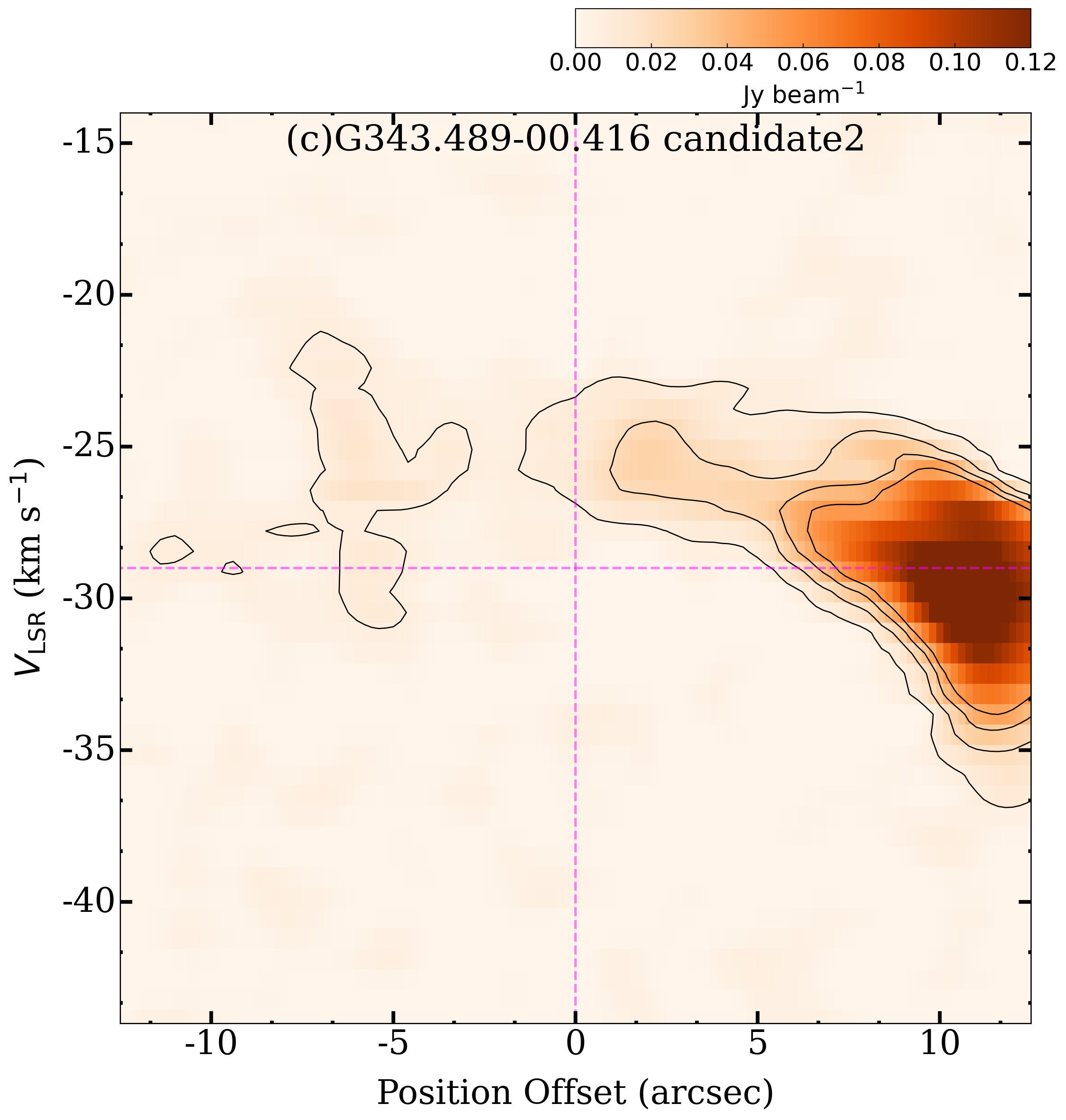}{0.3\textwidth}{}}
\caption{
Position-velocity diagrams of  H$_2$CO(3$_{0,3}$-3$_{0,2}$) for regions labeled as (a), (b), and (c) in Figure \ref{WING}.
The vertical and horizontal dashed magenta lines represent the positions and systemic velocities of the dense cores, respectively.
}
\label{PV} 
\end{figure*}
\begin{figure*}
\plotone{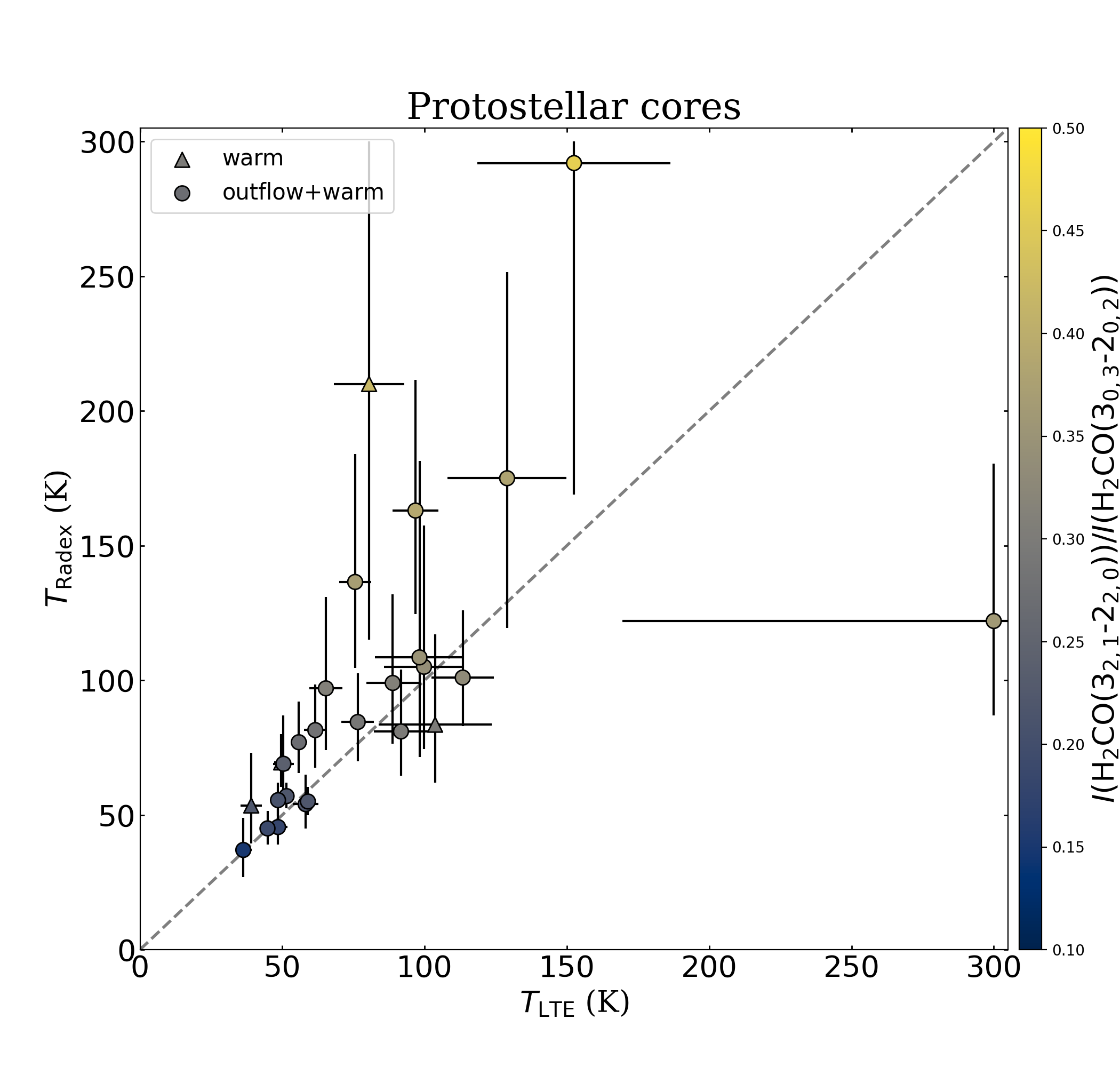}
\caption{
Relation between protostellar core temperatures derived under LTE assumption ($T_{\rm LTE}$) versus Radex modeling ($T_{\rm Radex}$).
The color indicates the H$_2$CO(3$_{2,1}$--2$_{2,0}$)/H$_2$CO(3$_{0,3}$--2$_{0,2}$) integrated intensity ratio
($I$(H$_2$CO(3$_{2,1}$--2$_{2,0}$))/$I$(H$_2$CO(3$_{0,3}$--2$_{0,2}$)))
derived from the Gaussian fitting.
The triangles and squares are the warm core and warm-and-outflow core groups, respectively.
}\label{LTE-Radex_core} 
\end{figure*}
\begin{figure*}
\gridline{\fig{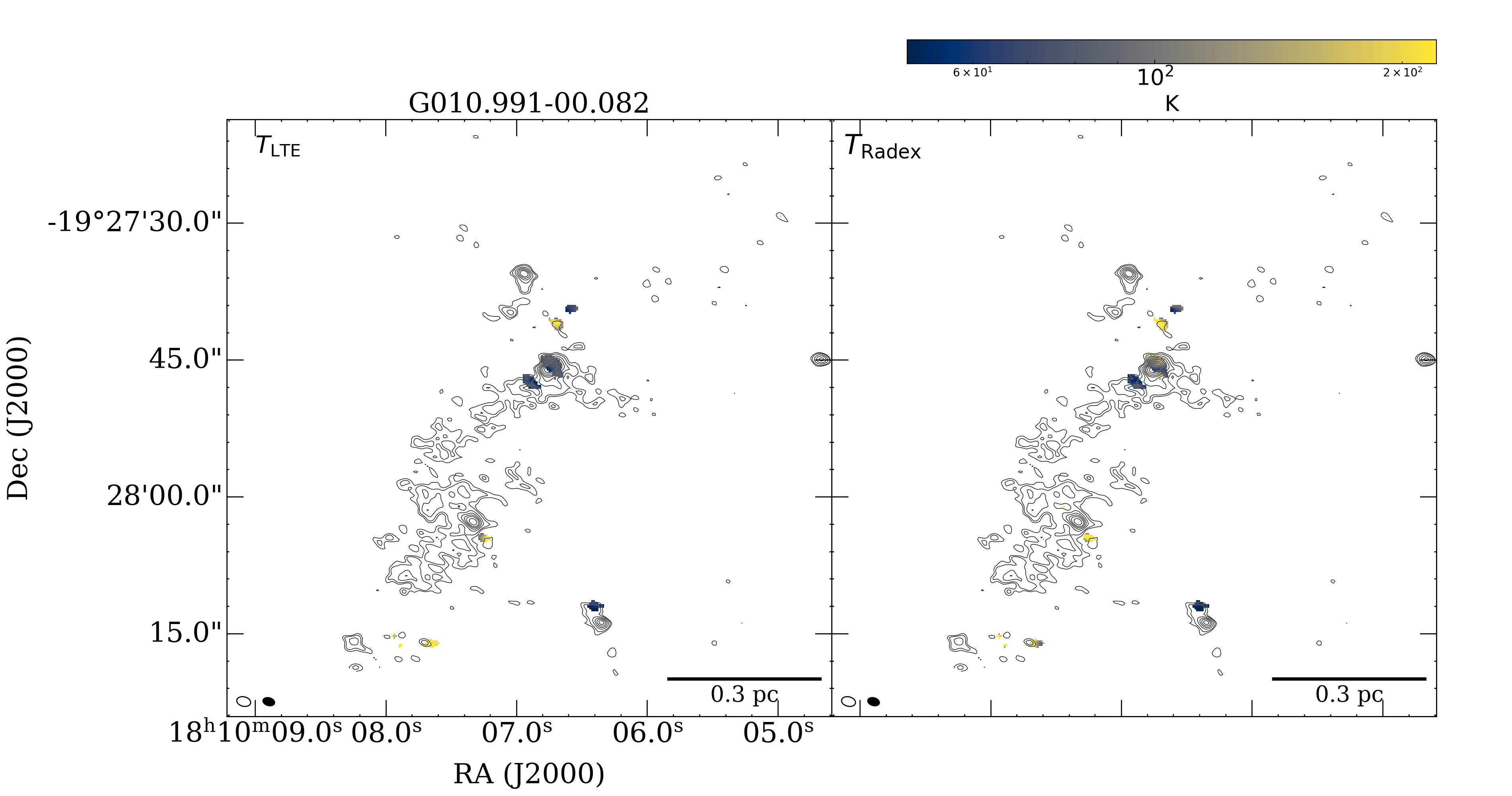}{0.9\textwidth}{}}
\vspace{-0.75cm}
\gridline{\fig{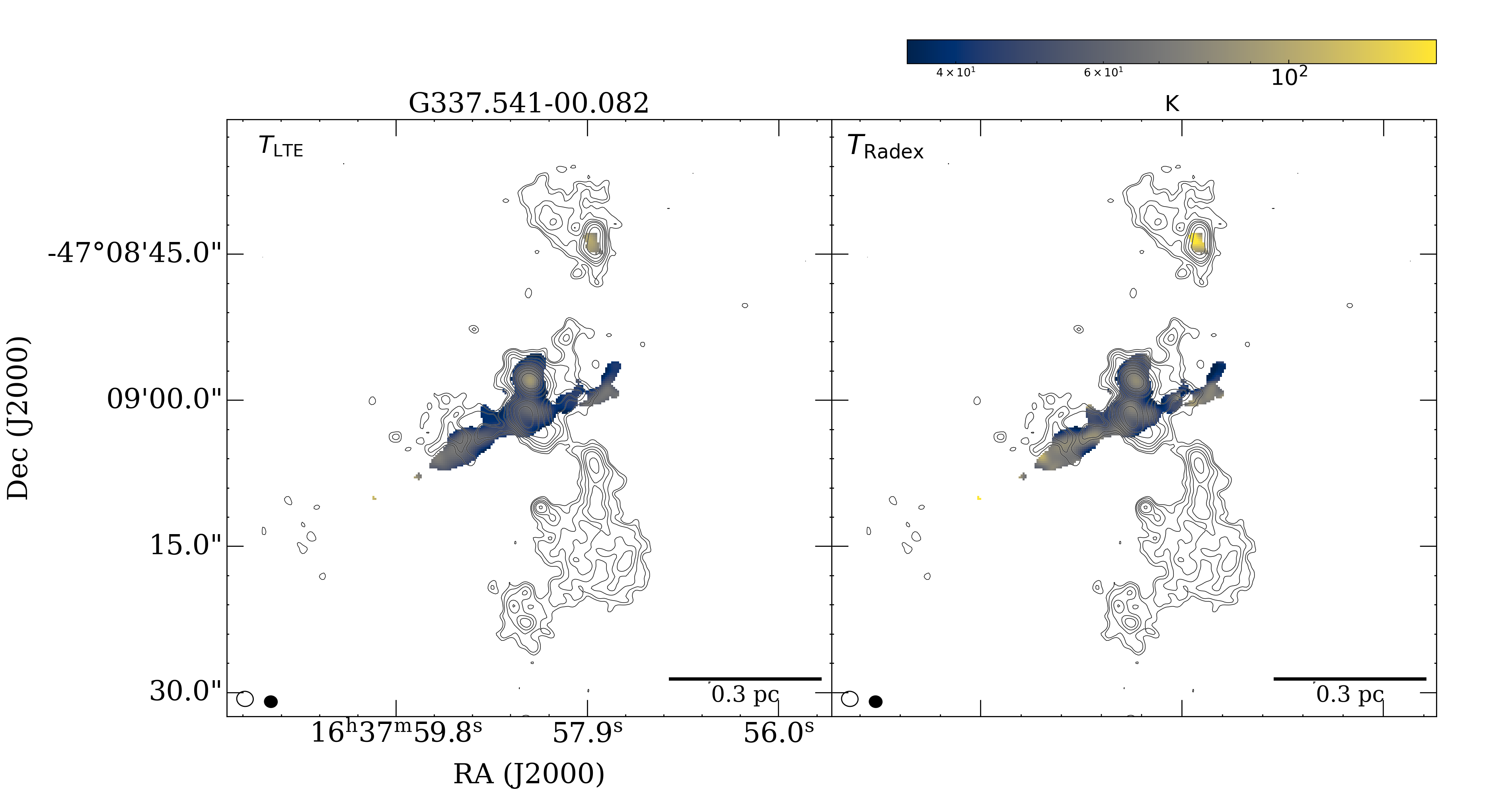}{0.9\textwidth}{}}
\vspace{-0.75cm}
\caption{
Kinetic temperature distributions (Left: LTE, Right: Radex) for two IRDC clumps.
The black contours are the 1.3mm dust continuum.
Contour levels are 3, 4, 5, 7, 10, 14, and 20 $\times$ $\sigma$, where
$\sigma$ = 0.115 mJy beam$^{-1}$ for G010.991-00.082
(1$\farcs$1 resolution);
and 3, 4, 6, 8, 10, 14, 20, 30, 45 and 75 $\times$ $\sigma$, where $
\sigma$ = 0.068 mJy beam$^{-1}$ for G337.541-00.082
(1$\farcs$2 resolution).
Synthesized beams are displayed at the bottom left in each panel
(open ellipses: H$_2$CO; filled ellipses: continuum).
The images of the other 7 IRDCs are presented in the appendix.
}
\label{TEMP-1} 
\end{figure*}
\begin{figure*}
\plotone{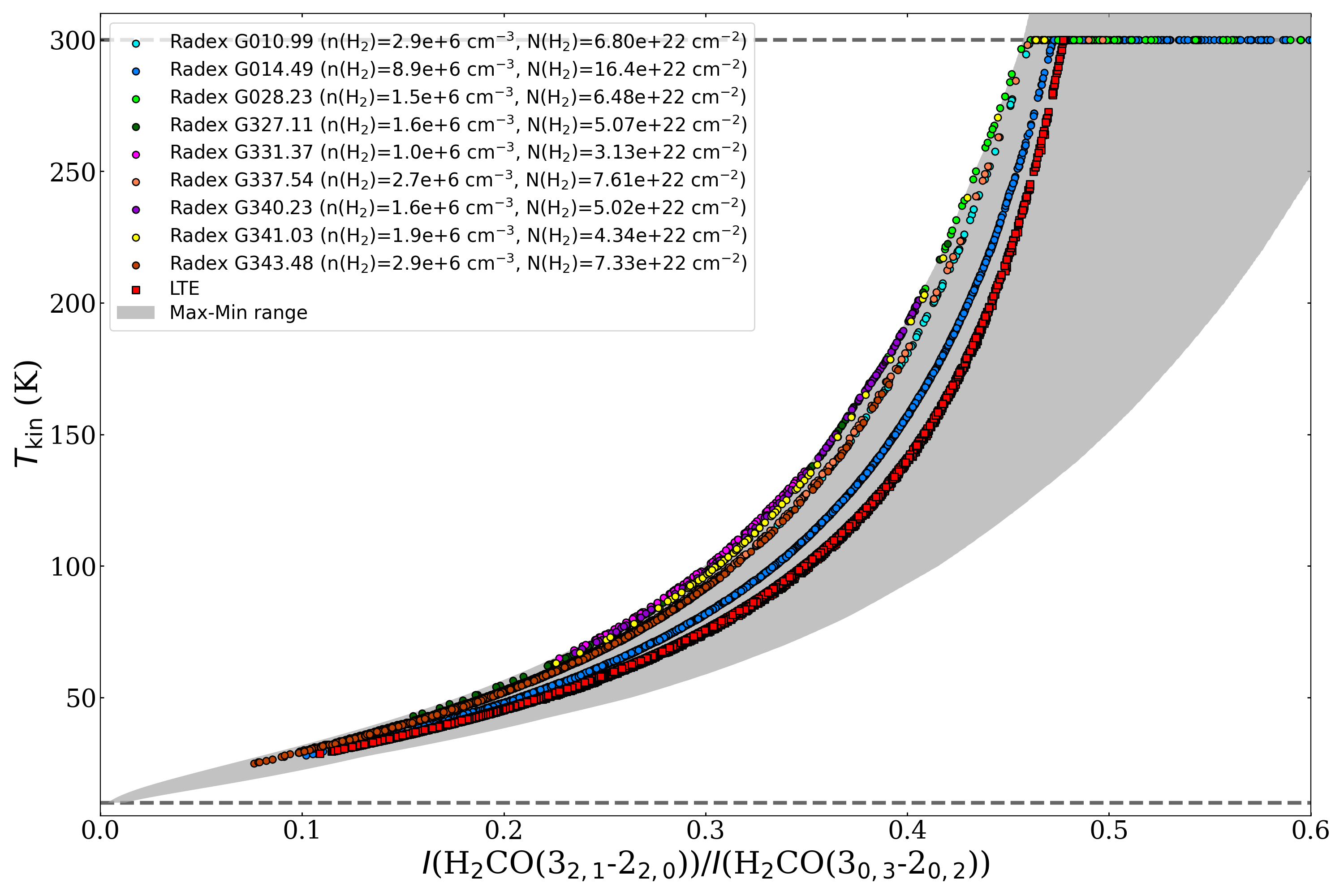}
\caption{
Pixel-by-pixel correlation between kinetic temperature ($T_{\rm kin}$)
and H$_2$CO(3$_{2,1}$--2$_{2,0}$)/H$_2$CO(3$_{0,3}$--2$_{0,2}$) integrated intensity ratio ($I$(H$_2$CO(3$_{2,1}$-2$_{2,0}$))/$I$(H$_2$CO(3$_{0,3}$-2$_{0,2}$))).
The circles and squires indicate the values based on the LTE assumption and with Radex modeling. $I$(H$_2$CO(3$_{2,1}$-2$_{2,0}$))/$I$(H$_2$CO(3$_{0,3}$-2$_{0,2}$) of the Radex temperature is derived from Gaussian fitting.
$I$(H$_2$CO(3$_{2,1}$-2$_{2,0}$))/$I$(H$_2$CO(3$_{0,3}$-2$_{0,2}$) of the LTE temperature is from the best fit under LTE assumption.
The gray-filled area shows the uncertainty, calculated from the possible n(H$_2$) and N(H$_2$) ranges:
n(H$_2$) = 10$^5$ -- 2.0 $\times$ 10$^7$ cm $^{-3}$ and 
N(H$_2$) = 10$^{22}$ -- 10$^{24}$ cm $^{2}$.
The black-dashed lines indicate the kinetic temperature range, 10 $\leq$ $T_{\rm kin}$ $\le$ 300 K, 
for both methods.
}
\label{LTE-Radex_ratio} 
\end{figure*}
\begin{deluxetable*}{cccccccc}
\tablecaption{Properties of H$_2$CO emission ($T_{\rm kin}$, $\sigma_{\rm T}$, and $\sigma_{\rm NT}$).\label{tab:pro_H2CO-1}}
\tablewidth{0pt}
\tablehead{
\colhead{IRDC} & \multicolumn2c{$\overline{T_{\rm kin}}$} &
\multicolumn2c{$\overline{\sigma_{\rm T}}$} &
\multicolumn2c{$\overline{\sigma_{\rm NT}}$}  \\
\colhead{} &\multicolumn2c{(K)} & 
\multicolumn2c{(km s$^{-1}$)} & 
\multicolumn2c{(km s$^{-1}$)}  \\
\colhead{} & \colhead{LTE} & \colhead{Radex} &
\colhead{LTE} & \colhead{Radex} &
\colhead{LTE} & \colhead{Radex} 
}
\startdata
G010.991-00.082 & 112 ($\pm$77)  & 127 ($\pm$75) & 0.17 ($\pm$0.05) & 0.18 ($\pm$0.05) & 1.8 ($\pm$0.9)  & 1.9  ($\pm$1.2)\\
G014.492-00.139 & 100 ($\pm$56)  & 120 ($\pm$64) & 0.16 ($\pm$0.04) & 0.18 ($\pm$0.04) & 2.4 ($\pm$1.3)  & 2.4  ($\pm$1.3)\\
G028.273-00.167 & 220 ($\pm$63)  & 281 ($\pm$30) & 0.24 ($\pm$0.04) & 0.28 ($\pm$0.03) & 1.7 ($\pm$0.2)  & 1.5  ($\pm$0.3)\\
G327.116-00.294 & 72 ($\pm$12)   & 104 ($\pm$34) & 0.14 ($\pm$0.01) & 0.17 ($\pm$0.03) & 2.1 ($\pm$0.3)  & 2.0  ($\pm$0.3)\\
G331.372-00.116 & 76 ($\pm$13)   & 100 ($\pm$19) & 0.14 ($\pm$0.01) & 0.17 ($\pm$0.02) & 2.3 ($\pm$0.2)  & 2.3  ($\pm$0.4)\\
G332.969-00.029 & --- & --- & --- & --- & --- & --- \\
G337.541-00.082 & 51 ($\pm$13)   & 62 ($\pm$32)  & 0.12($\pm$0.01)  & 0.13 ($\pm$0.03) & 1.7 ($\pm$0.4)  & 1.7 ($\pm$0.4)\\
G340.179-00.242 & --- & --- & --- & --- & --- & --- \\
G340.222-00.167 & --- & --- & --- & --- & --- & --- \\
G340.232-00.146 & 94 ($\pm$23)   & 136 ($\pm$41) & 0.16 ($\pm$0.02) & 0.19 ($\pm$0.03) & 2.3 ($\pm$0.1)  & 2.3 ($\pm$0.2)\\
G341.039-00.114 & 91 ($\pm$9)    & 132 ($\pm$63) & 0.16 ($\pm$0.01) & 0.19 ($\pm$0.04) & 2.2 ($\pm$0.2)  & 2.2 ($\pm$0.4)\\
G343.489-00.416 & 48 ($\pm$12)   & 58 ($\pm$22)  & 0.11 ($\pm$0.01) & 0.13 ($\pm$0.02) & 1.7 ($\pm$0.5)  & 1.7 ($\pm$0.5) \\
\enddata
\tablenotetext{}{To derive these values, the pixel scale was re-binned to 0$\farcs$6, approximately half of the synthesized beam, from the original value (0$\farcs$2).}
\end{deluxetable*}
\begin{deluxetable*}{cccccccc}
\tablecaption{Properties of H$_2$CO emission ($a_{\rm s}$, $R_{\rm p}$, and $\mathcal{M}$).\label{tab:pro_H2CO-2}}
\tablewidth{0pt}
\tablehead{
\colhead{IRDC} &
\multicolumn2c{$\overline{a_{\rm s}}$} &
\multicolumn2c{$\overline{R_{\rm p}}$} &
\multicolumn2c{$\overline{\mathcal{M}}$} \\
\colhead{} &
\multicolumn2c{(km s$^{-1}$)} & 
\colhead{} & \colhead{} &
\colhead{} & \colhead{} \\
\colhead{} & 
\colhead{LTE} & \colhead{Radex} &
\colhead{LTE} & \colhead{Radex} &
\colhead{LTE} & \colhead{Radex}
}
\startdata
G010.991-00.082 & 0.60 ($\pm$0.19) & 0.64 ($\pm$0.20) & 0.13 ($\pm$0.06) & 0.17 ($\pm$0.10) & 3.0 ($\pm$0.7) & 3.0 ($\pm$1.5)\\
G014.492-00.139 & 0.57 ($\pm$0.15) & 0.62 ($\pm$0.16) & 0.11 ($\pm$0.11) & 0.16 ($\pm$0.20) & 4.3 ($\pm$2.4) & 4.1 ($\pm$2.5)\\
G028.273-00.167 & 0.86 ($\pm$0.13) & 0.99 ($\pm$0.06) & 0.30 ($\pm$0.13) & 0.52 ($\pm$0.26) & 2.0 ($\pm$0.5) & 1.5 ($\pm$0.4)\\
G327.116-00.294 & 0.50 ($\pm$0.04) & 0.59 ($\pm$0.10) & 0.06 ($\pm$0.02) & 0.10 ($\pm$0.05) & 4.2 ($\pm$0.7) & 3.5 ($\pm$0.9)\\
G331.372-00.116 & 0.52 ($\pm$0.04) & 0.59 ($\pm$0.06) & 0.05 ($\pm$0.01) & 0.06 ($\pm$0.01) & 4.5 ($\pm$0.3) & 4.0 ($\pm$0.5)\\
G332.969-00.029 & --- & --- & --- & --- & --- & --- \\
G337.541-00.082 & 0.42 ($\pm$0.05) & 0.46 ($\pm$0.09) & 0.07 ($\pm$0.03) & 0.08 ($\pm$0.07) & 4.0 ($\pm$0.9) & 3.9 ($\pm$1.1)\\
G340.179-00.242 & --- & --- & --- & --- & --- & --- \\
G340.222-00.167 & --- & --- & --- & --- & --- & --- \\
G340.232-00.146 & 0.57 ($\pm$0.07) & 0.68 ($\pm$0.11) & 0.06 ($\pm$0.02) & 0.09 ($\pm$0.03) & 4.2 ($\pm$0.6) & 3.6 ($\pm$0.8)\\
G341.039-00.114 & 0.56 ($\pm$0.03) & 0.64 ($\pm$0.14) & 0.07 ($\pm$0.02) & 0.17 ($\pm$0.08) & 3.9 ($\pm$0.4) & 3.5 ($\pm$1.1)\\
G343.489-00.416 & 0.41 ($\pm$0.05) & 0.44 ($\pm$0.07) & 0.07 ($\pm$0.05) & 0.09 ($\pm$0.05) & 4.1 ($\pm$1.2) & 3.9 ($\pm$1.1)\\
\enddata
\tablenotetext{}{To derive these values, the pixel scale was re-binned to 0$\farcs$6, approximately half of the synthesized beam, from the original value (0$\farcs$2).}
\end{deluxetable*}
\maxdeadcycles=1000
\begin{deluxetable*}{ccccccccc}
\tablecaption{Relation between kinetic temperature and $\sigma_{\rm NT(H_2CO(3_{0,3}-2_{0,2})).}$ \label{Relation_Temp-SNT}}
\tablewidth{0pt}
\tablehead{
\colhead{Data} & \multicolumn4c{Spearman's correlation coefficients} & \multicolumn4c{Least-squares fitting results\tablenotemark{a}} \\
               & \multicolumn2c{$r$-value} &  \multicolumn2c{$p$-value}  & \multicolumn2c{$\alpha$-value} &  \multicolumn2c{$\beta$-value}\\
               & \colhead{LTE} & \colhead{Radex} & \colhead{LTE} & \colhead{Radex} & \colhead{LTE} & \colhead{Radex} & \colhead{LTE} & \colhead{Radex} 
}
\startdata
Core                   & 0.39  & 0.21 & 0.05     & 0.30     & 0.43 ($\pm$0.21) & 0.04 ($\pm$0.22) & 60.0 ($\pm$13.7) & 95.6 ($\pm$19.0)  \\ 
Core (except G014.49)  & 0.79  & 0.34 & 0.001    & 0.26     & 1.78 ($\pm$0.22) & 0.24 ($\pm$0.55) & 22.6 ($\pm$5.9)  & 84.4 ($\pm$38.0)\\ \hline
Pixel\tablenotemark{b} 
                       & 0.41  & 0.23 & 0.0\tablenotemark{c} & 0.0\tablenotemark{c} 
                       & 0.49 ($\pm$0.03) & 0.21 ($\pm$0.04) & 51.7 ($\pm$1.6)  & 75.5 ($\pm$2.4)\\
Pixel (except G014.49)\tablenotemark{b}  
                       & 0.34  & 0.26 & 0.0\tablenotemark{c} & 0.0\tablenotemark{c} 
                       & 0.69 ($\pm$0.05) & 0.34 ($\pm$0.07) & 39.5 ($\pm$1.5)  & 59.5 ($\pm$2.8) \\
\enddata
\tablenotetext{a}{
Format of least-squares fitting is $T_{\rm LTE}$ (or $T_{\rm Radex}$) = $\beta$ $\times$ $\sigma_{\rm NT(H_2CO(3_{0,3}-2_{0,2}))}$$^\alpha$. 
}
\tablenotetext{b}{
Pixels with $T_{\rm kin}$ = 300 K are discarded because these values are considered to be lower limits.
}
\tablenotetext{c}{
The actual calculated $p$-values are less than 1.0 $\times$ 10$^{-10}$, which for all purposes here is considered to be equivalent to zero.
}
\end{deluxetable*}
\begin{deluxetable*}{cccccc}
\tablecaption{Properties of warm lines.\label{warmline}}
\tablewidth{0pt}
\tablehead{
\colhead{IRDC} & \colhead{Core} & \multicolumn2c{$T_{\rm kin}$} & \colhead{$\sigma_{\rm H_2CO(3_{0,3}-2_{0,2}})$} &  \colhead{Detected lines}  \\
                &               &  \colhead{LTE} & \colhead{Radex} &                                            &    \\
                &               &  \colhead{(K)}   & \colhead{(K)}     &   \colhead{(km s$^{-1}$)}  &                    
}
\startdata
G010.991-00.082 & ALMA6         &  300 ($\pm$130) & 122 ($^{+59}_{-35}$)  & 4.9 ($\pm$0.3)  & HC$_3$N \\ 
                            & ALMA28       &  129 ($\pm$21)   & 171 ($^{+81}_{-51}$)  & 2.1 ($\pm$0.1)  & HC$_3$N \\  \hline
G014.492-00.139 & ALMA1         &  77  ($\pm$6)     & 85  ($^{+18}_{-15}$)  & 2.7 ($\pm$0.1)  & OHC$_3$N, CS     \\
                             & ALMA2         &  97  ($\pm$8)    & 163 ($^{+48}_{-39}$)  & 1.3 ($\pm$0.1)  & HC$_3$N, OCS  \\ 
                & ALMA4         &  113 ($\pm$11)  & 126 ($^{+25}_{-18}$)  & 5.2 ($\pm$0.1)  & HC$_3$N      \\ 
                & ALMA8         &  92  ($\pm$10)   & 81  ($^{+23}_{-17}$)  & 1.3 ($\pm$0.1)  & HC$_3$N         \\ 
                & ALMA14        &  98  ($\pm$16)  & 109 ($^{+73}_{-37}$)  & 5.0 ($\pm$0.2)  & HC$_3$N \\   
                & ALMA18        &  76  ($\pm$6)   & 136 ($^{+48}_{-32}$)  & 6.1 ($\pm$0.2)  & HC$_3$N, OCS \\   \hline
G028.273-00.167 & ALMA6         &  153 ($\pm$34)  & 292 ($^{+8}_{-123}$)  & 1.7 ($\pm$0.1)  & HC$_3$N \\   \hline
G337.541-00.082 & ALMA1         &  62  ($\pm$4)   & 55 ($^{+6}_{-5}$)     & 2.1 ($\pm$0.1)  & HC$_3$N, OCS \\    
                & ALMA3         &  66  ($\pm$6)   & 97 ($^{+34}_{-23}$)   & 2.2 ($\pm$0.1)  & OCS \\
                & ALMA17        &  49  ($\pm$2)   & 56 ($^{+7}_{-6}$)     & 1.6 ($\pm$0.1)  & HC$_3$N  \\ \hline
G341.039-00.114 & ALMA1         &  89  ($\pm$9)   & 99 ($^{+34}_{-23}$)   & 2.4 ($\pm$0.1)  & HC$_3$N  \\ \hline
G343.489-00.416 & ALMA1         &  45  ($\pm$2)   & 45 ($^{+7}_{-6}$)     & 1.8 ($\pm$0.1)  & HC$_3$N, OCS   \\   
\enddata
\end{deluxetable*}
\clearpage
\appendix
\section{Spatial distributions}\label{sec:a-1}
\begin{figure*}
\gridline{\fig{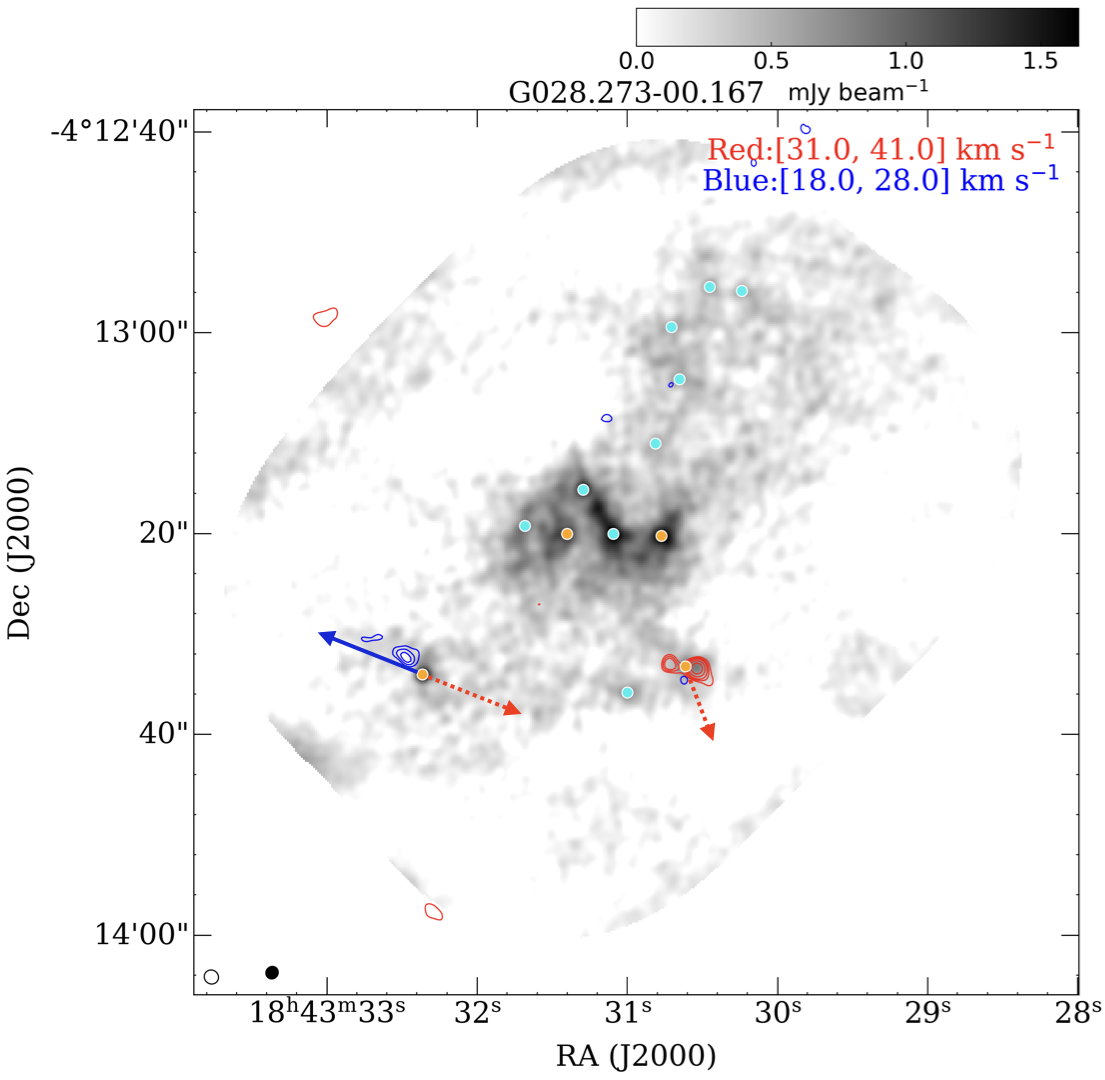}{0.4\textwidth}{}
\fig{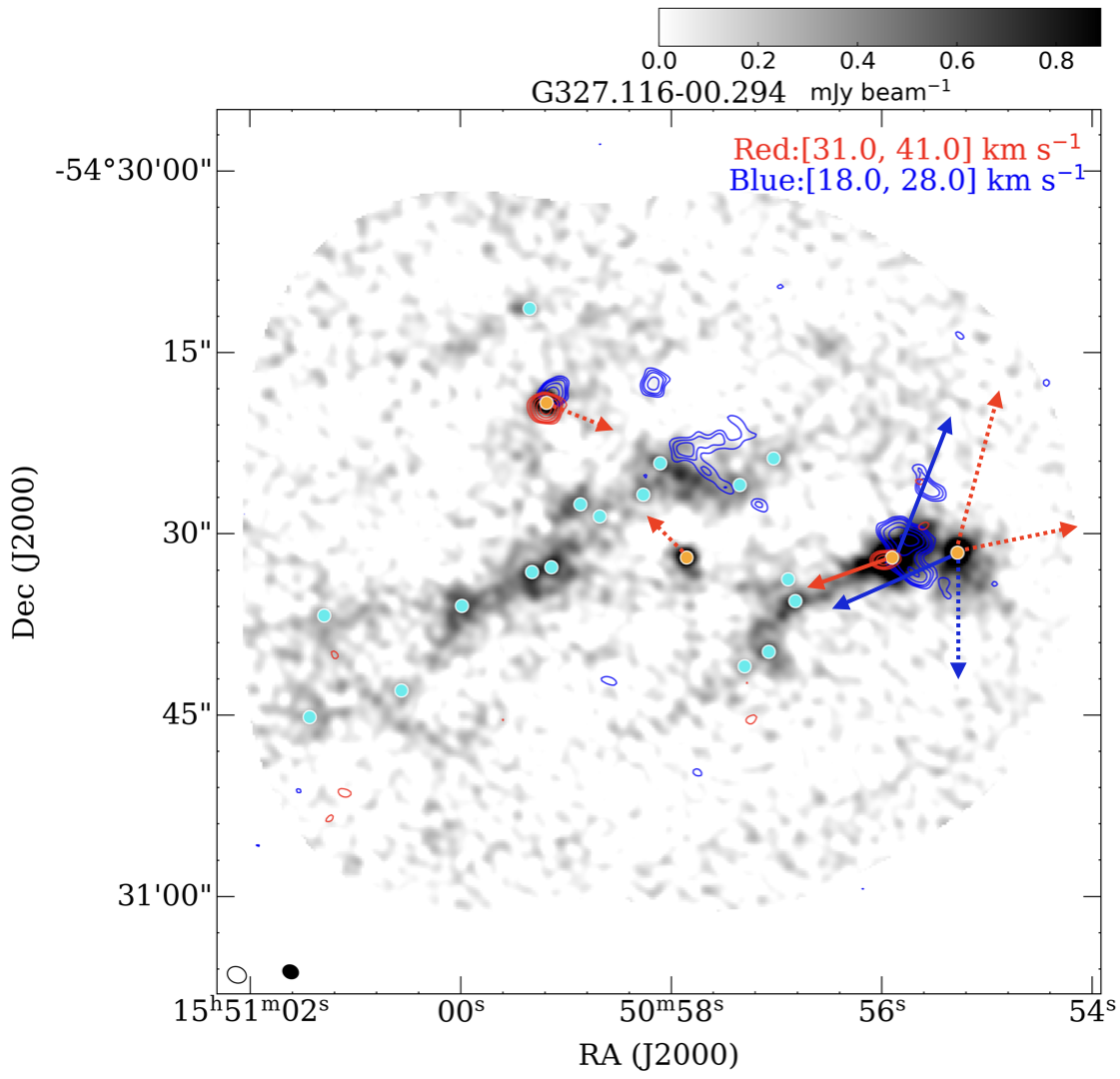}{0.4\textwidth}{}}
\vspace{-0.75cm}
\gridline{\fig{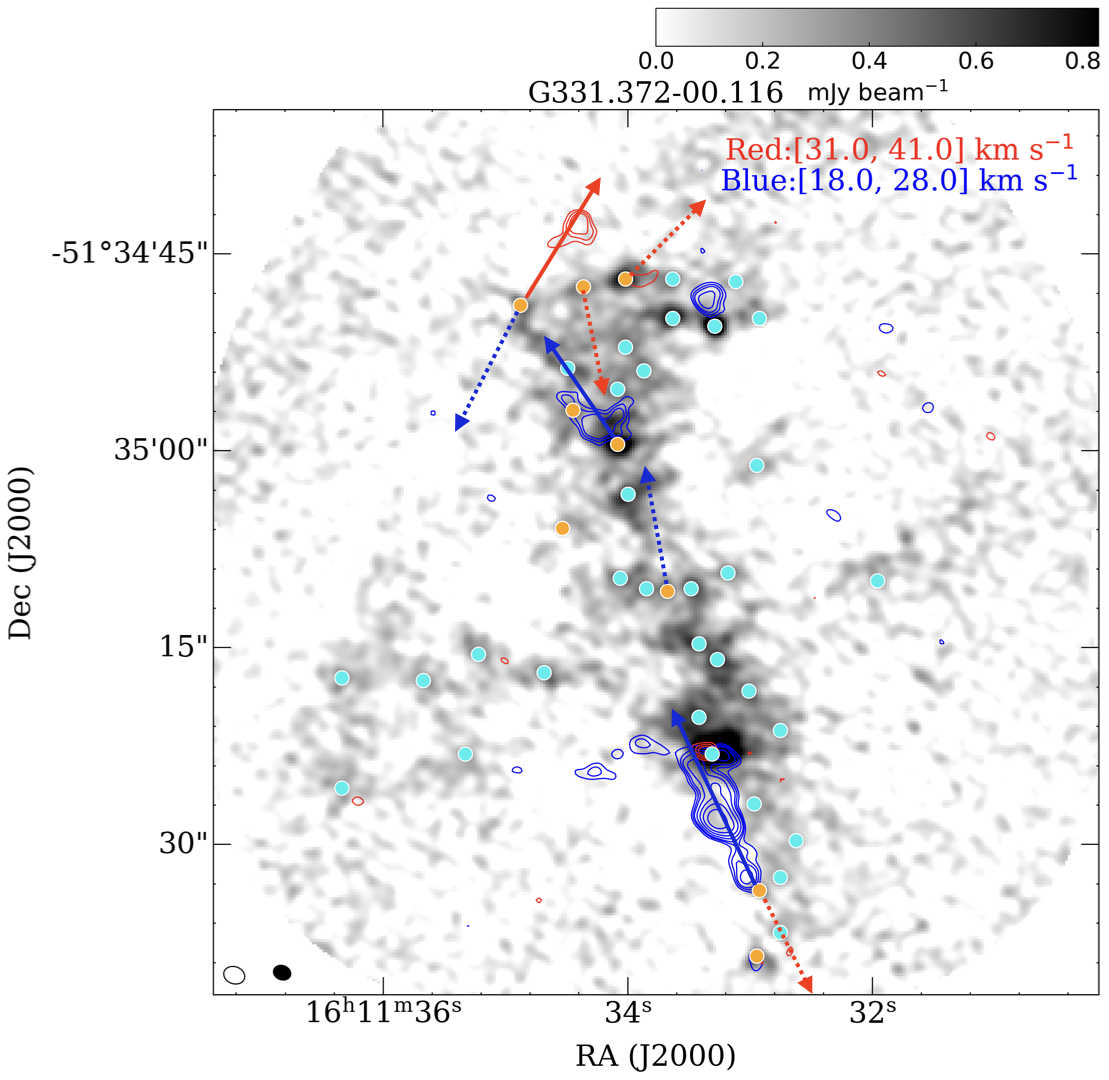}{0.4\textwidth}{}
\fig{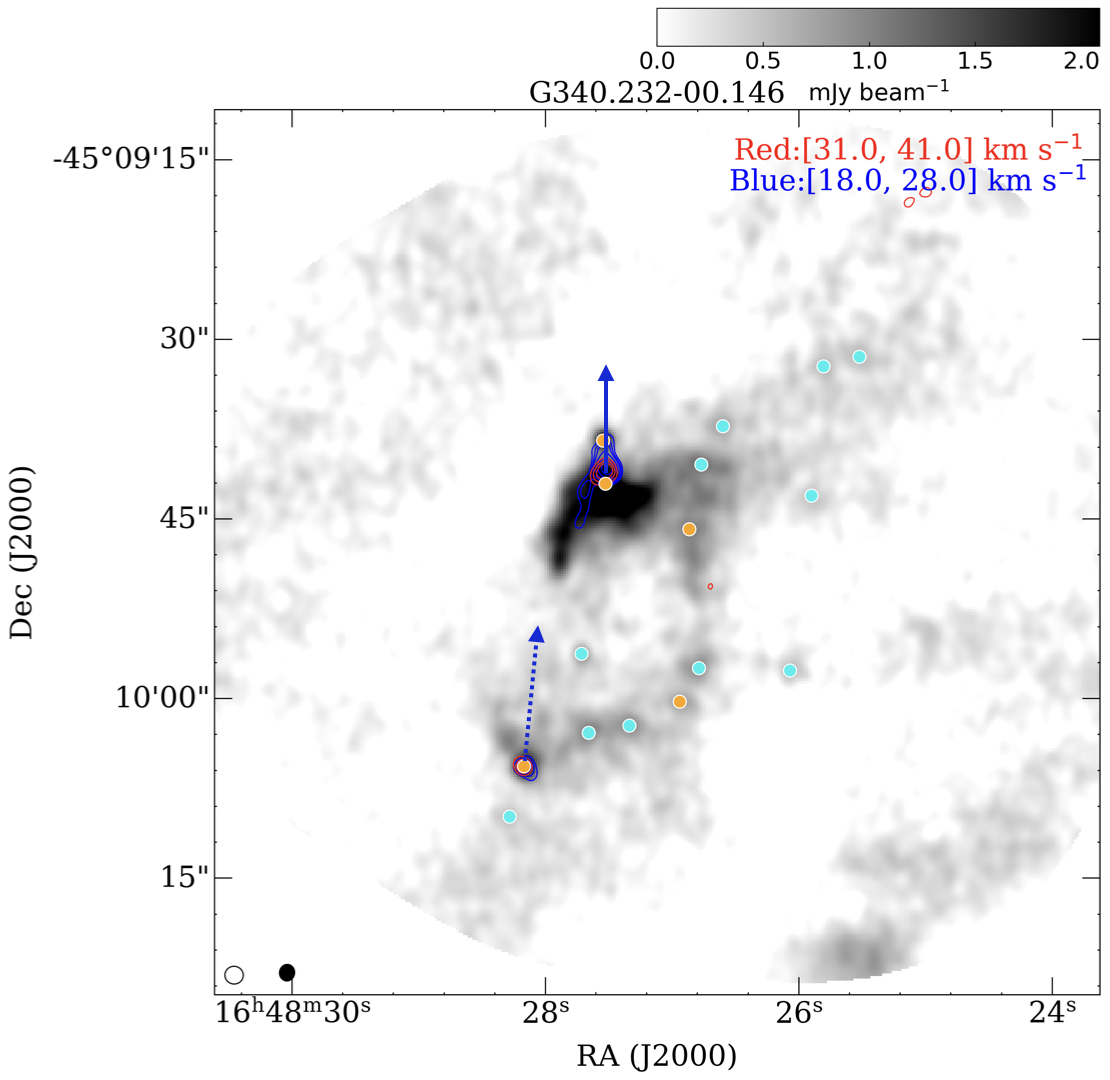}{0.4\textwidth}{}}
\vspace{-0.75cm}
\gridline{\fig{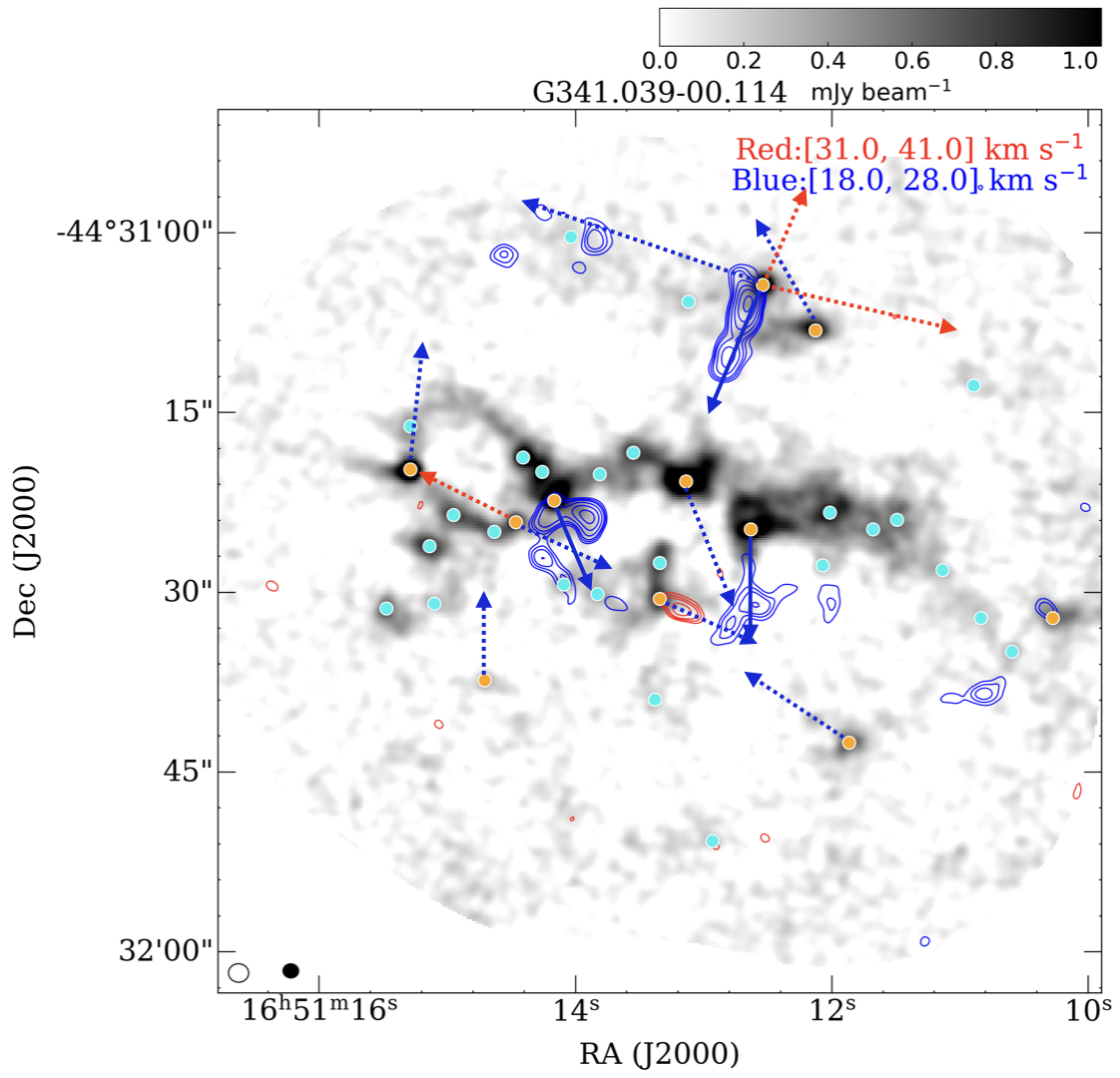}{0.4\textwidth}{}
}
\caption{
Blue- (blue contours) and red-shifted (red contours) components of H$_2$CO(3$_{0,3}$-3$_{0,2}$) for five IRDC clumps:
G028.273-00.167, G327.116-00.294, 331.372-00.116,
G340.232-00.146, and G341.039-00.114.
The gray-scale background shows the 1.3 mm dust continuum.
The cyan and orange circles indicate the positions of prestellar candidates and protostellar cores, respectively,
identified in ASHES \citep{Li2022}.
Synthesized beams are displayed at the bottom left of each panel
(open ellipses: H$_2$CO; filled ellipses: continuum).
The red and blue arrows indicate the outflow directions identified by CO and SiO emission lines in \citet{Li2020},
which investigates the outflow structure with the same ASHES data.
(solid arrows: also detected in H$_2$CO(3$_{0,3}$-3$_{0,2}$), dotted arrows: not detected in H$_2$CO(3$_{0,3}$-3$_{0,2}$)).
}
\label{A-WING} 
\end{figure*}

In this appendix, we report on the spatial distributions of IRDC clumps, which are not shown in the main text.
Figure \ref{A-MOM0} shows the integrated intensity distributions of the three transitions of H$_2$CO in 10 IRDC clumps.
Figure \ref{A-WING} presents the blue- and red-shifted components of H$_2$CO(3$_{0,3}$-3$_{0,2}$) in 5 IRDC clumps.
We do not show the images of G340.179-00.242 and G340.222-00.167 because blue- and red-shifted components are not detected (emission is less than 3$\sigma$)
in these two IRDC clumps.
Figure \ref{A-TEMP} shows the temperature distributions of 7 IRDC clumps.
We do not show the images of G332.969-00.029, G340.179-00.242, and G340.222-00.167 because 
H$_2$CO(3$_{2,2}$-2$_{2,1}$) and H$_2$CO(3$_{2,1}$-2$_{2,0}$) emission are not detected in these clumps. 

\begin{figure*}
\gridline{\fig{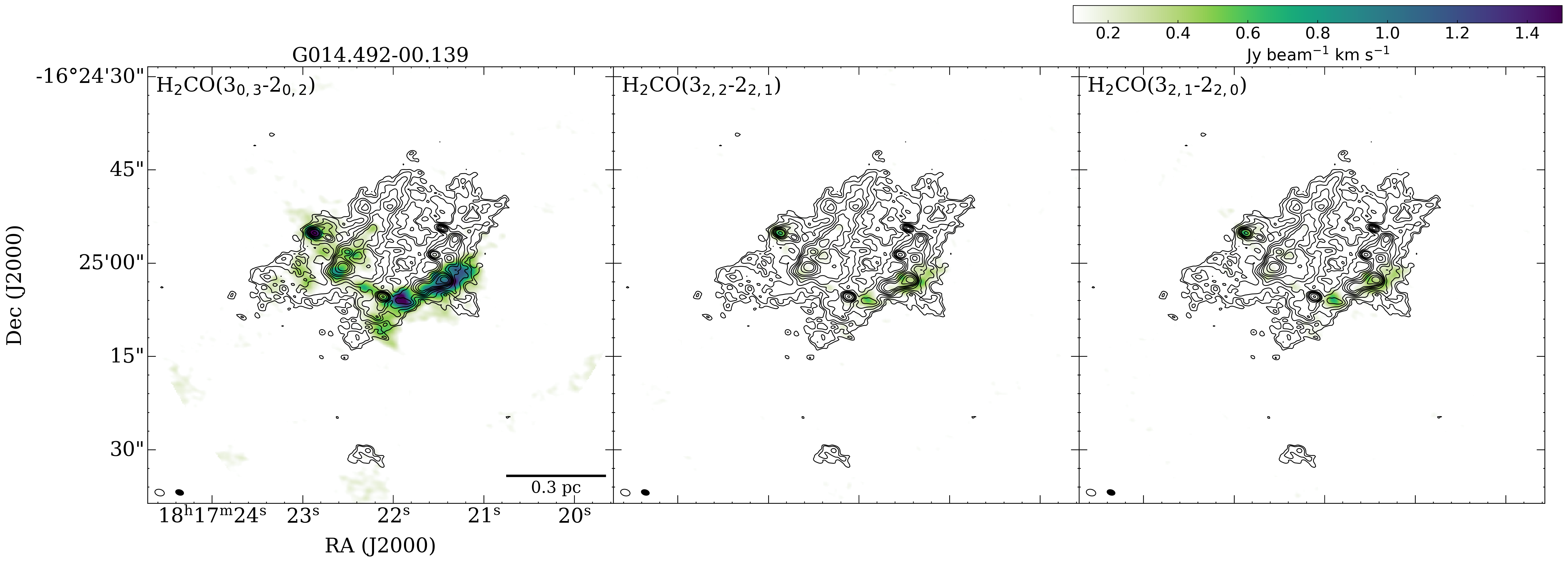}{0.95\textwidth}{}}
\vspace{-0.75cm}
\gridline{\fig{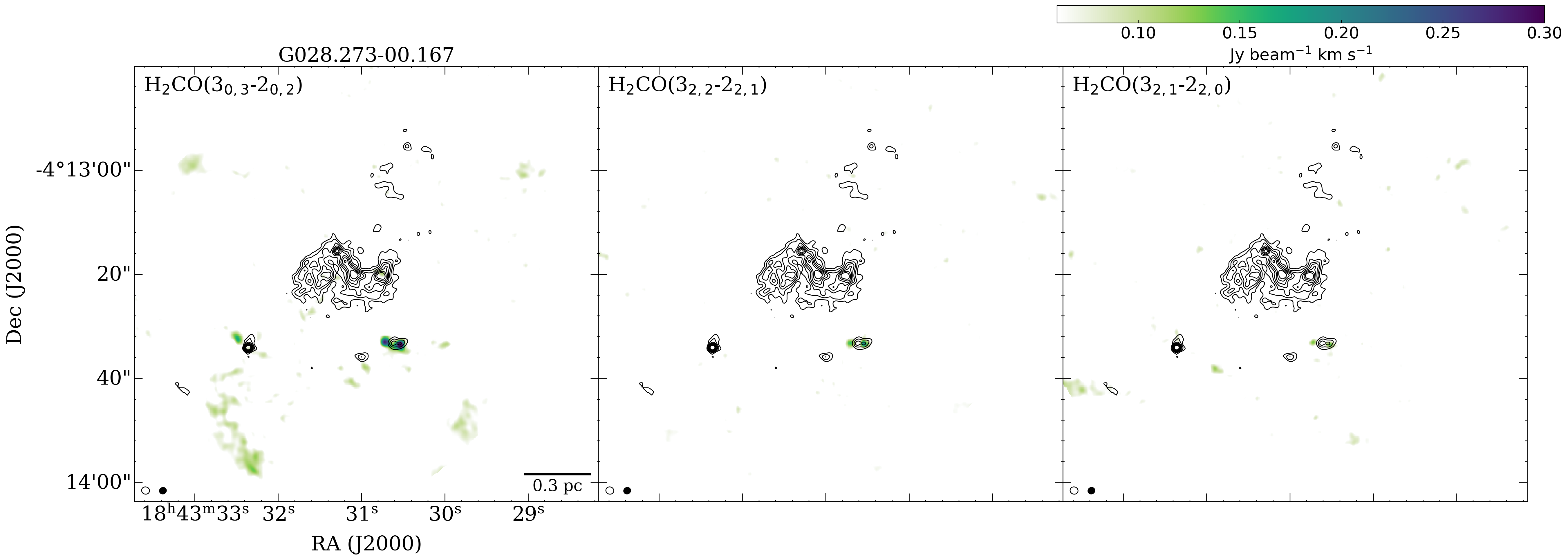}{0.95\textwidth}{}}
\vspace{-0.75cm}
\gridline{\fig{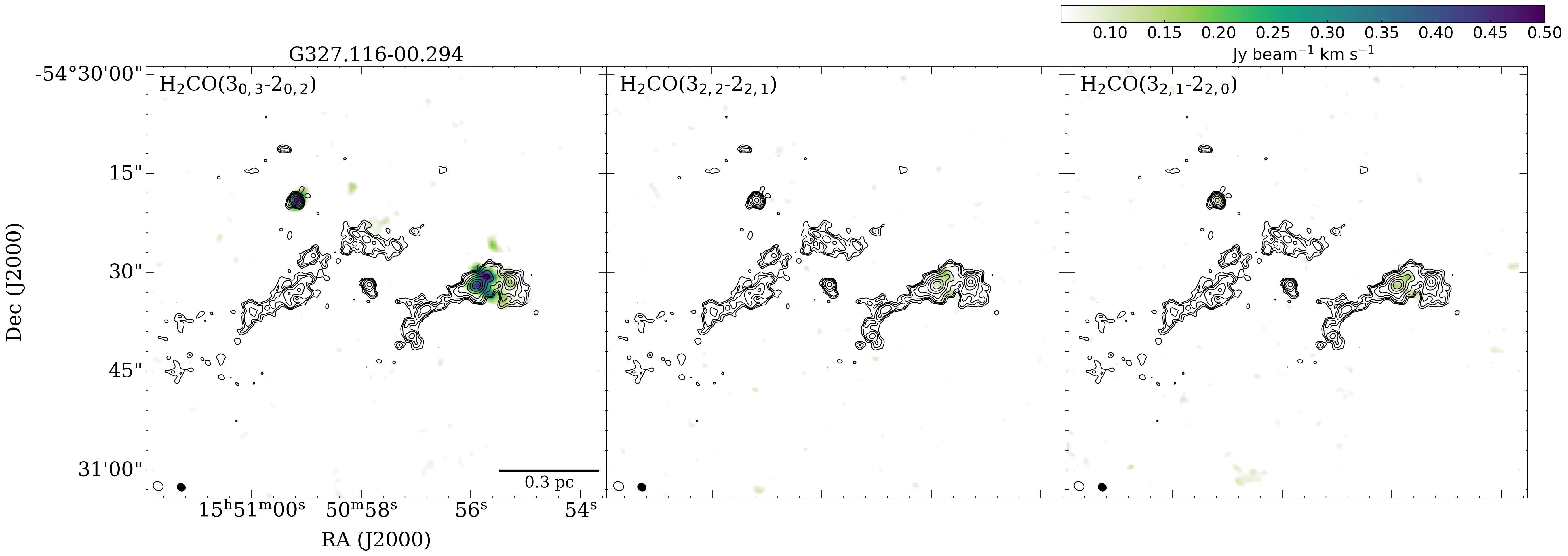}{0.95\textwidth}{}}
\vspace{-0.75cm}
\caption{
Integrated intensity distributions of three transitions of H$_2$CO
(Left: H$_2$CO(3$_{03}$-3$_{02}$),
Middle: H$_2$CO(3$_{22}$-3$_{21}$),
Right: H$_2$CO(3$_{21}$-3$_{20}$))
for 10 IRDC clumps: G014.492-00.139, G028.273-00.167, G327.116-00.294, G331.372-00.116, 
G332.969-00.029, G340.179-00.242, G340.222-00.167, G340.232-00.146, and G343.489-00.416.
The black contours show the 1.3mm dust continuum.
Contour levels are 3, 4, 6, 8, 10, 12, 15, 18, 25, and 35 $\times$ $\sigma$, with
$\sigma$ = 0.168 mJy beam$^{-1}$ for G014.492-00.139
(1$\farcs$1 angular resolution),
3, 4, 5, 7, 10, 14, and 20 $\times$ $\sigma$, with
$\sigma$ = 0.164 mJy beam$^{-1}$ for G028.273-00.167
(1$\farcs$2 angular resolution),
3, 4, 5, 7, 10, 15, 23, and 35 $\times$ $\sigma$, with
$\sigma$ = 0.089 mJy beam$^{-1}$ for G327.116-00.294
(1$\farcs$2 angular resolution),
3, 4, 5, 7, 9, 12 and 16 $\times$ $\sigma$, with $
\sigma$ = 0.083 mJy beam$^{-1}$ for G331.372-00.116
(1$\farcs$2 angular resolution),
3, 4, 5, 6, 7 and 8 $\times$ $\sigma$, with $
\sigma$ = 0.080 mJy beam$^{-1}$ for G332.969-00.029
(1$\farcs$1 angular resolution),
3, 4, 5 and 6 $\times$ $\sigma$, with $
\sigma$ = 0.094 mJy beam$^{-1}$ for G340.179-00.242
(1$\farcs$3 angular resolution),
3, 4, 5, 6, 7, 10, 14 and 18 $\times$ $\sigma$, with $
\sigma$ = 0.112 mJy beam$^{-1}$ for G340.222-00.167
(1$\farcs$3 angular resolution),
3, 4, 5, 7, 8, 11, 14, 18 and 23 $\times$ $\sigma$, with $
\sigma$ = 0.139 mJy beam$^{-1}$ for G340.232-00.146
(1$\farcs$3 angular resolution),
3, 4, 6, 9, 12, 16 and 22 $\times$ $\sigma$, with $
\sigma$ = 0.070 mJy beam$^{-1}$ for G341.039-00.114
(1$\farcs$2 angular resolution),
and 3, 4, 6, 8, 12, 20, 40 and 100 $\times$ $\sigma$, with $
\sigma$ = 0.068 mJy beam$^{-1}$ for G343.489-00.416
(1$\farcs$2 angular resolution).
Synthesized beams are displayed at the bottom left in each panel
(open ellipses: H$_2$CO, filled ellipses: continuum).
}
\label{A-MOM0} 
\end{figure*}
\begin{figure*}
\figurenum{\ref{A-MOM0}}
\gridline{\fig{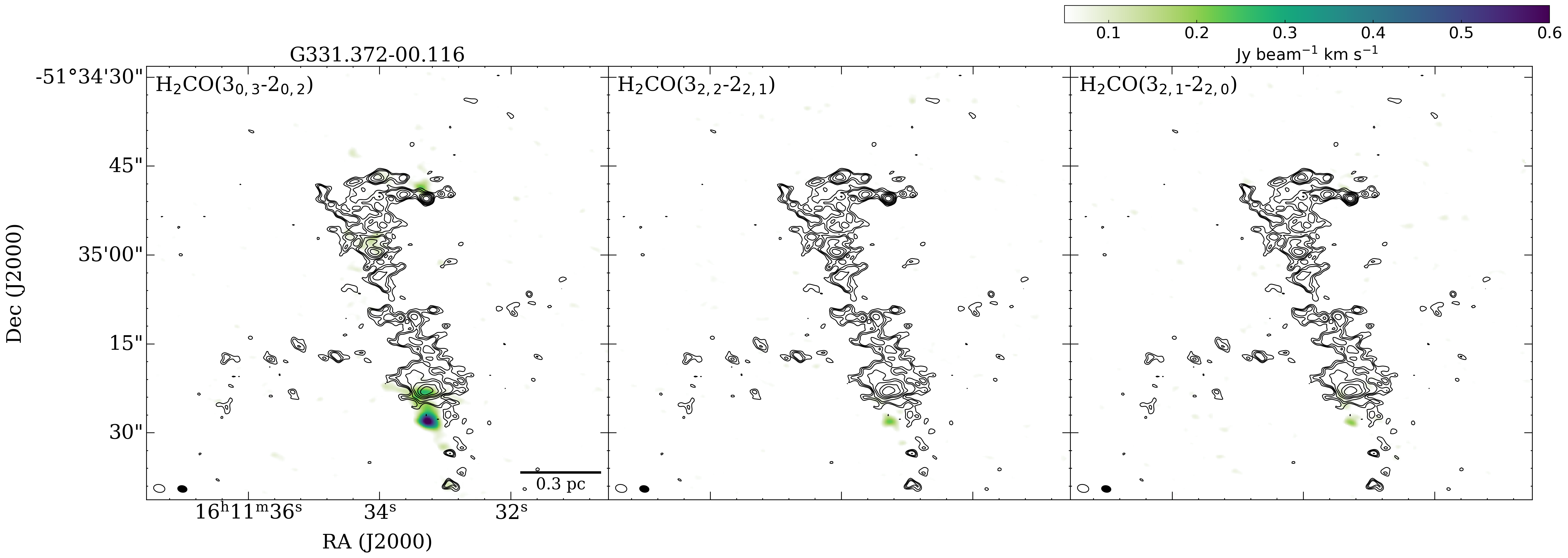}{0.95\textwidth}{}}
\vspace{-0.75cm}
\gridline{\fig{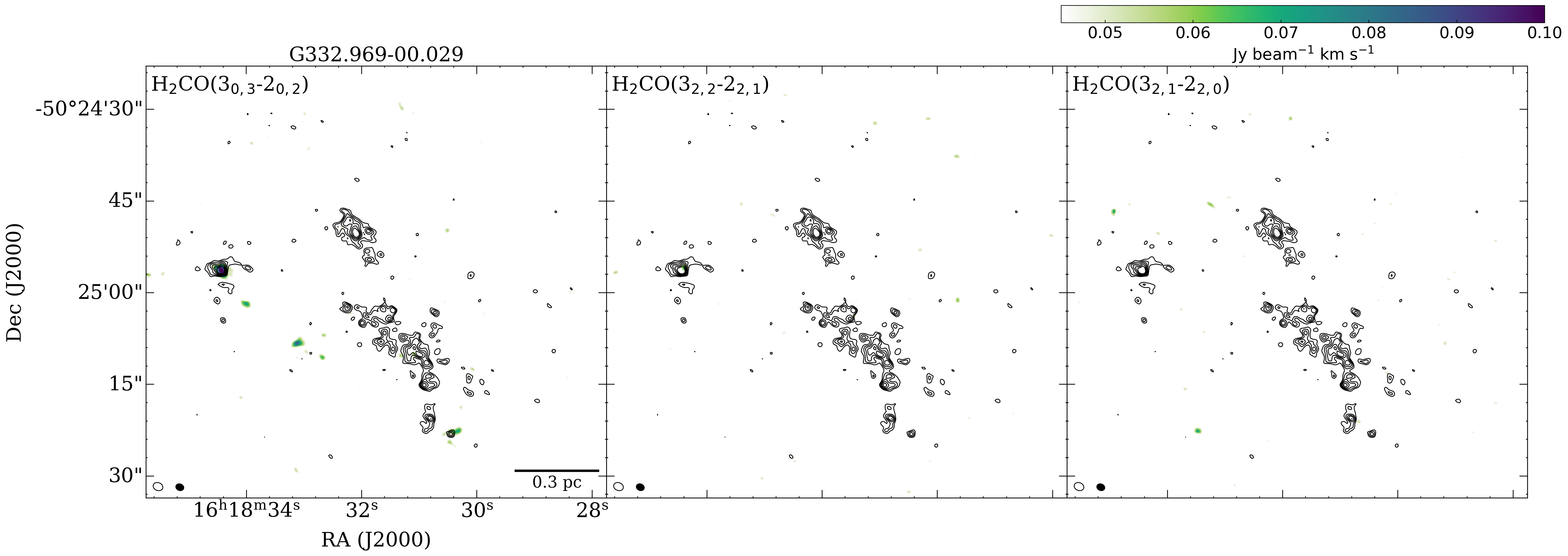}{0.95\textwidth}{}}
\vspace{-0.75cm}
\gridline{\fig{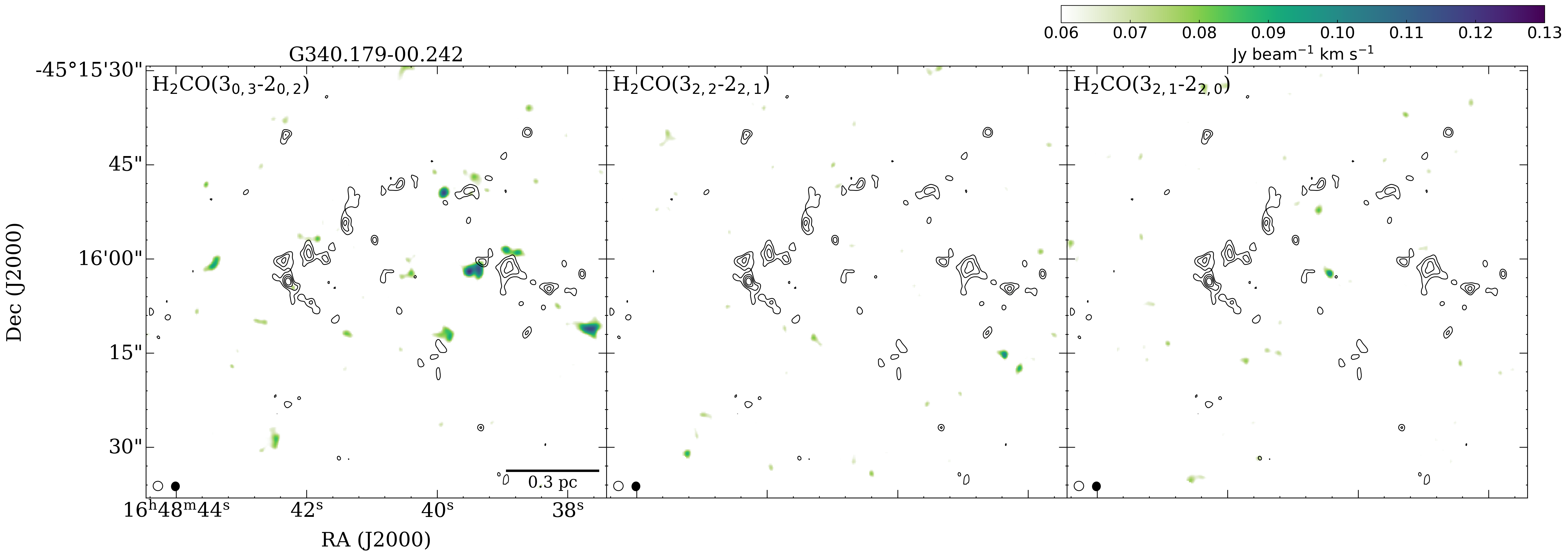}{0.95\textwidth}{}}
\vspace{-0.75cm}
\caption{
(Continued.)
}
\end{figure*}
\begin{figure*}
\figurenum{\ref{A-MOM0}}
\gridline{\fig{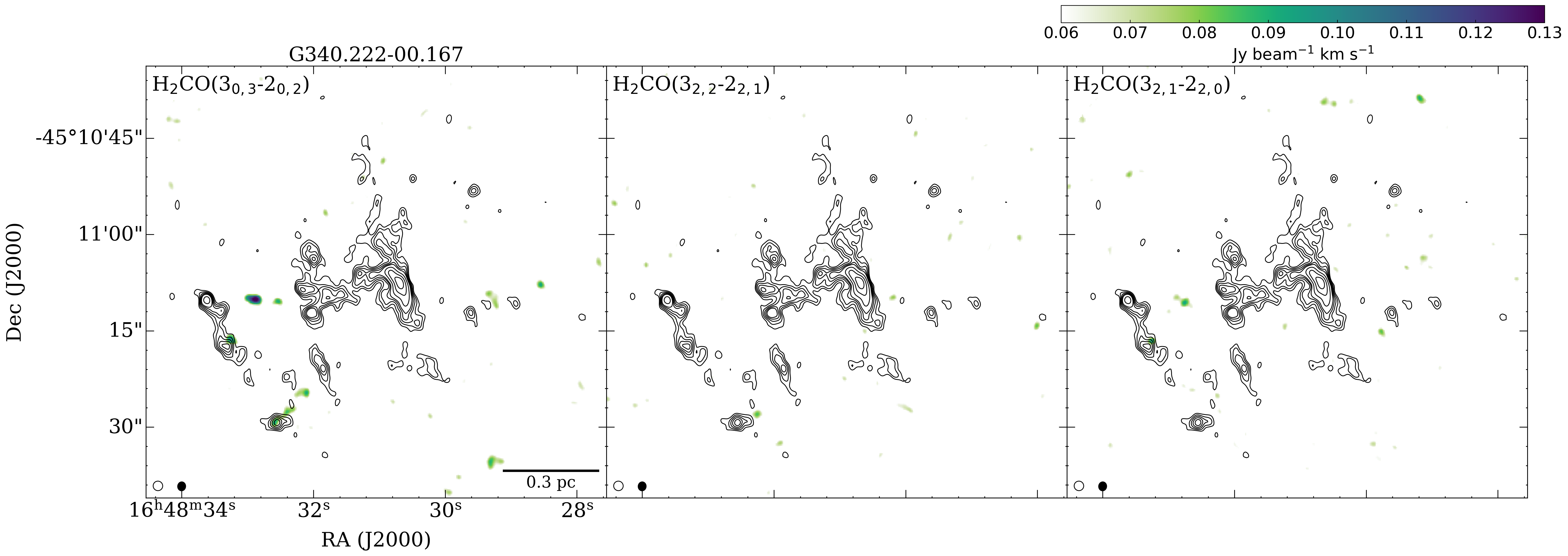}{0.95\textwidth}{}}
\vspace{-0.75cm}
\gridline{\fig{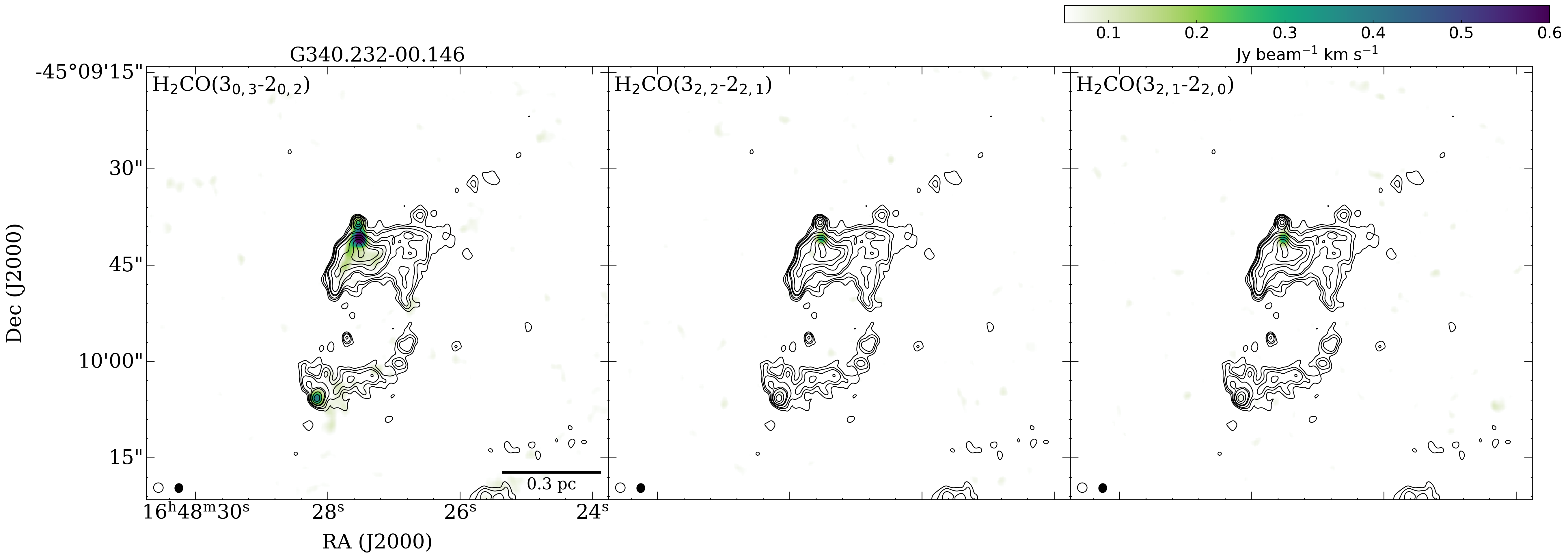}{0.95\textwidth}{}}
\vspace{-0.75cm}
\gridline{\fig{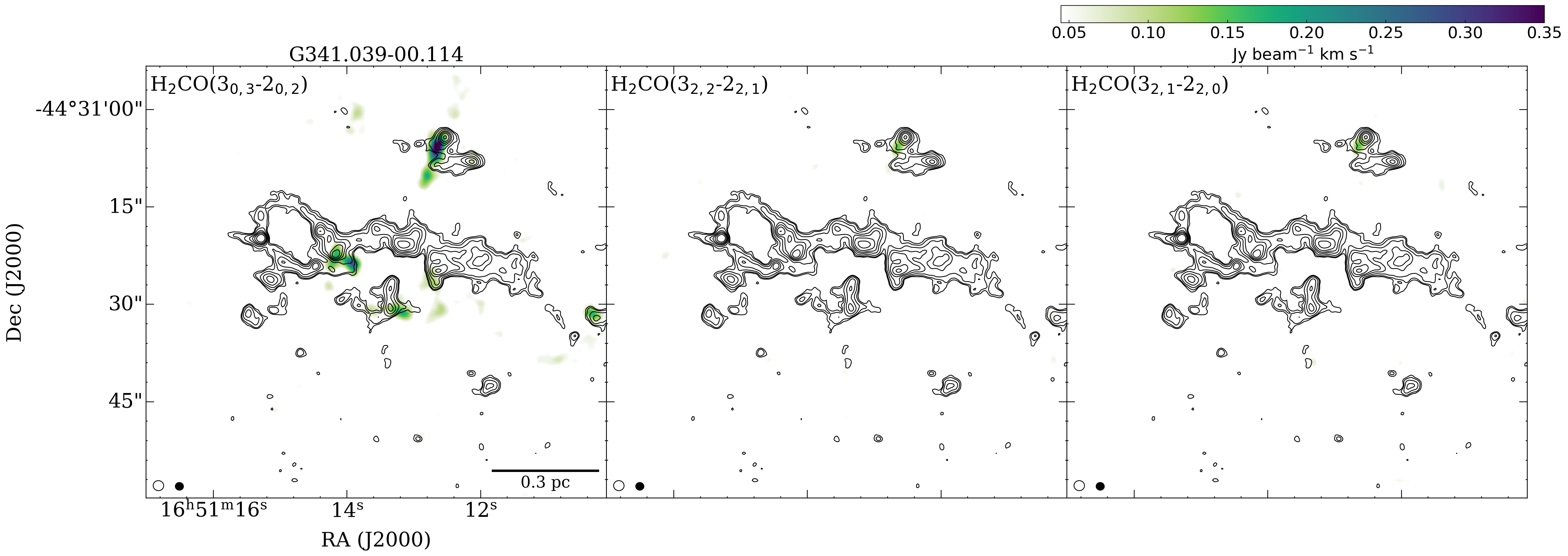}{0.95\textwidth}{}}
\vspace{-0.75cm}
\caption{
(Continued.)
}
\end{figure*}
\begin{figure*}
\figurenum{\ref{A-MOM0}}
\gridline{\fig{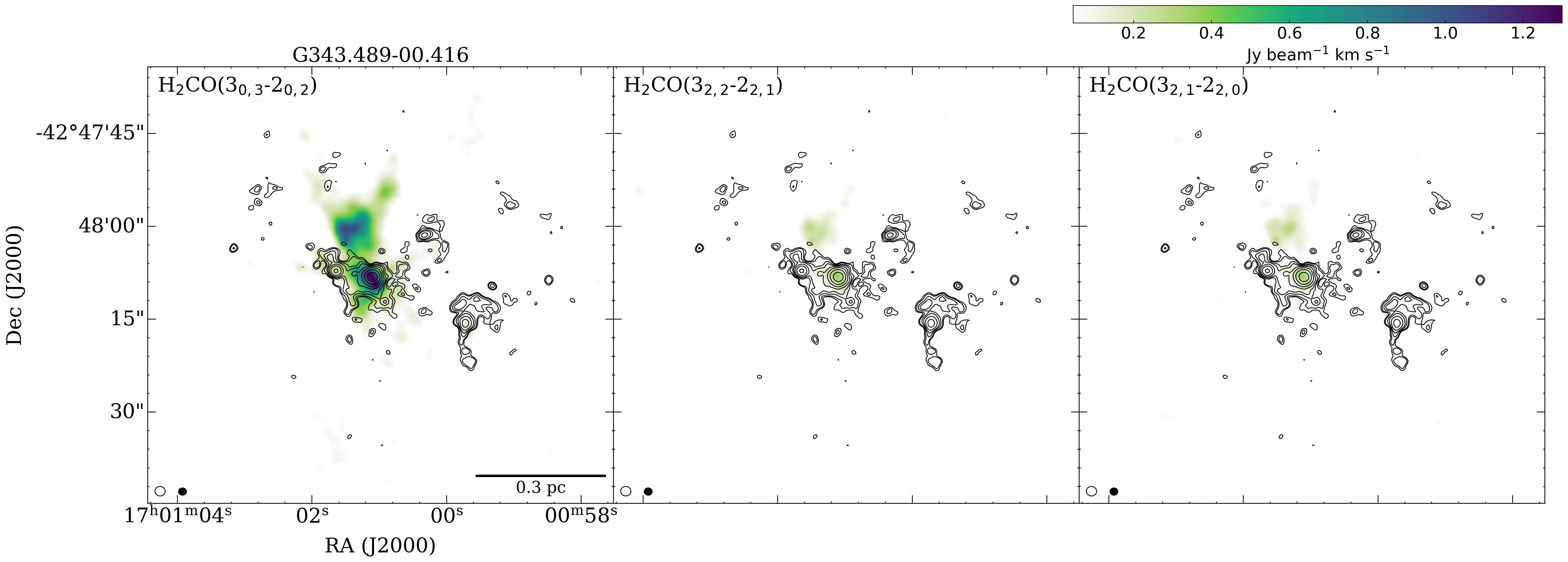}{0.95\textwidth}{}}
\vspace{-0.75cm}
\caption{
(Continued.)
}
\end{figure*}
\begin{figure*}
\gridline{\fig{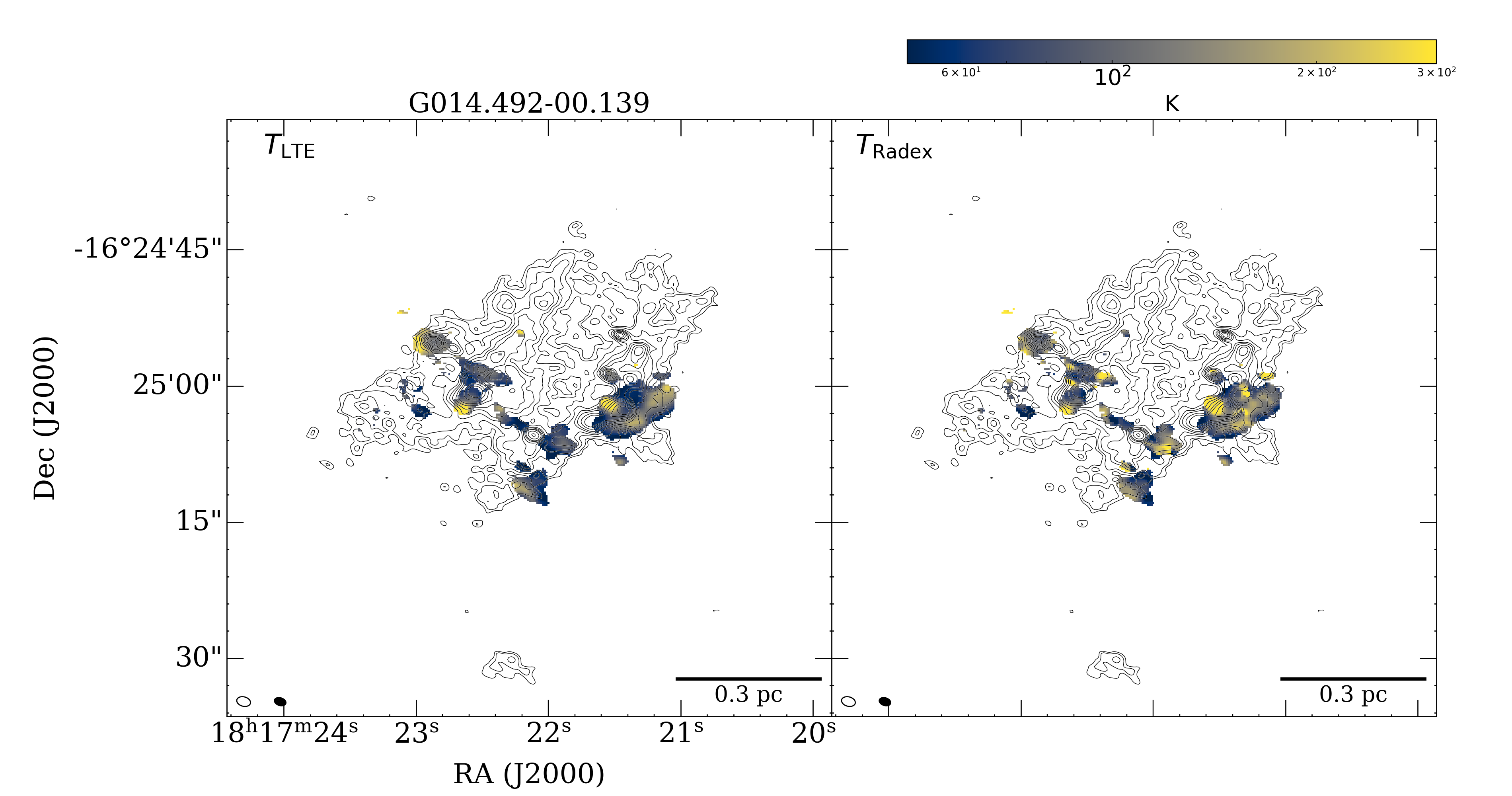}{0.9\textwidth}{}}
\vspace{-0.75cm}
\gridline{\fig{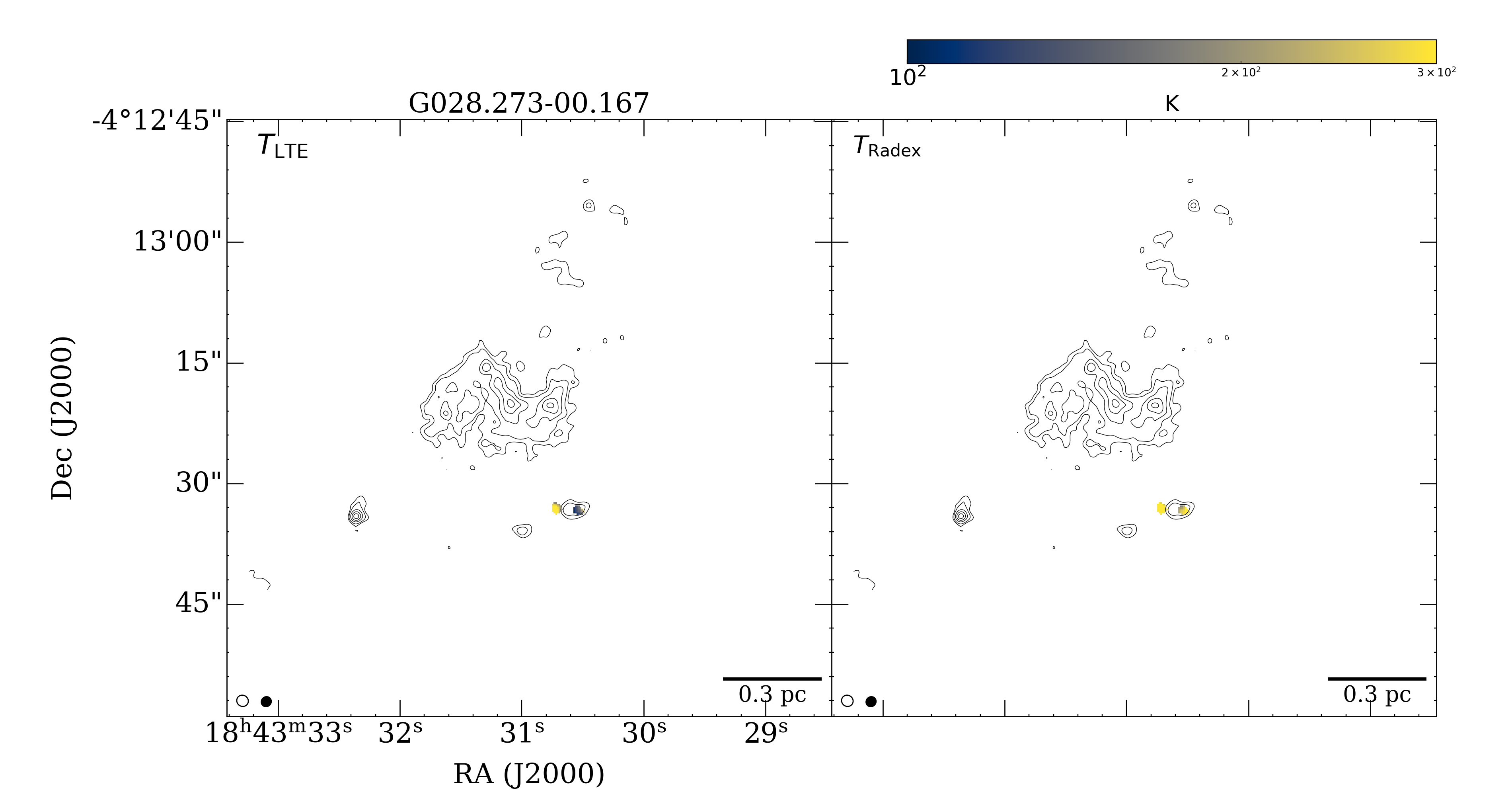}{0.9\textwidth}{}}
\vspace{-0.75cm}
\caption{
Kinetic temperature distribution (Left: LTE, Right: Radex) for 7 IRDC clumps:
G014.492-00.139, G028.273-00.167, G327.116-00.294, 331.372-00.116,
G340.232-00.146, G341.039-00.114, and G343.489-00.416.
The black contours are the 1.3mm dust continuum.
Contour levels are 3, 4, 6, 8, 10, 12, 15, 18, 25, and 35 $\times$ $\sigma$, with
$\sigma$ = 0.168 mJy beam$^{-1}$ for G014.492-00.139
(1$\farcs$1 angular resolution),
3, 4, 5, 7, 10, 14, and 20 $\times$ $\sigma$, with
$\sigma$ = 0.164 mJy beam$^{-1}$ for G028.273-00.167
(1$\farcs$2 angular resolution),
3, 4, 5, 7, 10, 15, 23, and 35 $\times$ $\sigma$, with
$\sigma$ = 0.089 mJy beam$^{-1}$ for G327.116-00.294
(1$\farcs$2 angular resolution),
3, 4, 5, 7, 9, 12 and 16 $\times$ $\sigma$, with $
\sigma$ = 0.083 mJy beam$^{-1}$ for G331.372-00.116
(1$\farcs$2 angular resolution),
3, 4, 5, 7, 8, 11, 14, 18 and 23 $\times$ $\sigma$, with $
\sigma$ = 0.139 mJy beam$^{-1}$ for G340.232-00.146
(1$\farcs$3 angular resolution),
3, 4, 6, 9, 12, 16 and 22 $\times$ $\sigma$, with $
\sigma$ = 0.070 mJy beam$^{-1}$ for G341.039-00.114
(1$\farcs$2 angular resolution),
and 3, 4, 6, 8, 12, 20, 40 and 100 $\times$ $\sigma$, with $
\sigma$ = 0.068 mJy beam$^{-1}$ for G343.489-00.416
(1$\farcs$2 angular resolution).
Synthesized beams are displayed at the bottom left in each panel
(open ellipses: H$_2$CO, filled ellipses: continuum).
}
\label{A-TEMP} 
\end{figure*}
\begin{figure*}
\figurenum{\ref{A-TEMP}}
\gridline{\fig{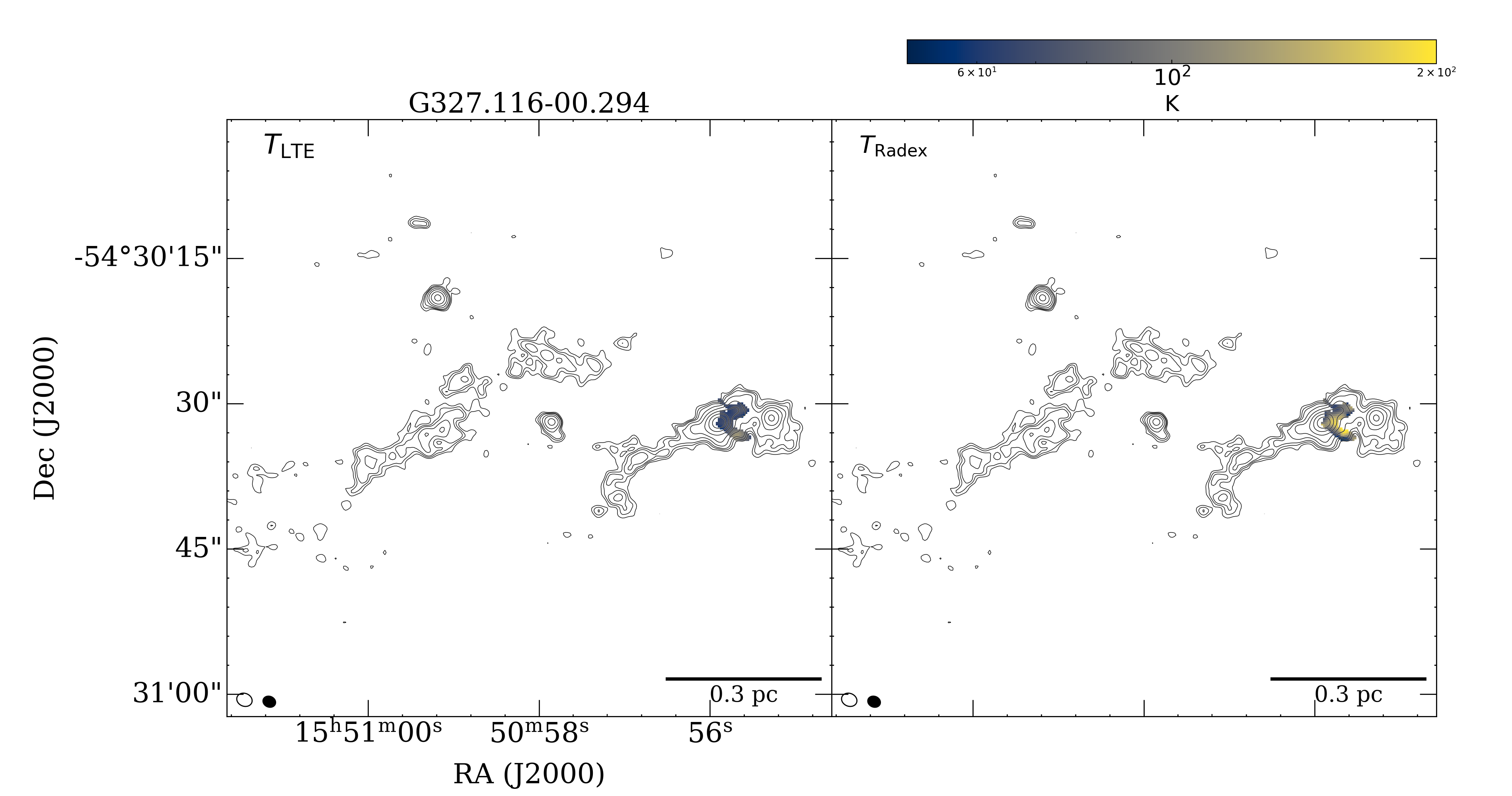}{0.9\textwidth}{}}
\vspace{-0.75cm}
\gridline{\fig{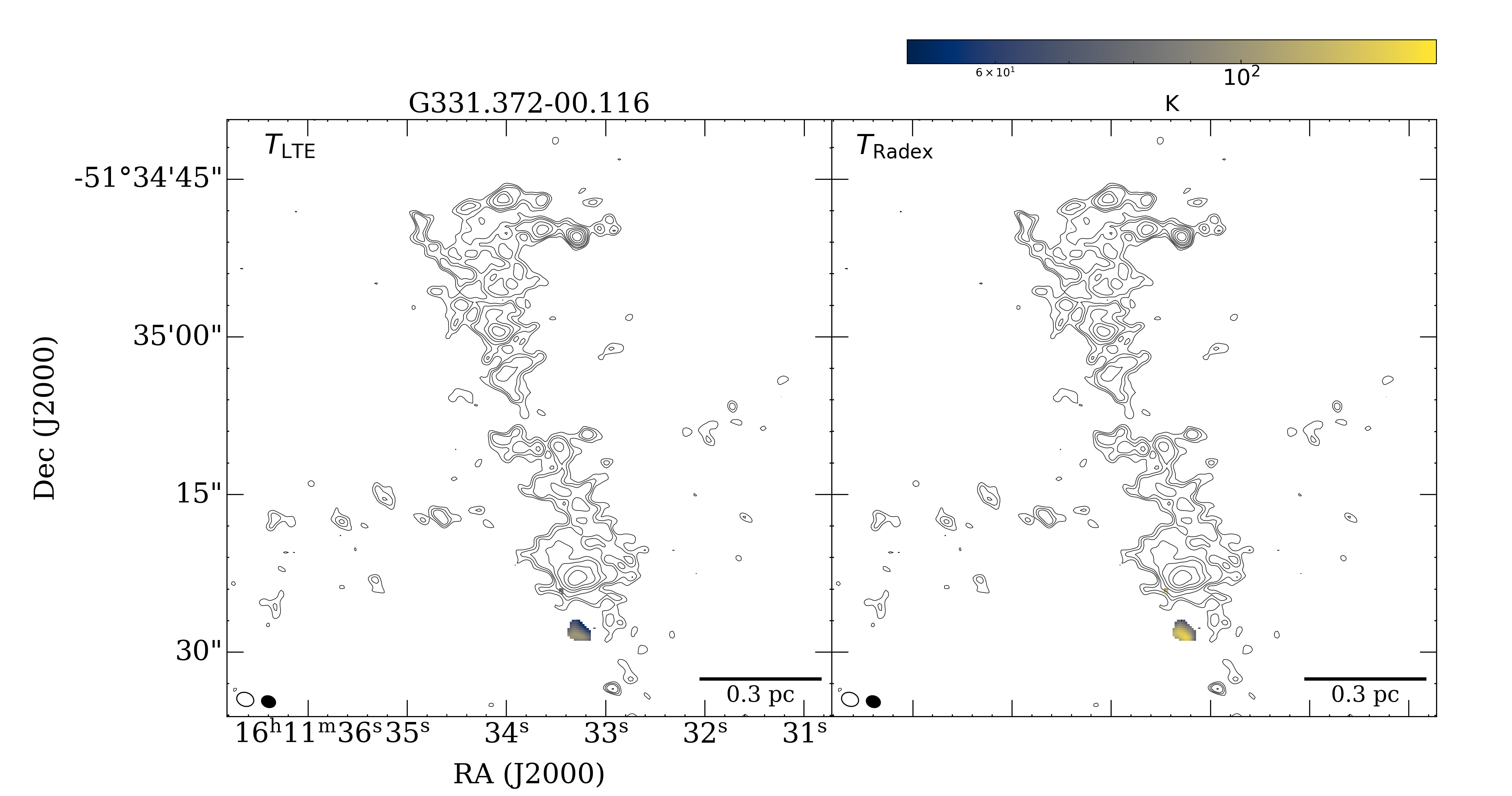}{0.9\textwidth}{}}
\vspace{-0.75cm}
\caption{
(Continued.)
}
\end{figure*}
\begin{figure*}
\figurenum{\ref{A-TEMP}}
\gridline{\fig{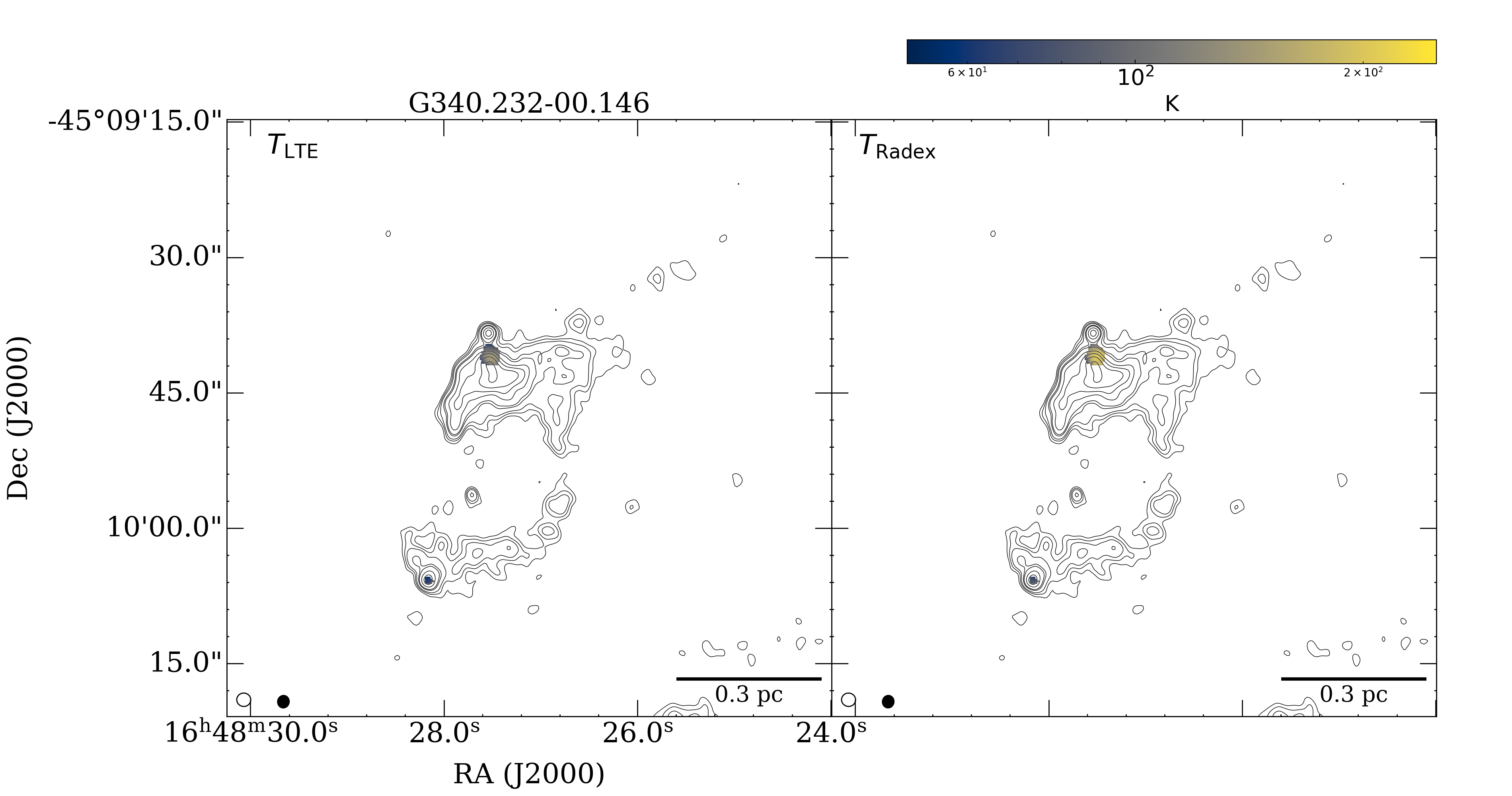}{0.9\textwidth}{}}
\vspace{-0.75cm}
\gridline{\fig{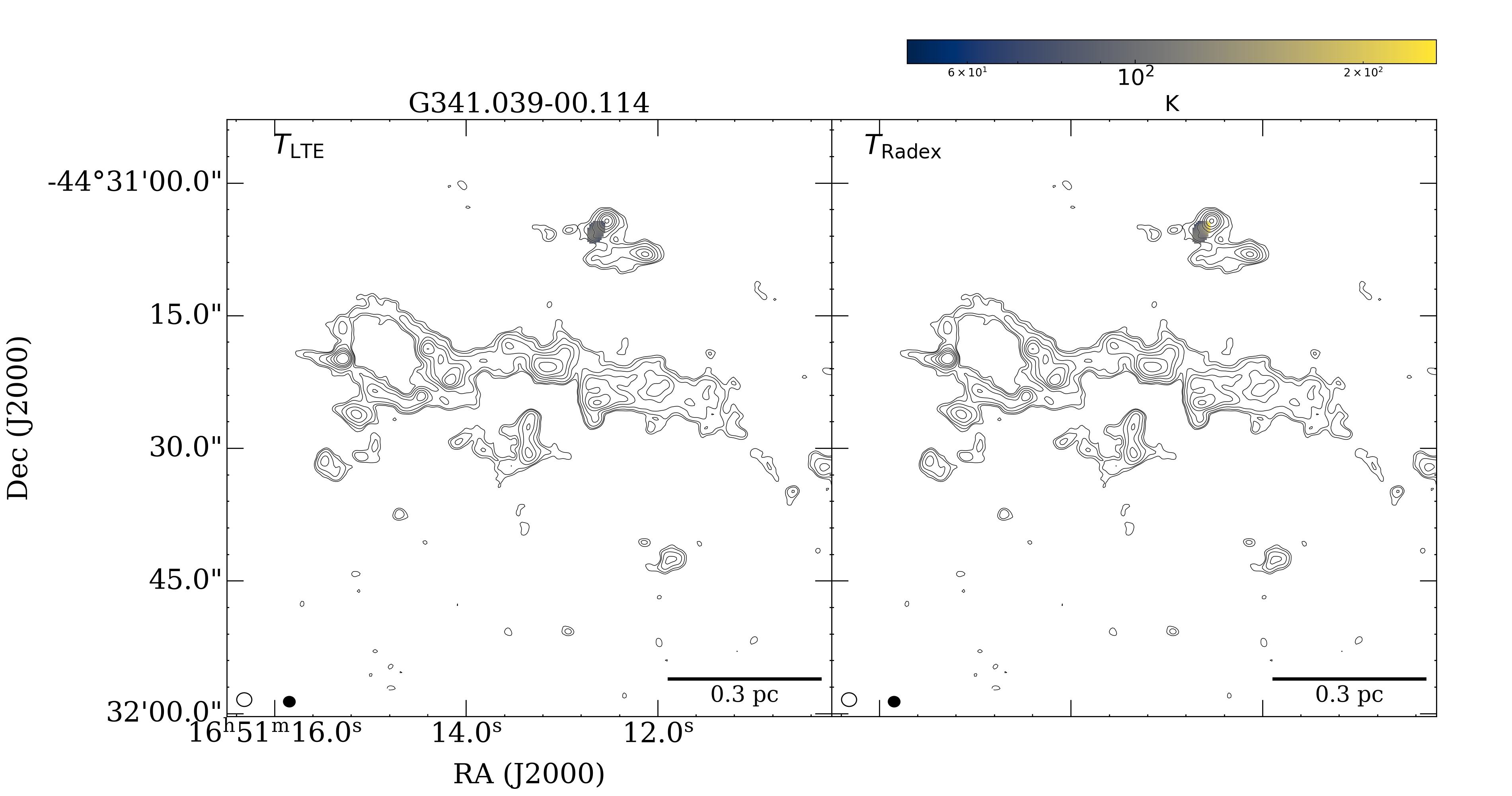}{0.9\textwidth}{}}
\vspace{-0.75cm}
\caption{
(Continued.)
}
\end{figure*}
\begin{figure*}
\figurenum{\ref{A-TEMP}}
\gridline{\fig{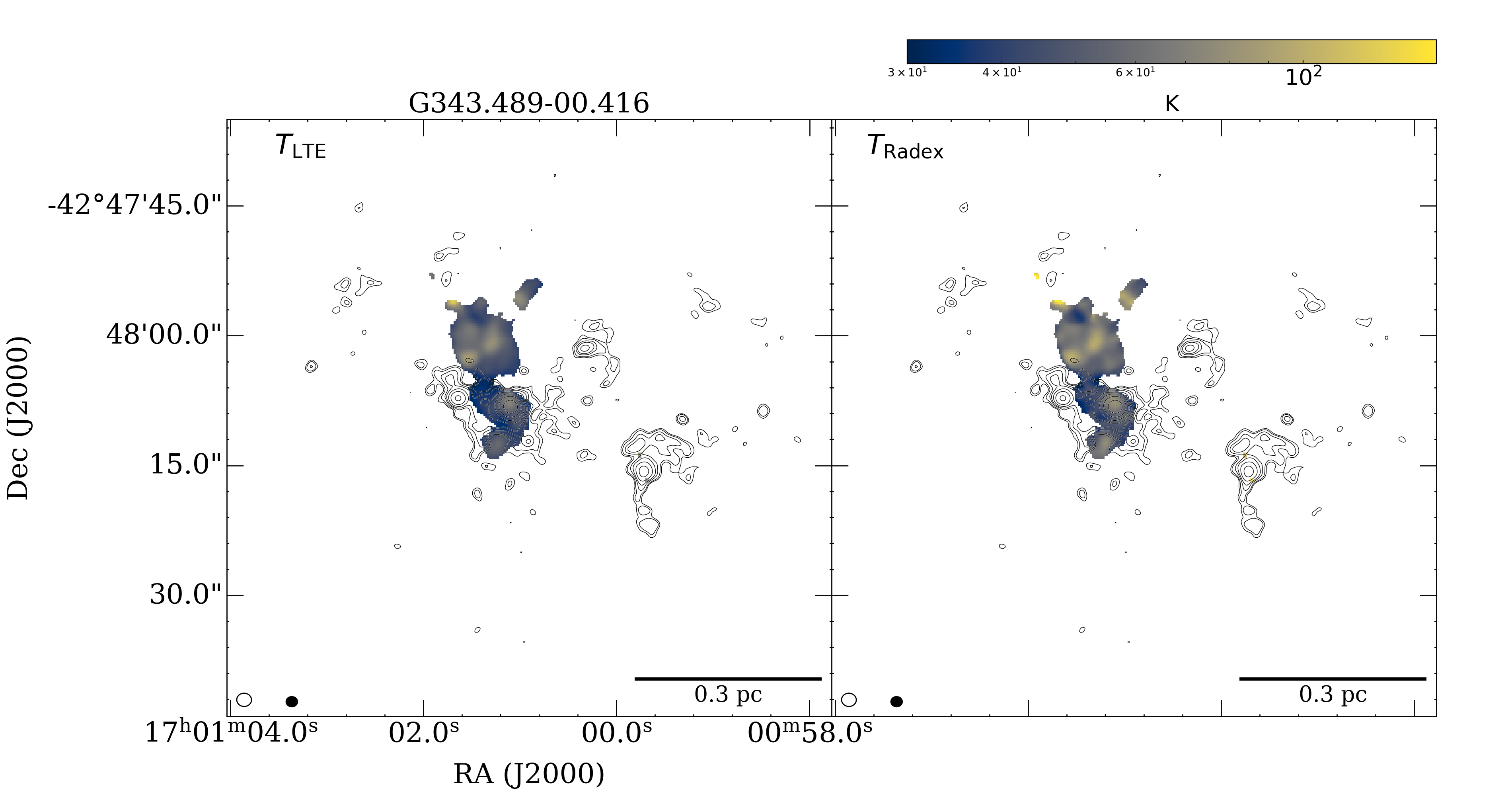}{0.9\textwidth}{}}
\vspace{-0.75cm}
\caption{
(Continued.)
}
\end{figure*}

\section{Radex modeling}\label{sec:a-2}
\begin{figure*}
\gridline{\fig{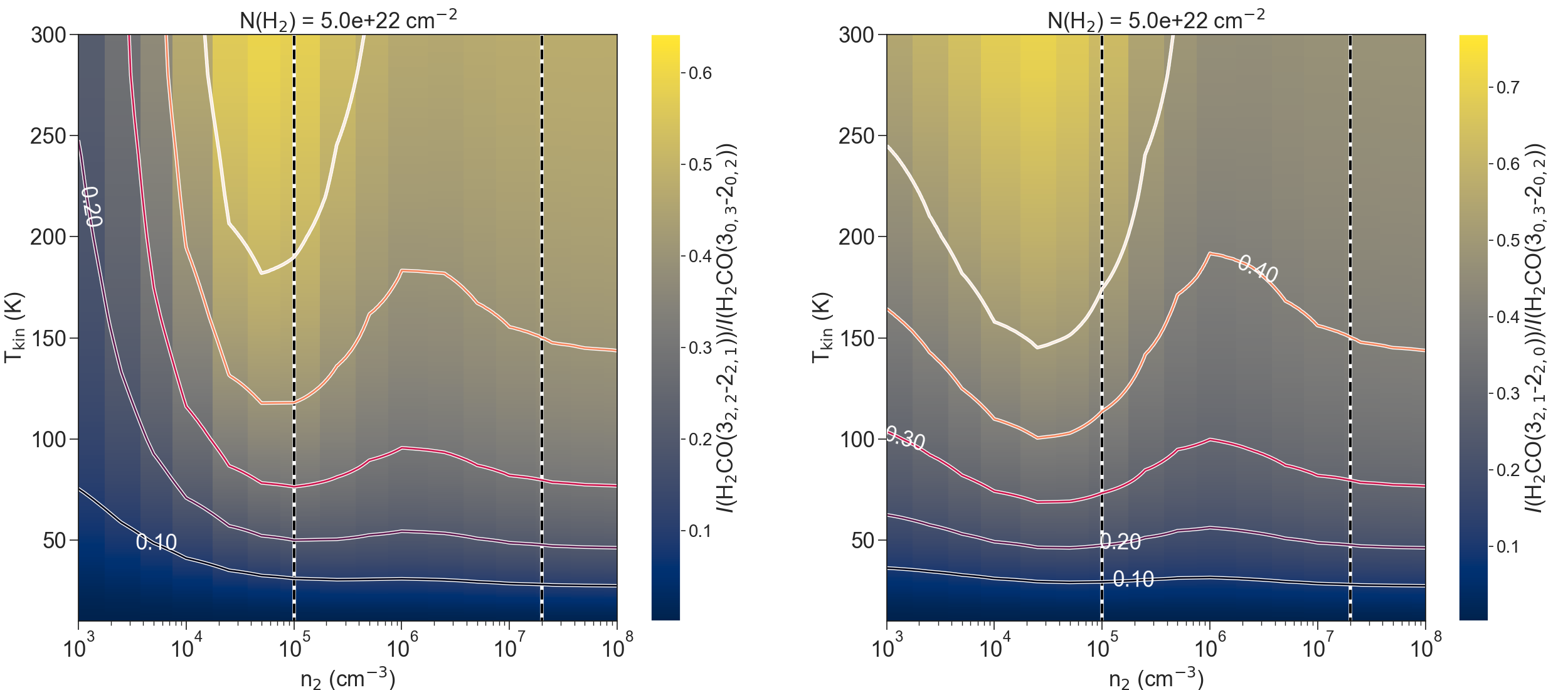}{0.95\textwidth}{}}
\vspace{-0.75cm}
\gridline{\fig{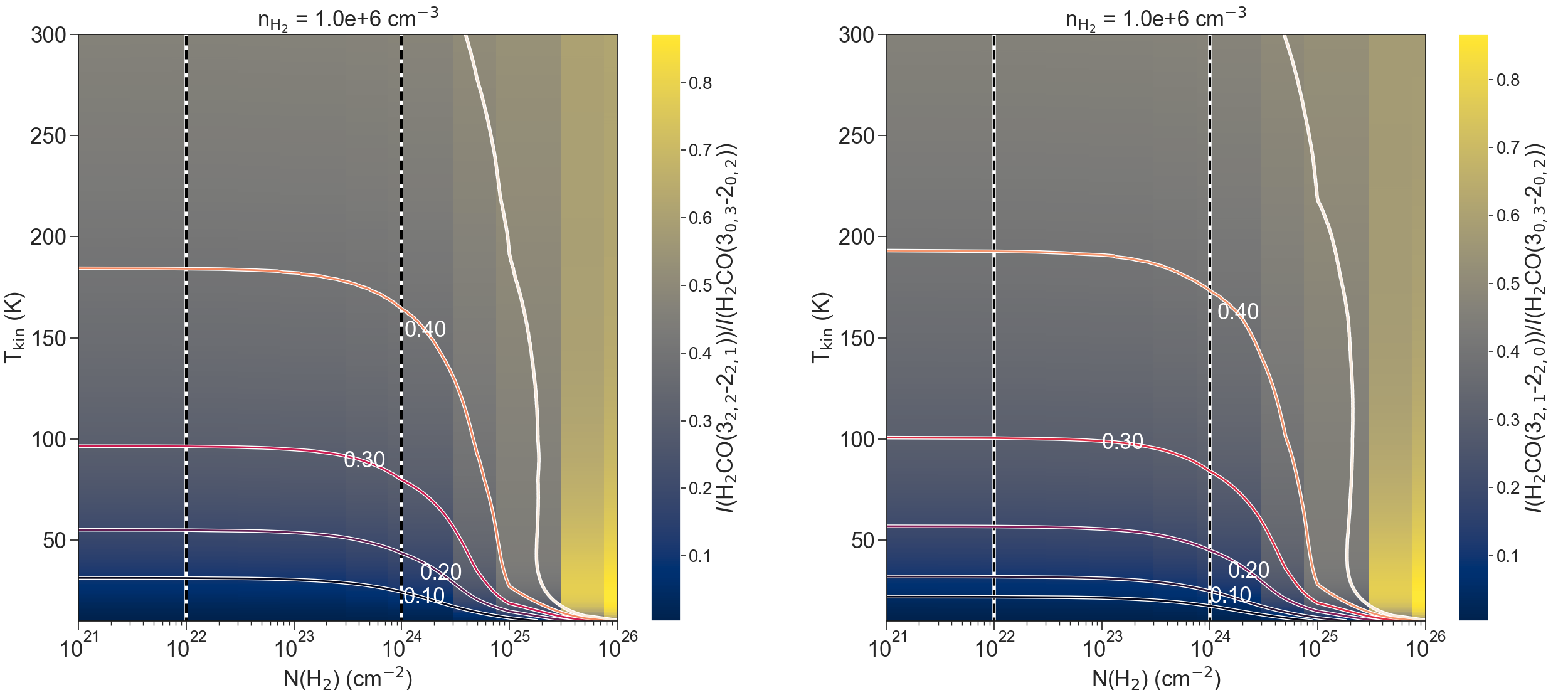}{0.95\textwidth}{}}
\caption{
Example of Radex modeling.
Top: 
H$_2$CO(3$_{2,2}$-2$_{2,1}$)/H$_2$CO(3$_{0,3}$-2$_{0,2}$) (left)
and H$_2$CO(3$_{2,1}$-2$_{2,0}$)/H$_2$CO(3$_{0,3}$-2$_{0,2}$) (right) integrated intensity ratios
as a function of $n_{\rm H_2}$ and $T_{\rm kin}$ assuming that N(H$_2$) = 5.0 $\times$ 10$^{22}$ cm$^{-2}$.
Bottom:
H$_2$CO(3$_{2,2}$-2$_{2,1}$)/H$_2$CO(3$_{0,3}$-2$_{0,2}$) (left)
and H$_2$CO(3$_{2,1}$-2$_{2,0}$)/H$_2$CO(3$_{0,3}$-2$_{0,2}$) (right) integrated intensity ratios
as a function of N(H$_2$) and $T_{\rm kin}$ assuming that $n_{\rm H_2}$ = 1.0 $\times$ 10$^{6}$ cm$^{-3}$.
The black dotted lines indicate possible $n_{\rm H_2}$ and N(H$_2$) ranges:
$n_{\rm H_2}$ = 10$^5$ -- 2.0 $\times$ 10$^7$ cm$^{-3}$ (top)
and 
N(H$_2$) = 10$^{22}$ -- 10$^{24}$ cm$^{-2}$ (bottom).
}
\label{Radex-Model} 
\end{figure*}

In this appendix, we illustrate the Radex modeling.
Figure \ref{Radex-Model} displays an example of the relation between 
$n_{\rm H_2}$, N(H$_2$), $T_{\rm kin}$, H$_2$CO(3$_{2,2}$-2$_{2,1}$)/H$_2$CO(3$_{0,3}$-2$_{0,2}$),
and H$_2$CO(3$_{2,1}$-2$_{2,0}$)/H$_2$CO(3$_{0,3}$-2$_{0,2}$) integrated intensity ratio.
This figure indicates that
H$_2$CO(3$_{2,1}$--2$_{2,0}$)/H$_2$CO(3$_{0,3}$--2$_{0,2}$) ratio
is more sensitive to densities than H$_2$CO(3$_{2,2}$--2$_{2,1}$)/H$_2$CO(3$_{0,3}$--2$_{0,2}$) ratio \citep{Ao2013}
and the effect from the assumption of the N(H$_2$) is negligible compared to the assumption of the n(H$_2$) for the Radex modeling.

\bibliographystyle{aasjournal}
\bibliography{Ref}

\end{document}